\documentclass[lettersize,journal]{IEEEtran}
\usepackage{amsmath,amsfonts}
\usepackage{algorithmic}
\usepackage{array}
\usepackage[caption=false,font=footnotesize,labelfont=rm,textfont=rm]{subfig}
\usepackage{textcomp}
\usepackage{stfloats}
\usepackage{url}
\usepackage{verbatim}
\usepackage{graphicx}
\def\BibTeX{{\rm B\kern-.05em{\sc i\kern-.025em b}\kern-.08em
    T\kern-.1667em\lower.7ex\hbox{E}\kern-.125emX}}
\usepackage{balance}
\usepackage{enumitem}

\usepackage[tikz]{bclogo}
\usepackage[edges]{forest}
\usepackage[framemethod=tikz]{mdframed}
\definecolor{bgblue}{RGB}{245,243,253}
\definecolor{ttblue}{RGB}{91,194,224}

\definecolor{vision_color}{cmyk}{0.45,0.2,0,0}
\definecolor{lang_color}{cmyk}{0,0.37,0.82,0}
\definecolor{action_color}{cmyk}{0.36,0,0.67,0}

\usepackage{amssymb}
\usepackage{makecell}
\usepackage[figuresright]{rotating}
\usepackage{graphicx}
\usepackage{multirow}

\begin{document}

\title{A Survey on DRL based UAV Communications and Networking: DRL Fundamentals, Applications and Implementations} 

\author{Wei Zhao, \IEEEmembership{Member,~IEEE}, Shaoxin Cui, Wen Qiu, \IEEEmembership{Member,~IEEE}, Zhiqiang He, \IEEEmembership{Student Member,~IEEE}, Zhi Liu, \IEEEmembership{Senior Member,~IEEE}, Xiao Zheng, Bomin Mao, \IEEEmembership{Senior Member,~IEEE}, and Nei Kato, \IEEEmembership{Fellow,~IEEE}
       
\thanks{Wei Zhao, Shaoxin Cui, Wen Qiu, Zhiqiang He and Xiao Zheng are with School of Computer Science and Technology, Anhui University of Technology, China. Emails: zhaowei@ahut.edu.cn, cuihshaoxin6@gmail.com, clorisqiu1@gmail.com, tinyzqh@gmail.com, 
xzheng@ahut.edu.cn.}
    
\thanks{Zhi Liu is with the Department of Computer and Network Engineering, the University of Electro-Communications, Japan. Email: liuzhi@uec.ac.jp.}

\thanks{Bomin Mao is  with the School of Cybersecurity, Northwestern Polytechnical University, China. Email: maobomin@nwpu.edu.cn.}
  
\thanks{Nei Kato is with  Graduate School of Information Sciences, Tohoku University, Sendai, Japan. Email: kato@it.is.tohoku.ac.jp.}

\thanks{The corresponding authors are Qiu Wen and Zhiqiang He.}
  }





\maketitle

\begin{abstract}
Unmanned aerial vehicles (UAVs) are playing an increasingly pivotal role in modern communication networks, offering flexibility and enhanced coverage for a variety of applications. However, UAV networks pose significant challenges due to their dynamic and distributed nature, particularly when dealing with tasks such as power allocation, channel assignment, caching, and task offloading. Traditional optimization techniques often struggle to handle the complexity and unpredictability of these environments, leading to suboptimal performance.
This survey provides a comprehensive examination of how deep reinforcement learning (DRL) can be applied to solve these mathematical optimization problems in UAV communications and networking. Rather than simply introducing DRL methods, the focus is on demonstrating how these methods can be utilized to solve complex mathematical models of the underlying problems. We begin by reviewing the fundamental concepts of DRL, including value-based, policy-based, and actor-critic approaches. Then, we illustrate how DRL algorithms are applied to specific UAV network tasks by discussing from problem formulations to DRL implementation.
By framing UAV communication challenges as optimization problems, this survey emphasizes the practical value of DRL in dynamic and uncertain environments. We also explore the strengths of DRL in handling large-scale network scenarios and the ability to continuously adapt to changes in the environment. In addition, future research directions are outlined, highlighting the potential for DRL to further enhance UAV communications and expand its applicability to more complex, multi-agent settings.
\end{abstract}


\section{Introduction}
\subsection{Background and Motivation}
\IEEEPARstart{T}{he} rapid advancement of UAVs has significantly expanded their use across various industries, including logistics, surveillance, disaster recovery, and 5/6G wireless communication \cite{10415630,10174680}. UAVs offer unprecedented flexibility in delivering services in remote and inaccessible areas, where traditional infrastructure is either unavailable or insufficient. Their ability to provide on-demand communication networks, especially for areas lacking conventional infrastructure, has positioned UAVs as crucial enablers in wireless communication systems \cite{10416899}.

Despite these advancements, managing UAV communication and networking remains highly challenging due to: (1) dynamic network topology: UAVs are highly mobile, leading to constantly changing network structures. The movement of UAVs necessitates continuous adaptation to varying communication conditions \cite{10341311}; (2) resource constraints: UAVs have limited battery life, processing power, and bandwidth, which need to be efficiently managed to maintain optimal communication performance \cite{10293163,liu2021robust}; (3) real-time decision making: UAVs must handle tasks such as resource allocation, channel assignment, power control, and task offloading in real-time, under dynamic and uncertain conditions \cite{10292756}.
Addressing these challenges often involves solving complex mathematical optimization problems. For example, determining optimal power allocation among UAVs, selecting the best communication channels, and managing data caching efficiently are all critical problems that require effective optimization strategies \cite{10680055}. In the context of 5/6G, these tasks become even more complex due to the need to manage dense and heterogeneous networks, as well as coordinate across diverse frequency bands and ultra-reliable low-latency communication services \cite{9806418}.

Traditional methods, such as linear programming and convex optimization, are commonly used to address these problems in communication networks. However, these methods face significant limitations:
They are often designed for static or slowly changing environments.
They rely on full system knowledge, which is difficult to obtain in fast-changing UAV environments.
They struggle to cope with the high-dimensional and non-linear nature of UAV network optimization problems, especially when dealing with the dynamic demands of 5/6G systems \cite{10529955}.
This is where DRL becomes essential \cite{sutton2018reinforcement, xu2014reinforcement}. DRL, which allows an agent to learn optimal actions through interaction with its environment, is well-suited for the dynamic and uncertain nature of UAV networks, particularly in the context of 5/6G. DRL offers a robust solution to real-time optimization problems, enabling UAVs to autonomously manage resources, allocate power, optimize channel assignments, and maintain high-quality communication performance, all while adapting to fluctuating network conditions.

\subsection{DRL Emergence in UAV Systems}
As UAVs become integral to communication networks, managing their resources and optimizing network performance require solving complex mathematical problems \cite{10040611}. These problems include power allocation, channel assignment, task offloading, and data caching. Traditionally, researchers have relied on well-established optimization techniques such as linear programming, convex optimization, and heuristic algorithms to address these challenges. While these methods have proven effective in static or less dynamic network environments, they encounter significant limitations in the rapidly changing and unpredictable settings in UAV networks.

There are some key limitations in traditional optimization methods.
(1) Static assumptions and limited adaptability:
Traditional optimization techniques often assume static or semi-static environments, where network parameters such as channel conditions, interference levels, and UAV positions are relatively constant. However, in UAV networks, the topology is highly dynamic, with UAVs frequently changing their positions and communication links being subject to fluctuating conditions. This mobility introduces a significant challenge for methods that require pre-defined or static inputs. Traditional approaches fail to adapt efficiently to dynamic environment, leading to suboptimal performance or the need for frequent re-optimization, which incurs high computational costs \cite{9354588}.

(2) Scalability issues:
As the number of UAVs and users in the network increases, the complexity of solving optimization problems grows exponentially. Traditional methods \cite{9023473} like exhaustive search and branch-and-bound can become computationally intractable as they require exploring all possible combinations of variables. In dense UAV networks with a large number of nodes and communication channels, these methods cannot scale effectively, leading to prohibitive computational delays that are unacceptable in real-time applications.

(3) Lack of real-time decision making:
Many traditional optimization techniques, particularly those that rely on deterministic models, are designed for offline analysis and optimization. These methods require extensive system knowledge and often cannot operate in real time. UAV networks, on the other hand, demand rapid decision-making to adapt to changing environmental factors, such as network congestion, battery depletion, or interference. The inability of traditional methods to make quick, real-time decisions under uncertain conditions limits their applicability in UAV networks \cite{10382630}.

(4) Difficulty in handling uncertainty:
UAV communication systems operate in highly uncertain environments, with factors like weather conditions, node failures, and unpredictable user demands affecting network performance. Traditional optimization methods often rely on deterministic models or require precise knowledge of system parameters, which is not feasible in real-world UAV networks. For example, methods based on convex optimization \cite{8247211} typically assume known, fixed parameters, which makes them less effective when dealing with the uncertainty and variability inherent in UAV operations. Additionally, these methods are ill-suited to handling the non-linearities and complex interdependencies between variables that characterize UAV communication problems.

(5) Inflexibility to multi-agent and cooperative scenarios:
In many UAV-assisted networks, multiple UAVs must cooperate to achieve common goals, such as optimizing coverage or minimizing network latency. Traditional optimization methods often focus on single-agent scenarios, where a central controller optimizes for a single UAV or a group of UAVs treated as a single entity. These methods do not effectively model the distributed decision-making process required in multi-agent environments, where UAVs must collaborate, share information, and make decentralized decisions in real time \cite{10415630}.

To overcome the limitations of traditional optimization techniques in UAV communication networks, DRL  has emerged as a powerful tool, which focuses on training an agent to make decisions based on interactions with its environment by reinforcement learning (RL), and excels in handling high-dimensional data and complex, non-linear relationships by deep learning (DL). The integration of these two techniques allows DRL to address the dynamic, uncertain, and large-scale optimization challenges present in UAV networks.

In UAV systems, DRL has proven particularly effective due to its ability to handle dynamic environments, uncertainty, and real-time decision-making. The following factors make DRL well-suited for UAV communication networks with advantage in 
adaptation to dynamic environments, 
scalability in complex systems,
handling uncertainty,
and real-time decision making. 
In addition, DRL has advantages in handling sequential decision-making tasks, where an agent learns how to make optimal decisions by interacting with an environment. Many UAV optimization problems can be naturally mapped into the DRL framework including concepts of states, actions, rewards, and policies. States represent the current status of the UAV network, including channel conditions, battery levels, task queues, or UAV positions.
Actions represent the set of possible decisions UAVs can make, such as allocating power, choosing communication channels, caching data, or offloading tasks.
Rewards quantify the immediate performance of a decision, such as reducing latency, saving energy, or increasing data throughput.
Policy is the strategy the UAV follows in choosing actions to maximize long-term rewards.
By structuring optimization problems as an Markov decision process (MDP), DRL allows UAVs to autonomously learn the best course of action based on the current network state and expected future states. This learning-based approach can dynamically adapt to changing network conditions, allowing for real-time optimization.

\subsection{Contributions and Organization of the Survey}
 
\begin{table}[h]
    \centering
    \caption{Acronyms and Abbreviations}\label{tab:my_label}
    \renewcommand{\arraystretch}{1.5} 
    \arrayrulewidth=1pt 
    \scalebox{0.72}{
    \begin{tabular}{|c|c|} 
        \hline
        \textbf{Acronym} & \textbf{Full form} \\
        \hline
         UAV & unmanned aerial vehicle \\
        \hline
         DRL & deep reinforcement learning \\
        \hline
         (MA)RL & (multi-agent) reinforcement learning\\
        \hline
         DL & deep learning \\
        \hline
        (PO)MDP & (partially observable) Markov decision process\\
        \hline
        TD & temporal difference\\
        \hline
        (D)DQN & (double) deep Q-network\\
        \hline
        NPG & natural policy gradient\\
        \hline
        KL & kullback-leibler\\
        \hline
        FIM & fisher information matrix\\
        \hline
        TRPO & trust region policy optimization\\
        \hline
        (MA)PPO & (multi-agent) proximal policy optimization\\
        \hline
        GAE & generalized advantage estimation\\
        \hline
        AC & actor critic\\
        \hline
        A3C & asynchronous advantage actor critic\\
        \hline
        A2C & advantage actor critic\\
        \hline
        (D)DPG & (deep) deterministic policy gradient\\
        \hline
        TD3& twin delayed deep deterministic\\
        \hline
        SAC & soft actor-critic\\
        \hline
        MADDPG & multi-agent DDPG\\
        \hline
        Dec-POMDP & decentralized partially observable MDP\\
        \hline
        POMG & partially observable Markov games\\
        \hline
        CTDE & centralized training with decentralized execution\\
        \hline
        IQL & independent Q-learning\\
        \hline
        QoS & quality of service \\
        \hline
        NOMA &  non-Orthogonal multiple access\\
        \hline
        (A)BS & (aerial) base station \\
        \hline
        RIS & reconfigurable intelligent surface\\
        \hline
        CUAV & cognitive radio technology-enhanced UAV\\
        \hline
        MBA &  multiple beam antennas \\
        \hline
        CSI & channel state information \\
        \hline
        SINR & signal to interference plus noise ratio \\
        \hline
        MINLP & mixed-integer nonlinear programming \\
        \hline
        RB & resource block \\
        \hline
        MDQN & mutual deep Q-network \\
        \hline
        UE & user equipment \\
        \hline
        US & group of user\\
        \hline
        LoS & line-of-sight\\
        \hline
        IIoT & ndustrial Internet of things\\
        \hline
        (A)AoI & (average) age of informatio\\
        \hline
        PD-NOMA & ower-domain non-orthogonal multiple acces\\
        \hline
        AAIoT & aerial access Internet of things network\\
        \hline
        HAP &  high-altitude platform\\
        \hline
        IoTDs &  Internet of things devices\\
        \hline
         JAPORA &  joint IoTD association, partial offloading, and communication resource allocation\\
        \hline
        JDCCO & joint data caching and computation offloading \\
        \hline
    \end{tabular}
    }
\end{table}

In this survey, we aim to provide a comprehensive review of how DRL can be applied to solve key optimization problems in UAV communications and networking. \textbf{The core focus of this survey is not merely to describe DRL algorithms but to emphasize how DRL can be employed to address specific optimization models within UAV systems, which is the first survey to discuss DRL implementation for optimization models, to the best of our knowledge.} Thus, we do not provide a comprehensive study on relative surveys that mainly focus on general description of DRL applications. Our key contributions are as follows:
\begin{itemize}
    \item \textbf{In-depth Coverage of DRL Fundamentals:} We provide a thorough introduction to the fundamental concepts and algorithms of DRL, including value-based, policy-based, and actor-critic methods. This section serves as a primer for readers to understand how these core DRL techniques work and how they can be adapted to address specific optimization problems in UAV communication networks. We explore key elements such as MDPs, the exploration-exploitation tradeoff, and the role of reward functions in learning optimal policies.
    \item \textbf{Implementation of DRL Techniques:} For each major UAV optimization problem, we review the DRL techniques most suited to solving them. We explain how specific DRL algorithms, such as value-based, policy-based, and actor-critic methods, can be adapted to address the unique challenges posed by these optimization problems.
    \item \textbf{Future Directions:} Finally, we identify emerging trends and future research directions where DRL is likely to play a pivotal role in UAV communications and networking.
\end{itemize}



\subsection{Organization of the Survey}
The remainder of this survey is organized as follows: We begin by introducing the core DRL concepts and algorithms, laying the foundation for understanding how DRL techniques operate and how they are applied in various optimization tasks. Following this, we explore the use of DRL in addressing specific optimization challenges in UAV communication networks, including power allocation, channel assignment, caching, and task offloading. For each problem, we examine the relevant DRL techniques, such as value-based, policy-based, actor-critic, and multi-agent methods, demonstrating how they are tailored to solve the mathematical optimization models underlying UAV networks. The survey also discusses emerging applications of DRL in other aspects of UAV communication, as well as future research directions where DRL can further enhance UAV system performance. We conclude by summarizing the key insights and providing a road-map for future advancements in DRL-based UAV communications and networking.

\section{A Tutorial of DRL Concepts and Algorithms}
\noindent

\subsection{RL Fundamentals} 
\label{sec:RL}


\subsubsection{Overview}

\begin{figure}[ht]
\centering
\includegraphics[width=3in]{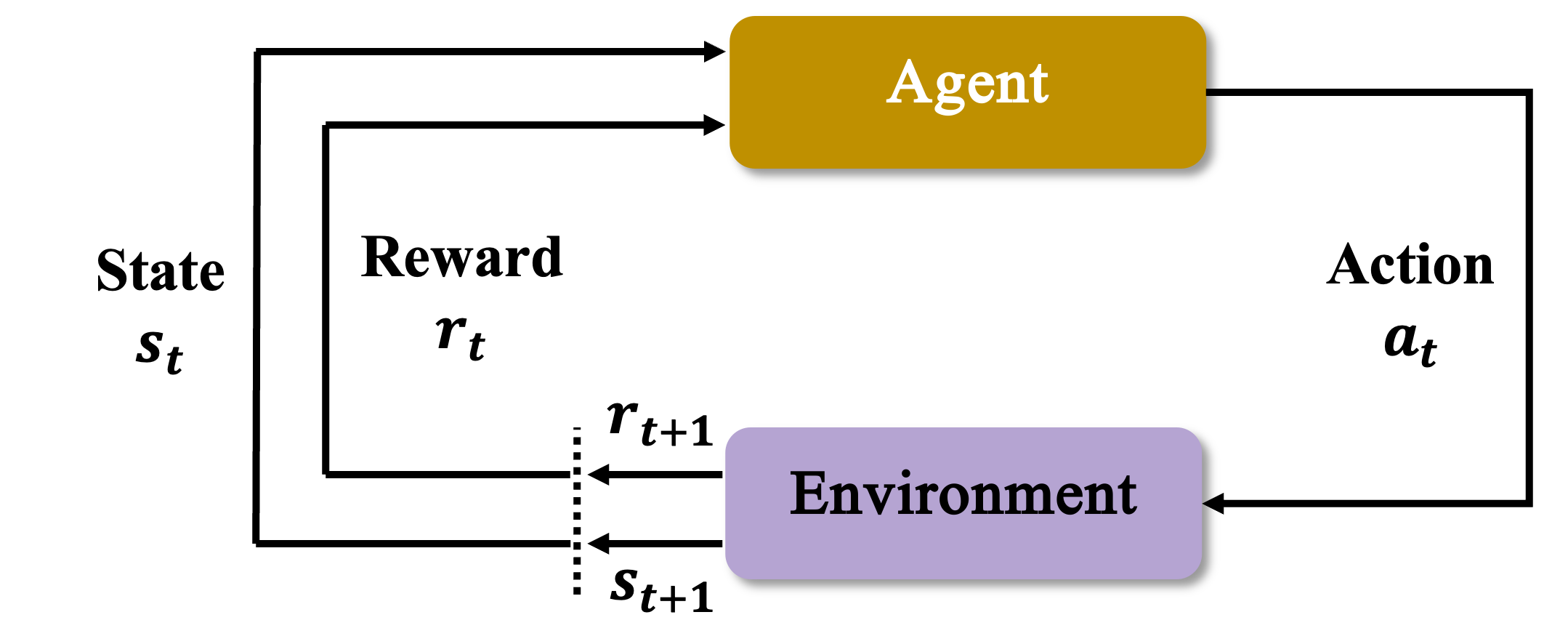}
\caption{Process of agent and environment interaction.}
\label{fig_rl}
\end{figure}

To effectively solve sequential decision optimization problems, a formal mathematical framework is essential. The MDP provides this framework by modeling the interaction between an agent and its environment, incorporating the Markov property where future states depend only on the current state and action. Once a problem is formulated as an MDP, RL serves as a powerful tool for finding optimal solutions. RL enables the agent to learn by interacting with the environment and receiving reward feedback, as shown in Fig. \ref{fig_rl}. Successfully solving MDPs with RL requires understanding key concepts such as the Markov decision framework (states, actions, and transitions), policy evaluation and the Bellman equation for strategy assessment, the Bellman optimality equation for optimal solutions, and methods like policy and value iteration to determine optimal policies systematically.


\tikzset{%
    parent/.style =          {align=center,text width=1cm,rounded corners=3pt, line width=0.3mm, fill=gray!10,draw=gray!80},
    child/.style =           {align=center,text width=2.3cm,rounded corners=3pt, fill=ttblue!10,draw=ttblue!80,line width=0.3mm},
    grandchild/.style =      {align=center,text width=2cm,rounded corners=3pt},
    greatgrandchild/.style = {align=center,text width=1.5cm,rounded corners=3pt},
    greatgrandchild2/.style = {align=center,text width=1.5cm,rounded corners=3pt},    
    referenceblock/.style =  {align=center,text width=1.5cm,rounded corners=2pt},
    value_based/.style =           {align=center,text width=1.5cm,rounded corners=3pt, fill=ttblue!10,draw=ttblue!80,line width=0.3mm},   
    value_based1/.style =           {align=center,text width=3cm,rounded corners=3pt, fill=ttblue!10,draw=ttblue!80,line width=0.3mm},   
    value_based2/.style =           {align=center, text width=10.5cm,rounded corners=3pt, fill=ttblue!10,draw=ttblue!0,line width=0.3mm},  
    policy_based/.style =           {align=center,text width=1.5cm,rounded corners=3pt, fill=violet!10,draw=violet!80,line width=0.3mm},   
    policy_based1/.style =           {align=center,text width=3cm,rounded corners=3pt, fill=violet!10,draw=violet!80,line width=0.3mm},   
    policy_secondary/.style =           {align=center,text width=2cm,rounded corners=3pt, fill=violet!10,draw=violet!80,line width=0.3mm},   
    policy_work/.style =           {align=center,text width=7.96cm,rounded corners=3pt, fill=violet!10,draw=violet!0,line width=0.3mm},    
    policy_based2/.style =           {align=center,text width=10.5cm,rounded corners=3pt, fill=violet!10,draw=violet!0,line width=0.3mm},    
    actor_critic/.style =           {align=center,text width=1.5cm,rounded corners=3pt, fill= cyan!10,draw= cyan!80,line width=0.3mm},   
    actor_critic1/.style =           {align=center,text width=3cm,rounded corners=3pt, fill= cyan!10,draw= cyan!80,line width=0.3mm},   
    actor_critic2/.style =           {align=center,text width=10.5cm,rounded corners=3pt, fill= cyan!10,draw= cyan!0,line width=0.3mm},      
    multi_agent/.style =           {align=center,text width=1.5cm,rounded corners=3pt, fill= orange!10,draw= orange!80,line width=0.3mm},   
    multi_agent1/.style =           {align=center,text width=3cm,rounded corners=3pt, fill= orange!10,draw= orange!80,line width=0.3mm},   
    multi_agent2/.style =           {align=center,text width=10.5cm,rounded corners=3pt, fill= orange!10,draw= orange!0,line width=0.3mm},   
}

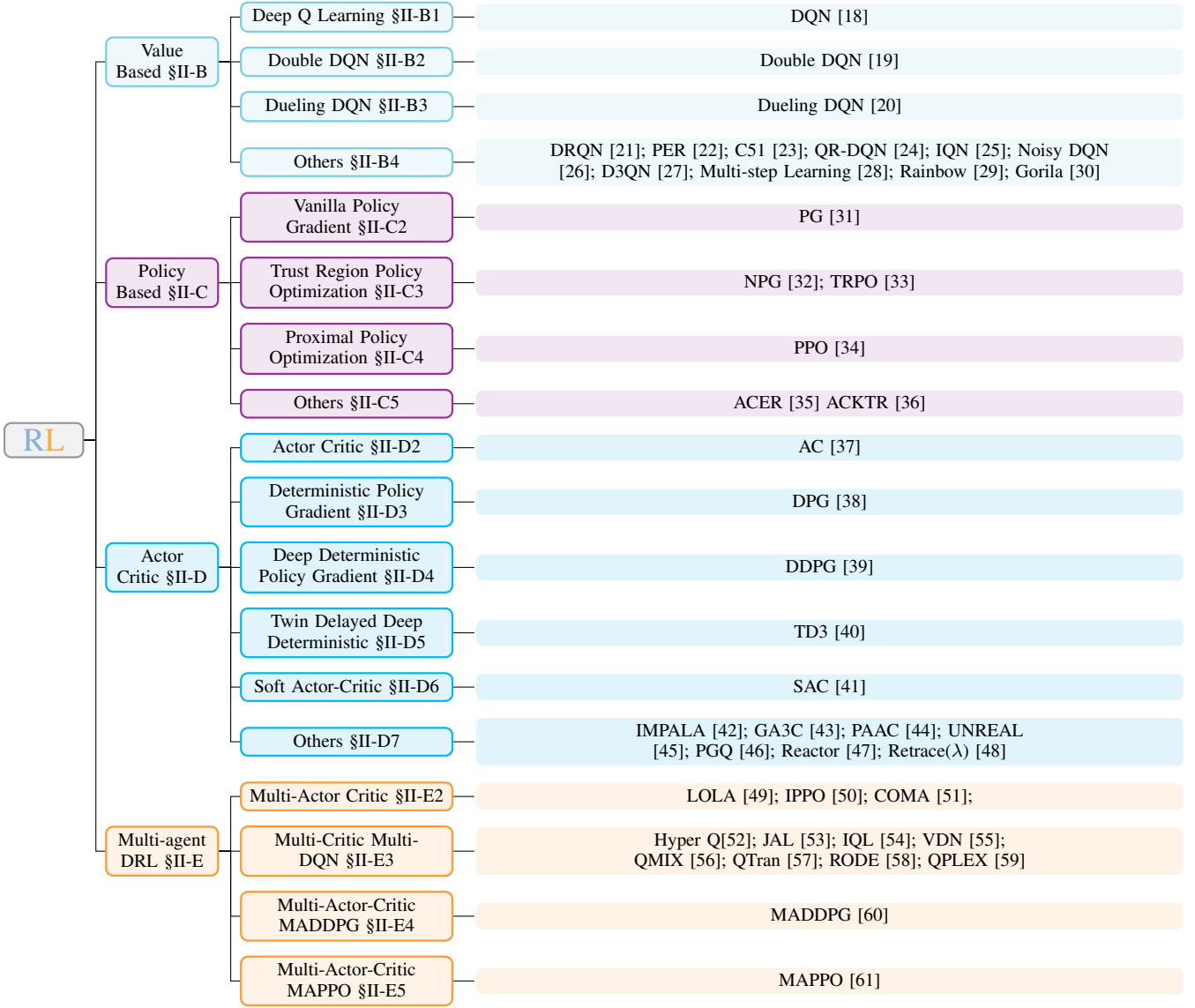
\begin{figure*}
\footnotesize
        \begin{forest}
            for tree={
                forked edges,
                grow'=0,
                draw,
                rounded corners,
                node options={align=center,},
                text width=2.7cm,
                s sep=6pt,
                calign=child edge, 
                calign child=(n_children()+1)/2,
            },
            [\textcolor{vision_color}{\Large R}\textcolor{lang_color}{\Large L}, fill=gray!45, parent, calign=center
                [Value Based \S\ref{sec:VB}, for tree={value_based}
                    [Deep Q Learning \S\ref{subsec:DQN},  value_based1
                        [DQN \cite{mnih2015human}
                        , value_based2]
                    ]
                    [Double DQN \S\ref{subsec:DoubleDQN}, value_based1
                        [Double DQN \cite{hasselt2010double}
                        , value_based2]
                    ]
                    [Dueling DQN \S\ref{subsec:DuelingDQN}, value_based1
                        [Dueling DQN \cite{wang2016dueling}
                        , value_based2]                        
                    ]
                    [Others \S\ref{subsec:DQNOthers}, value_based1
                        [DRQN \cite{hausknecht2015deep};
                        PER \cite{schaul2015prioritized};
                        C51 \cite{bellemare2017distributional};
                        QR-DQN \cite{dabney2018distributional};
                        IQN \cite{dabney2018implicit};
                        Noisy DQN \cite{fortunato2019noisy};
                        D3QN \cite{hu2020deep};
                        Multi-step Learning \cite{hernandez2019understanding};
                        Rainbow \cite{hessel2018rainbow};
                        Gorila \cite{parRL}
                        , value_based2]                        
                    ]
                ]
                [Policy Based \S\ref{sec:PB}, for tree={fill=violet!45,policy_based}
                    [Vanilla Policy Gradient \S\ref{subsec:vpg},  policy_based1
                        [PG \cite{sutton1999policy}
                        , policy_based2]                            
                    ]
                    [Trust Region Policy Optimization \S\ref{subsec:trpo},  policy_based1
                        [NPG \cite{NIPS2001_4b86abe4};
                        TRPO \cite{schulman2015trust} 
                        , policy_based2]
                    ]
                    [Proximal Policy Optimization \S\ref{subsec:ppo},  policy_based1,
                        [PPO \cite{schulman2017proximal}
                        , policy_based2]
                    ]
                    [Others \S\ref{subsec:pgothers},  policy_based1,
                        [ACER \cite{DBLP:journals/corr/WangBHMMKF16} 
                        ACKTR \cite{wu2017scalable}
                        , policy_based2]
                    ]
                ]
                [Actor Critic \S\ref{sec:AC}, for tree={fill=ttblue!45, actor_critic}
                    [Actor Critic \S\ref{subsec:ac}, actor_critic1
                        [AC \cite{konda1999actor}
                        , actor_critic2]  
                    ]
                    [Deterministic Policy Gradient \S\ref{subsec:dpg}, actor_critic1
                        [DPG \cite{silver2014deterministic}
                        , actor_critic2]  
                    ]
                    [Deep Deterministic Policy Gradient \S\ref{subsec:ddpg}, actor_critic1
                        [DDPG \cite{DBLPLillicrapHPHETS15}
                        , actor_critic2]  
                    ]
                    [Twin Delayed Deep Deterministic \S\ref{subsec:td3}, actor_critic1
                        [TD3 \cite{fujimoto2018addressing} 
                        , actor_critic2]  
                    ]
                    [Soft Actor-Critic \S\ref{subsec:sac}, actor_critic1
                        [SAC \cite{haarnoja2018soft}
                        , actor_critic2]  
                    ]
                    [Others \S\ref{subsec:ac_others}, actor_critic1
                        [IMPALA \cite{espeholt2018impala};
                        GA3C \cite{babaeizadeh2016reinforcement};
                        PAAC \cite{alfredo2017efficient};
                        UNREAL \cite{jaderberg2016reinforcement};
                        PGQ \cite{o2016combining};
                        Reactor \cite{gruslys2017reactor};
                        Retrace($\lambda$) \cite{munos2016safe}
                        , actor_critic2]  
                    ]
                ]
                [Multi-agent DRL \S\ref{sec:MA}, for tree={fill=ttblue!45, multi_agent}
                    [Multi-Actor Critic \S\ref{subsec:mac}, multi_agent1
                        [LOLA \cite{foerster2017learning};
                        IPPO  \cite{de2020independent};
                        COMA \cite{foerster2018counterfactual};
                        , multi_agent2]  
                    ]
                    [Multi-Critic Multi-DQN \S\ref{subsec:mdqn}, multi_agent1
                        [Hyper Q\cite{tesauro2003extending};
                        JAL \cite{claus1998dynamics};
                         IQL \cite{tan1993multi};
                         VDN \cite{sunehag2017value};
                         QMIX \cite{rashid2020monotonic};
                         QTran \cite{son2019qtran};
                         RODE \cite{wang2021rode};
                         QPLEX \cite{wang2021qplex}
                        , multi_agent2]  
                    ]
                    [Multi-Actor-Critic MADDPG \S\ref{subsec:maddpg}, multi_agent1
                        [ MADDPG \cite{lowe2017multi}
                        , multi_agent2]  
                    ]
                    [Multi-Actor-Critic MAPPO \S\ref{subsec:mappo}, multi_agent1
                        [MAPPO \cite{yu2022surprising}
                        , multi_agent2]  
                    ]
                ]  
            ]
        \end{forest}
        \caption{The taxonomy of DRL algorithms. Policy-based methods are not strictly devoid of critics, but policy-based approaches focus more on policy optimization, while Actor Critic methods emphasize systematic optimization, involving more efficient and parallel sampling.}
        \label{fig:taxonomy}
\end{figure*}

\subsubsection{Markov Decision Process}

An MDP is formally defined by a tuple $(\mathcal{S}, \mathcal{A}, \mathcal{P}, \mathcal{R}, \mathcal{\gamma})$ \cite{puterman2014markov}. Here, the state space $\mathcal{S}$ represents all possible situations that the decision-maker might encounter, capturing all relevant information needed for decision-making, while the action space $\mathcal{A}$ encompasses all possible decisions available to the agent. Both $\mathcal{S}$ and $\mathcal{A}$ are finite sets. The state transition function $\mathcal{P}: \mathcal{S} \times \mathcal{A} \rightarrow \Delta(\mathcal{S})$ describes the dynamics of the environment, where $\Delta(\mathcal{S})$ denotes the space of probability distributions over $\mathcal{S}$. Specifically, at time step $t$, given the current state $s_t$ and action $a_t$, the probability of transitioning to the next state $s_{t+1}$ is defined as $p(s_{t+1} \mid s_t, a_t)$. The reward function $\mathcal{R}: \mathcal{S} \times \mathcal{A} \rightarrow [0, R_{\max}]$ quantifies the immediate feedback, where $r(s_t, a_t)$ represents the immediate reward for taking action $a_t$ in state $s_t$. The discount factor $\gamma \in [0,1)$ determines how much the agent values future rewards compared to immediate ones. In the context of RL, the agent interacts with this MDP environment by following a policy $\pi$, which can be either deterministic as $\pi: \mathcal{S} \rightarrow \mathcal{A}$ where $a_t = \pi(s_t)$, or stochastic as $\pi: \mathcal{S} \rightarrow \Delta(\mathcal{A})$ where actions are sampled according to $a_t \sim \pi(s_t)$. These sequential interactions generate a trajectory $\tau_{\pi} = \left(s_1, a_1, r_1, s_2, \ldots, s_{H+1}\right)$, where $H$ is the length of trajectory. The initial state distribution is defined as $\rho: \mathcal{S} \rightarrow \Delta(\mathcal{S})$, completing the MDP description as a tuple $(\mathcal{S}, \mathcal{A}, \mathcal{P}, \mathcal{R}, \mathcal{\gamma}, \rho)$. The original optimization problem is formulated through the reward structure, where the agent aims to find a policy that maximizes the expected cumulative discounted reward $J(\pi)$, defined in eq. (\ref{optimal_objective_1}),

\begin{equation}
    J(\pi) = \mathbb{E}\left[\sum_{t=1}^{\infty} \gamma^{t-1} r_t \mid \pi, s_1\right].
    \label{optimal_objective_1}
\end{equation}
When the maximum expected rewards for all states in $\mathcal{S}$ are known and the environment model is available, the optimal policy can be derived through greedy action selection: at each state $s_t$, the action that leads to the next state $s_{t+1}$ with the highest expected reward should be selected. To obtain these state values across the state space, three primary approaches have been developed \cite{sutton2018reinforcement}. Value function-based methods, or dynamic programming, require a complete environment model, whereas Monte Carlo methods estimate values through sampling. Temporal difference (TD) methods also avoid needing a full model and allow incremental updates. The next section addresses two key tasks: evaluating state value functions for a given policy $\pi$ and finding the globally optimal state value function.



\subsubsection{Policy Evaluation and Bellman Equations}

In RL, the evaluation of a fixed policy $\pi$ involves computing value functions that quantify the expected long-term performance. The state value function $V^\pi(s_1): \mathcal{S} \rightarrow \mathbb{R}$ is defined in eq. (\ref{policy_evaluation_v}), which measures the expected cumulative discounted rewards when starting from an initial state $s_1$ and following policy $\pi$,

\begin{equation}
    V^\pi(s)=\mathbb{E}\left[\sum_{t=1}^{\infty} \gamma^{t-1} r_t \mid \pi, s=s_{1}\right].
    \label{policy_evaluation_v}
\end{equation}

To accurately measure the impact of specific actions, we define the action-state value function $Q^\pi: \mathcal{S} \times \mathcal{A} \rightarrow \mathbb{R}$ in eq. (\ref{policy_evaluation_q}). This function calculates the expected returns when starting from state $s_1$, taking action $a_1$, and then following policy $\pi$,

\begin{equation}
    Q^\pi(s, a)=\mathbb{E}\left[\sum_{t=1}^{\infty} \gamma^{t-1} r_t \mid \pi, s=s_{1}, a=a_{1}\right].
    \label{policy_evaluation_q}
\end{equation}

The value function $V^\pi$ and action-value function $Q^\pi$ are closely linked through the Bellman equations. These equations establish a recursive relationship between $V^\pi$ and $Q^\pi$ for any state $s_t \in \mathcal{S}$ and action $a_t \in \mathcal{A}$, based on dynamic programming principles. This relationship is expressed in eq. (\ref{bellman_equations}).

\begin{equation}
    \begin{split}
        &V^{\pi}(s_{t}) = Q^{\pi}(s_{t}, \pi(s_{t})), \\
        &Q^{\pi}(s_{t}, a_{t}) = r(s_{t}, a_{t}) + \gamma \mathbb{E}_{s_{t+1} \sim p(s_{t},a_{t})} \left[ V^{\pi}(s_{t+1}) \right].
    \end{split}
    \label{bellman_equations}
\end{equation}

For MDPs with finite state and action spaces, the recursive equations in eq. (\ref{bellman_equations}) enable the computation of $V^\pi(s)$ and $Q^\pi(s,a)$ through iterative methods. By repeatedly applying these updates until convergence, we obtain accurate estimates of the value functions, which serve as quantitative measures of policy $\pi$'s effectiveness across different states and actions.

\subsubsection{Bellman Optimality Equations}

An optimal policy $\pi^{\star}$ maximizes both the value function $V^{\pi}(s)$ for all states $s_t \in \mathcal{S}$ and the action-value function $Q^{\pi}$ for all state-action pairs $(s_t, a_t) \in \mathcal{S} \times \mathcal{A}$ \cite{puterman2014markov}. The relationship between these optimal value functions is described by the Bellman optimality equations in eq. (\ref{bellman_optimality_equations}),

\begin{equation}
    \begin{split}
        &V^{\pi^{\star}}(s_{t}) = \max_{a_{t} \in \mathcal{A}} Q^{\pi^{\star}}(s_{t}, a_{t}), \\
        &Q^{\pi^{\star}}(s_{t}, a_{t}) = r(s_{t}, a_{t}) + \gamma \mathbb{E}_{s_{t+1} \sim p(s_{t},a_{t})} \left[ V^{\pi^{\star}} (s_{t+1}) \right],
    \end{split}
    \label{bellman_optimality_equations}
\end{equation}
where $V^{\pi^{\star}}(s_{t})$ denotes the optimal value function, representing the maximum expected return achievable from state $s_t$. Similarly, $Q^{\pi^{\star}}(s_{t}, a_{t})$ represents the optimal action-value function, indicating the maximum expected return when taking action $a_t$ in state $s_t$. Given these optimal value functions, the optimal policy $\pi^{\star}$ can be derived through greedy action selection as shown in eq. (\ref{greedy_policy}).

\begin{equation}
    \pi^{\star}(s_{t})=\underset{a_{t} \in \mathcal{A}}{\arg \max } Q^{\pi^{\star}}(s_{t}, a_{t}), \forall s_{t} \in \mathcal{S} .
    \label{greedy_policy}
\end{equation}

Computing the optimal policy directly from eq. (\ref{greedy_policy}) presents a fundamental challenge, as it requires knowledge of the optimal action-value function $Q^{\pi^{\star}}$. For MDPs with tabular, this challenge can be addressed through two primary methods: policy iteration and value iteration. Policy iteration involves alternating between policy evaluation, which computes the value function for the current policy, and policy improvement, which updates the policy by selecting actions greedily based on the current value function. On the other hand, value iteration directly approximates the optimal value function by iteratively applying Bellman optimality until convergence.

\subsubsection{Policy Iteration}
The policy iteration algorithm begins by selecting an arbitrary policy, denoted as $\pi_{0}$, and then proceeds to repeatedly perform policy evaluation and policy improvement. In this process, the policy $\pi \left(a_{t} \mid s_{t} \right)$ is evaluated using the conditions in eq. (\ref{policy_evaluation_v}) and eq. (\ref{policy_evaluation_q}), which can be interpreted as probabilities. The state value function $V$ and state-action value function $Q$ can be expressed in eq. (\ref{v_q_evaluation}),
\begin{equation}
    \begin{split}
        &V_{k+1}(s_{t}) =  \\
        & \!\sum_{a_{t} \in \mathcal{A}} \pi_{k+1}(a_{t} \mid s_{t})\!\left(r_{t} + \gamma \!\sum_{s_{t+1} \!\in \!\mathcal{S}} \!\mathcal{P}_{s_{t+1} \mid s_{t}, a_{t}} V_{k}\left(s_{t+1} \right)\!\right), \\
        &Q_{k+1}\!\left(s_{t}, a_{t} \right) = \!\sum_{s_{t+1} \!\in \mathcal{S}} p\left(s_{t+1} \mid s_{t}, a_{t} \right)\left[r_{t} + \!\gamma V_{k}\left(s_{t+1}\right)\right],
    \end{split}
    \label{v_q_evaluation}
\end{equation}
where $k+1$ represents the next round evaluation based on the $k$-th policy. The policy in the $k+1$-th round can be obtained by policy improvement in eq. 
 (\ref{policy_improvement}),

\begin{equation}
    \pi_{k+1}(s_{t})=\underset{a_{t} \in \mathcal{A}}{\arg \max }  Q_{k} (s_{t}, a_{t}).
    \label{policy_improvement}
\end{equation}
For any $s_{t} \in \mathcal{S}$ and $a_{t} \in \mathcal{A}$, it follows that deduce $V_{k}(s_{t}) = \sum_{a_{t} \in \mathcal{A}} \pi_{k}(a_{t} \mid s_{t}) Q_{k}(s_{t}, a_{t}) \leq \sum_{a_{t} \in \mathcal{A}} \pi_{k+1}(a_{t} \mid s_{t}) Q_{k}(s_{t}, a_{t})$. By expanding this inequation, we obtain eq. (\ref{policy_improve_monotonically}) for any $s_{t} \in \mathcal{S}$ and $a_{t} \in \mathcal{A}$:
\begin{align}
    & Q_{k}(s_{t}, a_{t}) = \mathbb{E}_{p, r} \left[ \sum_{t=1}^{\infty}\gamma^{t-1} r_{t} + \gamma^{t} V_{k}(s_{t}) \right] \nonumber \\
    & = \mathbb{E}_{p, r} \left[ \sum_{t=1}^{\infty}\gamma^{t-1} r_{t} + \gamma^{t} \sum_{a_{t} \in \mathcal{A}} \pi_{k}(a_{t} \mid s_{t}) Q_{k}(s_{t}, a_{t})  \right] \nonumber \\
    & \leq \mathbb{E}_{p, r} \left[ \sum_{t=1}^{\infty}\gamma^{t-1} r_{t} + \gamma^{t} \sum_{a_{t} \in \mathcal{A}} \pi_{k+1}(a_{t} \mid s_{t}) Q_{k}(s_{t}, a_{t})  \right]  \nonumber \\
    & \leq \mathbb{E}_{p, r, \pi_{k+1}} \left[ \sum_{t=1}^{\infty}\gamma^{t-1} r_{t} \right] = Q_{k+1}(s_{t}, a_{t}),
    \label{policy_improve_monotonically}
\end{align}
where $\mathbb{E}_{p, r}$ represents the expectation of state transitions and rewards. This iterative process ensures that policy improvement is monotonically achieved in all states until the optimal policy $\pi^*$ is reached.

\subsubsection{Value Iteration}

Policy iteration requires policy evaluation to converge before each policy improvement step, necessitating multiple traversals of the entire state space. Value iteration addresses this inefficiency by integrating policy improvement with policy evaluation.
The expansion of eq. (\ref{policy_improvement}) yields eq. (\ref{policy_improvement_expand}),

\begin{equation}
    \begin{split}
        & \pi_{k+1}(s_{t}) = \underset{a_{t} \in \mathcal{A}}{\arg \max } Q_{k}(s_{t}, a_{t}) \\
        & = \underset{a_{t} \in \mathcal{A}}{\arg \max } \sum_{s_{t+1}} p \left(s_{t+1} \mid s_{t}, a_{t} \right)\left[r_{t} + \gamma  V_{k} \left(s_{t+1}\right) \right].
    \end{split}
    \label{policy_improvement_expand}
\end{equation}

Incorporating eq. (\ref{policy_improvement_expand}) into eq. (\ref{v_q_evaluation}) leads to the value iteration formula in eq. (\ref{value_iteration}),

\begin{equation}
    V_{k+1}(s_{t})=\max_{a_{t} \in \mathcal{A}} \!\sum_{s_{t+1}} p \left(s_{t+1} \!\mid \!s_{t}, \!a_{t} \right)\left[r_{t} + \gamma  V_{k}\left(s_{t+1}\right)\right].
    \label{value_iteration}
\end{equation}
The corresponding iterative equation for the action value function is expressed in eq. (\ref{action_value_iteration}),

\begin{equation}
    \begin{split}
        & Q_{k+1}(s_{t},a_{t}) = \\
        & \quad \max _{a_{t+1}} \sum_{s_{t+1}} p \left(s_{t+1} \mid s_{t}, a_{t} \right)\left[r_{t} + \gamma  Q_{k}\left(s_{t+1}, a_{t+1} \right)\right].
    \end{split}
    \label{action_value_iteration}
\end{equation}

In value iteration, the value function estimates are refined by minimizing TD errors from observed trajectories. The policy improves simultaneously through greedy action selection based on the updated value functions. This interleaved approach of policy evaluation and improvement, rather than completing each step sequentially, enhances computational efficiency and facilitates faster convergence.


The estimation of next-state values $V_{s_{t+1}}$ often faces challenges due to uncertain state transition probabilities in real systems. Given the typically fixed action space, algorithms commonly utilize state-action value functions. Based on eq. (\ref{action_value_iteration}) and discrete sampled data $({s_{t}, a_{t}, r_{t}, s_{t+1}, a_{t+1}}) \in \tau_{\pi}$, the SARSA algorithm update rule is derived as shown in eq. (\ref{sarsa}),

\begin{equation}
    \begin{split}
        & Q_{k+1}(s_{t}, \!a_{t}) \!\leftarrow Q_{k}(s_{t}, \!a_{t})+\alpha(t) \!\left(Q_{\text{target}} - Q_{k} (s_{t}, \!a_{t})\right), \\
        & Q_{\text{target}} = r_{t}+\gamma Q_{k}\left(s_{t+1}, a_{t+1}\right).
    \end{split}
    \label{sarsa}
\end{equation}

The learning rate schedule $\alpha(t)$ controls update magnitudes in RL, while the target value $Q_{\text{target}}$ combines the immediate reward $r_{t}$ with the discounted future reward for a more accurate estimate than $Q_{k}(s_{t}, a_{t})$. SARSA computes $Q_{\text{target}}$ using the action $a_{t+1}$ from the current policy, making it an on-policy algorithm. In contrast, off-policy methods like Q-learning use separate policies for experience generation, allowing data reuse. Q-learning employs an $\epsilon$-greedy sampling policy and computes $Q_{\text{target}}$ with a greedy policy, updating as shown in eq. (\ref{q-learning}),

\begin{equation}
    \begin{split}
        & Q_{k+1}\!\left(s_{t}, \!a_{t} \right) \!\leftarrow \!Q_{k}\left(s_{t}, \!a_{t}\right)+\alpha(t)\! \left(Q_{\text{target}} - \!Q_{k}\!\left(s_{t}, \!a_{t}\right)\right), \\
        & Q_{\text{target}} = r_{t} +\gamma \max _{a_{t+1} \in \mathcal{A}} Q\left(s_{t+1}, a_{t+1}\right).
    \end{split}
    \label{q-learning}
\end{equation}


The main difference between Q-learning and SARSA is their method for estimating the target Q-value. There are several approaches available for this estimation:

\begin{enumerate}
    \item SARSA: $Q_{\text{target}} = r_{t}+\gamma Q\left(s_{t+1}, a_{t+1}\right)$,
    \item Q-Learning: $Q_{\text{target}} = r_{t}+\gamma \max _{a_{t+1}} Q\left(s_{t+1}, a_{t+1}\right)$,
    \item Monte Carlo: $Q_{\text{target}} = r_{t} + \gamma r_{t+1} + \gamma^{2} r_{t+2} + \cdots$,
    \item $n$-step RL: $r_{t} + \gamma r_{t+1} + \gamma^{2} r_{t+2} + \cdots + r^{t+n}Q(s_{t+n}, a_{t+n})$.
\end{enumerate}

\subsection{Value-based DRL}
\label{sec:VB}


In RL, traditional Q-learning and SARSA algorithms \cite{sutton2018reinforcement, watkins1992q} use tabular storage for the state-action value function, but this approach becomes impractical in continuous state-action spaces. Discretizing continuous actions simplifies the problem but reduces precision and may prevent reaching global optima. To address this, parameterized models \cite{xu2014reinforcement}, particularly neural networks, are preferred due to their strong generalization abilities, facilitating more efficient learning of state-action value functions.

In this section, we discuss scenarios where the state space $\mathcal{S}$ is infinite while the action space $\mathcal{A}$ remains finite. To address the high dimensionality of the state space and enable generalization, researchers use parametric function approximation, such as neural networks, represented by $V^\theta(s_{t}) \simeq V^\pi(s_{t})$ and $Q^\theta(s_{t}, a_{t}) \simeq Q^\pi(s_{t}, a_{t})$, where $\theta$ are the approximation parameters updated through reinforcement learning. The objective is to find $\theta$ minimizing the mean-squared error between the approximate $Q^\theta(s_{t}, a_{t})$ and true $Q^\pi(s_{t}, a_{t})$ state-action values, formalized by the loss function $J(\theta)=\mathbb{E}\left[\frac{1}{2}\left(Q^\pi(s_{t}, a_{t})-Q^\theta(s_{t}, a_{t})\right)^2\right]$. Using stochastic gradient descent on a single sample, the update rule of the SARSA algorithm for approximating $\theta$ is given by eq. (\ref{single_sample_update}),

\begin{align}\label{single_sample_update}
    & \theta \leftarrow \theta-\alpha(t) \frac{\partial J(\theta)}{\partial \theta}  \\
    & =\theta+\alpha(t)\left(Q^\pi(s_{t},a_{t})-Q^{\theta}(s_{t},a_{t})\right) \frac{\partial Q^{\theta}(s_{t}, a_{t})}{\partial \theta} \nonumber \\
    & =\theta+\alpha(t)\left(r_{t} + \!\gamma Q^{\theta}\left(s_{t+1}, \!a_{t+1}\right) - \!Q^{\theta}(s_{t}, \!a_{t})\right) \frac{\partial Q^{\theta}(s_{t}, \!a_{t})}{\partial \theta}, \nonumber
\end{align}
where $r_{t}$ represents the actual reward, and $Q^{\theta}\left(s_{t+1}, a_{t+1}\right)$ and $Q^{\theta}(s_{t}, a_{t})$ are the estimates provided by the function approximator. The target value $Q_{\text{target}}$ is typically used instead of the expected cumulative reward $Q^{\pi}$ due to the complexity of computing the latter.

The approximation of $Q^{\theta}\left(s_{t+1}, a_{t+1}\right)$ and $Q^{\theta}(s_{t}, a_{t})$ aims for relatively unbiased update losses. However, achieving stable training in practice requires techniques like experience replay, target networks, and learning rate scheduling to address issues like sample correlation and update instability. Specific algorithms integrating these techniques will be reviewed in the following sections to achieve robust and efficient learning performance.

\subsubsection{Deep Q Learning}
\label{subsec:DQN}

Deep Q-Network (DQN) \cite{mnih2015human} uses convolutional neural networks for powerful feature representation, enabling state-of-the-art performance in complex game environments. However, DQN faces challenges like instability from policy changes altering the data distribution. To mitigate this, DQN implements experience replay, maintaining a buffer $D_{t} = \{e_{1}, \cdots, e_{t}\}$ of experience tuples $e_{t} = (s_{t}, a_{t}, r_{t}, s_{t+1})$ to stabilize training, improve sample efficiency, and enable parallel processing.

DQN addresses the instability caused by simultaneous changes to $Q^{\theta}(s_{t+1}, a_{t+1})$ and $Q^{\theta}(s_{t}, a_{t})$ by implementing a target network. While the network computing current state-action values $Q^{\theta}(s_{t}, a_{t})$ updates continuously, the network estimating target values $Q^{\theta}(s_{t+1}, a_{t+1})$ updates periodically. This reduces update bias and enhances algorithmic stability. The DQN loss function at iteration $k$ is formulated as eq. (\ref{dqn_loss}),

\begin{equation}\label{dqn_loss}
    \begin{split}
        &J_{k}(\theta_{k}) =  \\
        &\mathbb{E}_{e_{t} \!\sim D_{t}} \!\left[ \!\left(\!r_{t} \!+ \!\gamma \!\max_{a_{t+1}} \!Q^{\theta^{\text{target}}}(s_{t+1}, \!a_{t+1}) \!- \!Q^{\theta_{k}}(\!s_{t}, \!a_{t}\!) \!\right)^{2} \!\right],
    \end{split}
\end{equation}
where $\theta^{\text{target}}$ represents the target network parameters at iteration $k$, updated by periodically copying $\theta_{k}$.

\subsubsection{Double DQN}
\label{subsec:DoubleDQN}

The Double Deep Q-Network (DDQN) addresses the overestimation issue in DQN by decoupling action selection and value evaluation \cite{hasselt2010double}. In DQN, the max operator for the target state-action value couples these two processes, leading to suboptimal action choices and overly optimistic value estimates. Specifically, when looking at the term $\max_{a_{t+1}} Q^{\theta^{\text{target}}}(s_{t+1}, a_{t+1})$ in eq. (\ref{dqn_loss}), it becomes apparent that both the selection of the next action $a_{t+1}$ and the evaluation of $Q_{\text{target}}$ depend on the same target network parameters $\theta^{\text{target}}$. This coupling is a major cause of value overestimation.

DDQN introduces a two-step process: it selects actions using the current network parameters $\theta$ to identify the action that maximizes the Q-value, and then evaluates this selected action using the target network parameters $\theta^{\text{target}}$  in $Q_{\text{target}} = r_{t} +\gamma   Q^{\theta^{\text{target}}}\left(s_{t+1},\underset{{a_{t+1} \in \mathcal{A}}}{\arg \max} Q^{\theta}(s_{t+1}, a_{t+1}) \right)$. This separation mitigates Q-value overestimation, reduces training time, and enhances learning stability, leading to more reliable value estimation and better overall performance \cite{azar2021drone}.

\subsubsection{Dueling DQN}
\label{subsec:DuelingDQN}

Dueling DQN introduces a novel decomposition of the state-action value function $Q^{\theta_{1}, \theta_{2}, \theta_{3}}(s_{t}, a_{t}) = V^{\theta_{1}, \theta_{2}}(s_{t}) + A^{\theta_{1}, \theta_{3}}(s_{t}, a_{t})$ \cite{wang2016dueling}. This architecture splits the network into two streams: one approximating the state value function $V(s_{t})$ and the other estimating the advantage function $A(s_{t}, a_{t})$. The key insight is that in many situations, it is unnecessary to estimate the value of each action separately. For instance, in gaming environments, knowing whether to move left or right may be sufficient for avoiding collisions without requiring precise value estimates for each action. This observation motivates the dueling architecture's efficiency in learning state values independently from action advantages. The advantage function follows the property that $\mathbb{E}_{a_{t} \sim \pi(s_{t})}\left[A^{\pi}(s_{t}, a_{t}) \right] = 0$, leading to a refined formulation in $Q^{\theta_{1}, \theta_{2}, \theta_{3}}(s_{t}, a_{t}) = V^{\theta_{1}, \theta_{2}}(s_{t}) + \!\left( A^{\theta_{1}, \theta_{3}}(s_{t}, a_{t}) - \frac{1}{|\mathcal{A}|} \sum_{a_{t+1}} A^{\theta_{1}, \theta_{2}}(s_{t}, a_{t}) \right)$. By explicitly separating the estimation of state values from action advantages, Dueling DQN enables more robust and efficient learning, particularly in scenarios where understanding the relative importance of different actions is crucial for optimal decision-making.



\subsubsection{Others}
\label{subsec:DQNOthers}

To address partially observable MDP (POMDP), DRL models have incorporated historical data into the state representation and used LSTM networks \cite{hausknecht2015deep}. This allows agents to learn more effective policies in partially observable environments. To handle large action spaces, research has used Deep Monte-Carlo methods \cite{pmlr-v139-zha21a} for efficient Q-function evaluation.

Structural improvements to DQN include prioritized experience replay \cite{schaul2015prioritized}, distributional RL C51 \cite{bellemare2017distributional}, and other innovations like Noisy DQN \cite{fortunato2019noisy} and multi-step learning  \cite{hernandez2019understanding}. QR-DQN \cite{dabney2018distributional} strengthened the theoretical foundations, and IQN \cite{dabney2018implicit} introduced flexible quantile regression for risk-sensitive policies. D3QN's combined approach to addressing overestimation bias \cite{hu2020deep}. Rainbow \cite{hessel2018rainbow} ultimately integrated these various improvements into a unified algorithm.

Training techniques like Pop-Art \cite{van2016learning} address the issue of varying target magnitudes, while Averaged-DQN \cite{anschel2017averaged} reduces the variance of approximation errors by averaging Q-value estimates. DQfD \cite{hester2018deep} incorporates human demonstrations to improve sample efficiency and fast reward propagation \cite{s.he2017learning} optimizes the utilization of experience replay. Other notable innovations include Soft DQN by entropy regularization \cite{schulman2017equivalence}, LS-DQN integration of linear least squares integration \cite{levine2017shallow}, and DQV approach of using a value network \cite{sabatelli2018deep}. Distributed architectures like Gorila \cite{parRL} enable large-scale DRL.

\subsection{Policy-based DRL}
\label{sec:PB}

\subsubsection{Overview}

Value-based RL approaches start by evaluating value functions across the state space, then derive policies to maximize those values. While DQN handles infinite state spaces through neural approximations, and variants can manage large discrete action spaces, they are inadequate for continuous (infinite) action spaces. Policy-based methods directly optimize the policy parameters $\omega$ to maximize the objective function $J(\omega)$, rather than deriving policies from value functions. This is achieved by reformulating optimization objective eq. (\ref{optimal_objective_1})  into eq. (\ref{optimal_objective_2}),

\begin{equation}\label{optimal_objective_2}
    \begin{split}
        & J(\omega) = \\
        & \mathbb{E} \left[ \mu(s_{1}) \prod_{t=0}^H \gamma^{t} p\left(s_{t+1} \mid s_t, a_t\right) \pi^{\omega}\left(a_t \mid s_t\right) r(s_{t}, a_{t}) \right].
    \end{split}
\end{equation}

A fundamental property of policy execution emerges when a policy $\pi^{\omega}$ is executed for a sufficient duration. The frequency of visiting each state-action pair $(s_t, a_t)$ stabilizes to fixed proportions, forming a probability distribution. This stationary distribution is mathematically represented in $d^{\pi^{\omega}}(s_{1}) =  \mu(s_{1}) \prod_{t=0}^H \gamma^{t} p\left(s_{t+1} \mid s_t, a_t\right) \pi^{\omega}\left(a_t \mid s_t\right)$. Based on this stationary distribution, the optimization objective can be reformulated in eq. (\ref{optimal_objective_3}),



\begin{equation}\label{optimal_objective_3}
    \begin{split}
        J(\pi^{\omega}) &= \mathbb{E}_{(s_{t},a_{t}) \sim d^{\pi^{w}}(s)} \left[r(s_{t},a_{t})\right] \\
        &= \mathbb{E}_{(s_{t}, a_{t}) \sim d^{\pi^{\omega}}(s)} \left[V(s_{1})\right],
    \end{split}
\end{equation}
where the policy is parameterized by $\omega$ and represented through $\pi^{\omega}(a_{t} \mid s_{t})$, the gradient $\nabla_\omega J(\pi^{\omega})$ exhibits a complex dependency structure. This is because it depends on both the actions generated by $\pi^{\omega}$ and the stationary state distribution induced by the policy. However, estimating how policy updates affect the state distribution becomes particularly challenging given that the environment dynamics are typically unknown. This fundamental challenge has motivated the development of various policy-based algorithms designed to effectively optimize policies effectively in such uncertain environments.

\subsubsection{Vanilla Policy Gradient}
\label{subsec:vpg}

Eq. (\ref{optimal_objective_2}) presents a challenge in computing its derivative due to the product of terms involving the policy $\pi^{\omega}$. To simplify this, Sutton proposed assuming the stationarity of $d^{\pi^{\omega}}$, resulting in a more elegant formulation of the derivative of the objective function. This reformulation eliminates the need to compute the derivative of the state distribution $d^{\pi}$, making the computation of the gradient $\nabla_\omega J(\pi^{\omega})$ significantly simpler.

The derivation of the policy gradient involves several important steps. First, we express the objective function in terms of the state-action value function $Q^{\pi^{\omega}}$. Then, by carefully manipulating the terms and applying the log-derivative trick, we arrive at the following formulation:

\begin{equation}
    \begin{split}
        & \nabla_\omega J(\pi^{\omega})  = \\
        & \nabla_\omega \sum_{s_{1} \in \mathcal{S}} d^{\pi^{\omega}}(s_{1}) \sum_{a_{t} \in \mathcal{A}} Q^{\pi^{\omega}}(s_{t}, a_{t}) \pi^{\omega}(a_{t} \mid s_{t}) \\
        & \propto \sum_{s_{1} \in \mathcal{S}} d^{\pi^{\omega}}(s_{1}) \sum_{a_{t} \in \mathcal{A}} Q^{\pi^{\omega}}(s_{t}, a_{t}) \nabla_\omega \pi^{\omega}(a_{t} \mid s_{t}) \\
        & = \!\sum_{s_{1} \!\in \mathcal{S}} \!d^{\pi^{\omega}}\!(s_{1}\!) \!\sum_{a_{t} \!\in \!\mathcal{A}} \pi^\omega(a_{t} \!\mid s_{t}) \!Q^{\pi^{\omega}}(s_{t}, \!a_{t}) \frac{\nabla_\omega \pi^{\omega}(a_{t} \!\mid s_{t})}{\pi^{\omega}(a_{t} \!\mid s_{t})} \\
        & = \mathbb{E}_{\pi^{\omega}}\left[Q^{\pi^{\omega}}(s_{t}, a_{t}) \nabla_{\omega} \ln \pi^{\omega}(a_{t} \mid s_{t})\right].
    \end{split}
\end{equation}

The theoretical foundation for this derivation, established in Sutton's work \cite{sutton1999policy}, highlights the challenge of high-variance gradients caused by reliance on empirical returns from trajectories. To mitigate this, subtracting a baseline value allows updates to be weighted by an advantage estimate instead of raw returns. While the vanilla policy gradient remains unbiased, its high variance has inspired algorithms to reduce variance while preserving unbiasedness. Approaches range from using average returns over episodes \cite{williams1992simple} to more advanced methods \cite{Schulmanetal_ICLR2016}, with equivalent expressions for the policy gradient balancing bias and variance as shown in $\nabla_\omega J(\pi^{\omega}) =\mathbb{E}\left[\sum_{t=0}^{\infty} \Psi_t \nabla_\omega \log \pi^{\omega}\left(a_t \mid s_t\right)\right]$. The term $\Psi_t$ can take various forms, each corresponding to different algorithmic approaches:


\begin{enumerate}
    \item $\sum_{t=0}^{\infty} r_t$: represent the total trajectory reward, measuring the cumulative return from the entire episode.
    \item $\sum_{t^{\prime}=t}^{\infty} r_t$: capture the reward following action $a_t$, forming the basis of the REINFORCE algorithm \cite{williams1992simple}.
    \item $\sum_{t^{\prime}=t}^{\infty} r_{t^{\prime}}-b\left(s_t\right)$: introduce a baseline-adjusted version, where $b\left(s_t\right)$ serves as a variance-reducing baseline function.
    \item $Q^{\theta} \left(s_t, a_t\right)$: utilize the state-action value function (critic), fundamental to actor-critic architectures \cite{sutton2018reinforcement}.
    \item $A^\theta \left(s_t, a_t\right)$: employ the advantage function to measure the relative benefit of action $a_t$ compared to the expected return.
    \item $r_t+V^\theta \left(s_{t+1}\right)-V^\theta \left(s_t\right)$: incorporate the TD residual, central to TD learning approaches \cite{sutton2018reinforcement}.
\end{enumerate}

\subsubsection{Trust Region Policy Optimization}
\label{subsec:trpo}

Vanilla policy gradient methods assume a stationary state-action visitation frequency $(s_t, a_t)$, requiring resampling after each policy update and preventing sample reuse. As the policy $\pi^{\omega}(a_t \mid s_t)$ is generated by a neural network, any parameter change $\Delta \omega$ directly alters the distribution. Natural Policy Gradient (NPG) \cite{NIPS2001_4b86abe4} treats the parameter space as a Riemannian manifold, using KL-divergence as a metric and the Fisher Information Matrix (FIM) to define curvature. This geometric perspective frames probability distributions as points in Riemannian space, with KL-divergence dictating distance and FIM characterizing local curvature. To control drastic policy shifts, NPG imposes a constraint $D_{\text{KL}}(\pi^{\omega}(a_t \mid s_t) | \pi^{\omega + \Delta \omega}(a_t \mid s_t)) = \epsilon$, ensuring a fixed distance $\epsilon$ between consecutive policies. Building on this, Trust Region Policy Optimization (TRPO) \cite{schulman2015trust} constrains optimization steps within a region where the surrogate objective remains an accurate approximation of the true cost function. This approach ensures policy improvement with substantial step sizes. 

The expected reward under a new policy $\tilde{\pi}$ relative to the old policy $\pi$ is expressed in eq (\ref{trpo_new_old}),

\begin{equation}\label{trpo_new_old}
    \begin{split}
        & \underset{\tau \sim \tilde{\pi}}{\mathbb{E}} \left[\sum_{t=1}^{\infty} \gamma^{t-1} A^\pi \left(s_t, a_t\right)\right] \\
        & =\underset{\tau \sim \tilde{\pi}}{\mathbb{E}}\!\left[\sum_{t=1}^{\infty} \!\gamma^{t-1}\left(r\left(s_t, \!a_t\right)\!+\!\gamma \!V^\pi\left(s_{t+1}\right)\!-\!V^\pi\left(s_t\right)\right)\!\right] \\
        & =\eta(\tilde{\pi})+\underset{\tau \sim \tilde{\pi}}{\mathrm{E}}\left[\sum_{t=1}^{\infty} \gamma^{t} V^\pi\left(s_{t+1}\right)-\sum_{t=1}^{\infty} \gamma^{t-1} V^{\pi} \left(s_t\right)\right] \\
        & =\eta(\tilde{\pi})+\underset{\tau \sim \tilde{\pi}}{\mathrm{E}}\left[\sum_{t=2}^{\infty} \gamma^{t-1} V^\pi\left(s_t\right)-\sum_{t=1}^{\infty} \gamma^{t-1} V^\pi\left(s_t\right)\right] \\
        & =\eta(\tilde{\pi})-\underset{\tau \sim \tilde{\pi}}{\mathrm{E}}\left[V^\pi\left(s_0\right)\right] =\eta(\tilde{\pi})-\eta(\pi).
    \end{split}
\end{equation}

Through algebraic manipulation of eq. (\ref{trpo_new_old}), we obtain the following relationship:

\begin{equation}
    \eta(\tilde{\pi})=\eta(\pi)+\sum_{s} d^{\tilde{\pi}}(s_{1}) \sum_{a_{t}} \tilde{\pi}(a_{t} \mid s_{t}) A^{\pi}(s_{t}, a_{t}),
    \label{trpo_new_old_reorganing}
\end{equation}
where $d^{\tilde{\pi}}$ represents the discounted visitation frequencies. A key insight is that if we can ensure $\sum_{s} d^{\tilde{\pi}}(s) \sum_{a_{t}} \tilde{\pi}(a_{t} \mid s_{t}) A^{\pi}(s_{t}, a_{t}) \geq 0$, then the expected discounted reward under the new policy will consistently improve. This means that if we start with an initial policy $\pi$ and find a policy $\tilde{\pi}$ that satisfies this condition, we can guarantees that $\eta(\tilde{\pi}) \geq \eta(\pi)$, establishing monotonic improvement. However, the challenge lies in practically ensuring this condition, as $d^{\tilde{\pi}}(s)$ requires sampling from the new policy $\tilde{\pi}$. This involves extensive sampling for each candidate policy, making the approach computationally prohibitive.
TRPO addresses this challenge by introducing an approximate form $\mathcal{L}$:

\begin{equation}
    \mathcal{L}(\tilde{\pi}) = \eta(\pi) + \sum_{s} d^{\pi}(s_{1}) \sum_{a_{t}} \tilde{\pi}(a_{t} \mid s_{t}) A^{\pi}(s_{t}, a_{t}).
    \label{approximate_form}
\end{equation}

The only difference between eq. (\ref{approximate_form}) and eq. (\ref{trpo_new_old_reorganing}) lies in the discounted visitation frequencies $d(s_{1})$. Their gradients maintain directional consistency, allowing eq. (\ref{approximate_form}) for policy updates when the difference between $d^{\tilde{\pi}}$ and $d^{\pi}$ remains small. TRPO provides a precise quantification of their performance difference. Given that $\mathbb{E}_{a\sim\pi}\left[A^{\pi}(s_{t}, a_{t})\right] = 0$, we have $\mathbb{E}_{\tilde{a}\sim\tilde{\pi}}\left[A^{\pi}(s,\tilde{a})\right] = \mathbb{E}_{(a,\tilde{a})\sim(\pi,\tilde{\pi})}\left[A^{\pi}(s,\tilde{a})-A^{\pi}(s,a)\right] =P(a\neq\tilde{a}|s)\mathbb{E}_{(a,\tilde{a})\sim(\pi,\tilde{\pi})|a\neq\tilde{a}}\left[A^{\pi}(s,\tilde{a})-A^{\pi}(s,a)\right] \leq \xi \cdot2\max_{s,a}|A^{\pi}(s,a)|$, where $P(a\neq\tilde{a}|s) \leq \xi$ defines a joint distribution $(a, \tilde{a} \mid s)$. For each timestep $t$, $n_{t}$ represents the count of instances where $a_i\neq\tilde{a}_i$ for $i < t$. Then $P\left(n_t>0\right) \leq 1-(1-\xi)^t$, as shown in the expression, $\mathbb{E}_{s_{t} \sim \tilde{\pi}, a_{t} \sim\tilde{\pi}} \left[A_{\pi}(s_{t}, a_{t})\right] = P\left(n_{t}=0\right) \mathbb{E}_{s_{t} \sim \tilde{\pi}, a_{t} \sim \tilde{\pi} \mid n_{t}=0} \left[ A^{\pi}(s_{t}, a_{t})\right] + P\left(n_{t} > 0\right) \mathbb{E}_{s_{t} \sim \tilde{\pi}, a_{t} \sim \tilde{\pi} \mid n_{t}>0} \left[ A^{\pi}(s_{t}, a_{t})\right]$. 

Similarly, we have
$\mathbb{E}_{s_{t} \sim \pi, a_{t} \sim\tilde{\pi}} \left[A^{\pi}(s_{t}, a_{t})\right] = P\left(n_{t}=0\right) \mathbb{E}_{s_{t} \sim \pi, a_{t} \sim \tilde{\pi} \mid n_{t}=0} \left[ A^{\pi}(s_{t}, a_{t})\right] + P\left(n_{t} > 0\right) \mathbb{E}_{s_{t} \sim \pi, a_{t} \sim \tilde{\pi} \mid n_{t}>0} \left[ A^{\pi}(s_{t}, a_{t})\right]$. 
By subtracting the expressions and defining $\varepsilon = \max _{s, a}\left|A_\pi(s, a)\right|$:

\begin{equation}
    \begin{split}
        &\!\sum_{t=0}^{\infty} \!\gamma^t \!\cdot \!|\mathbb{E}_{s_t \sim \tilde{\pi}, a_t \sim \tilde{\pi}}\left[\!A^{\pi}\!\left(s_t, \!a_t\right)\right]\!-\!\mathbb{E}_{s_t \!\sim \!\pi, \!a_t \sim \!\tilde{\pi}}\!\left[\!A^{\pi}\!\left(s_t, \!a_t\right)\right]\!| \\
        &=\sum_{t=0}^{\infty} \gamma^t \cdot P\left(n_t>0\right)|\mathbb{E}_{s_t \sim \tilde{\pi}, a_t \sim \tilde{\pi} \mid n_t>0}\left[A^{\pi}\left(s_t, a_t\right)\right] \\
        & \quad -\mathbb{E}_{s_t \sim \pi, a_t \sim \tilde{\pi} \mid n_t>0}\left[A^{\pi}\left(s_t, a_t\right)\right]| \\
        & \leq 4 \xi \left(1-(1-\xi)^t\right) \max _{s_{t}, a_{t}}\left|A^{\pi} (s_{t}, a_{t})\right| \\
        & = \sum_{t=0}^{\infty} \gamma^t \cdot 4 \varepsilon \xi\left(1-(1-\xi)^t\right) \leq \frac{4 \xi^2 \gamma \varepsilon}{(1-\gamma)^2}. \\
    \end{split}
\end{equation}
Therefore, the quantified of performance difference between eq. (\ref{approximate_form}) and eq. (\ref{trpo_new_old_reorganing}) is expressed in $\left|\eta(\tilde{\pi}) - L(\tilde{\pi})\right| \leq \frac{4 \xi^2 \gamma \varepsilon}{(1-\gamma)^2}$. Therefore, $\eta(\tilde{\pi}) \geq L(\tilde{\pi}) - \frac{4 \xi^2 \gamma \varepsilon}{(1-\gamma)^2}$. Replacing $\xi$ by the total variation divergence, denoted as $D_{\text{TV}}$. Thus, $\max _s D_{\text{TV}}(\pi(\cdot \mid s) \| \tilde{\pi}(\cdot \mid s)) \leq \xi$ and applying $D_{\text{TV}}(p | q)^2 \leq D_{\text{TV}}(p | q)$, we obtain $\eta(\tilde{\pi}) \! \geq \! L_\pi(\tilde{\pi}) \!- \!  C* D_{\mathrm{KL}}^{\max }(\pi, \tilde{\pi}), \text { where } C \! = \! \frac{4 \varepsilon \gamma}{(1-\gamma)^2}$.

The final objective of TRPO optimization is to calculate the overall advantage of state visitation distribution and actions. This is achieved by using importance sampling to account for any discrepancies between the distribution of training data and the actual state distribution of the policy, as shown in eq. (\ref{trpo_optimization}),

\begin{equation}\label{trpo_optimization}
    \begin{split}
        & J(\omega)=\sum_{s \in \mathcal{S}} d^{\pi} (s_{1}) \sum_{a_{t} \in \mathcal{A}}\left( \tilde{\pi}^{\omega}(a_{t} \mid s_{t}) A^{\pi}(s_{t}, a_{t})\right) \\
        & =\!\sum_{s \in \mathcal{S}} d^{\pi} \!(s_{1}\!) \!\sum_{a_{t} \in \mathcal{A}}\left(\pi(a_{t} \!\mid \!s_{t}) \frac{\tilde{\pi}^{\omega}(a_{t} \!\mid \!s_{t})}{\pi(a_{t} \!\mid \!s_{t})} A^{\pi} (s_{t}, \!a_{t})\right) \\
        & =\mathbb{E}_{s \sim d^{\pi}, a_{t} \sim \pi}\left[\frac{\tilde{\pi}^{\omega} (a_{t} \mid s_{t})}{\pi(a_{t} \mid s_{t})} A^{\pi} (s_{t}, a_{t}) \right].
    \end{split}
\end{equation}

This optimization is constrained by a trust region limiting the KL-divergence between the old policy $\pi$ and the new policy $\tilde{\pi}$, expressed as $\mathbb{E}_{s \sim d^{\pi}(s)}\left[D_{\mathrm{KL}}\left(\tilde{\pi}^{\theta}(\cdot \mid s) \| \pi (\cdot \mid s) \right) \right]  \leq \Delta_{KL}$, where $d^{\pi} = d^{\pi^{\omega}}$ represents the discounted state-visitation frequencies of policy $\pi^{\omega}$, and $\Delta_{\text{KL}}$ controls the maximum policy change per step.

\subsubsection{Proximal Policy Optimization}
\label{subsec:ppo}

Proximal Policy Optimization (PPO) \cite{schulman2017proximal} combines the performance benefits of TRPO with simpler implementation by using a clipped surrogate objective. The core idea is to calculate a probability ratio between the current policy and the behavior policy, defined as $r(\omega)=\frac{\tilde{\pi}^\omega(a \mid s)}{\pi (a \mid s)}$, where $\tilde{\pi}^\omega$ represents the current policy and $\pi$ denotes the behavior policy.
The initial objective function is formulated as $J(\omega) = \mathbb{E} \left[ r(\omega) A_{\pi}(s, a) \right]$, where $A_{\pi}(s, a)$ is the advantage function. Rather than employing TRPO's KL divergence constraint, PPO introduces a simpler constraint mechanism by restricting $r(\omega)$ to a narrow interval $[1-\delta, 1+\delta]$, where $\delta$ is a hyperparameter. This leads to the clipped objective function in eq. (\ref{obj_PPO_1}),

\begin{align} \label{obj_PPO_1}
    & J(\omega) = \\
    & \mathbb{E}\left[ \text{min} (r(\omega)A^{\pi}(s_{t}, a_{t}), \text{clip}(r(\omega), 1-\delta, 1+\delta)A^{\pi}(s_{t}, a_{t})) \right]. \nonumber
\end{align}

One of the main innovations of PPO is its capability to conduct multiple epochs of mini-batch updates, which greatly improves sample efficiency in comparison to traditional policy gradient methods. PPO maintains the advantages of TRPO while also providing simpler implementation and better sample complexity. Empirical studies have consistently shown that the clipped version of PPO outperforms its KL-divergence counterpart in a variety of benchmark tasks.

The stochastic policy $\pi^{\omega}(a_{t} \mid s_{t})$ is implemented using different probability distributions depending on the action space: a categorical distribution for discrete actions and a Gaussian distribution for continuous actions. This flexibility, combined with its robust performance and algorithmic simplicity, has established PPO as a reliable choice for RL applications.

\subsubsection{Others}
\label{subsec:pgothers}

The actor critic with experience replay (ACER) algorithm incorporates various advancements such as a stochastic dueling network, truncated importance sampling, and a novel trust region method that enhances stability and sample efficiency \cite{DBLP:journals/corr/WangBHMMKF16}. Similarly, the actor-critic using Kronecker-factored Trust Region (ACKTR) approach utilizes the NPG and employs Kronecker-factored approximate curvature with trust region to optimize both the actor and critic components. Compared to other actor-critic methods, ACKTR has shown to be more sample efficient \cite{wu2017scalable}.



\subsection{Actor Critic}
\label{sec:AC}


\subsubsection{Overview}

The actor-critic framework combines value-based and policy-based RL methods. It has an actor that determines actions and a critic that evaluates those actions. The actor's policy optimization is influenced by the critic's value estimations, balancing the high variance of policy gradients with the bias of value approximation.

The strength of the framework lies in its ability to seamlessly integrate the benefits of policy search methods and learned value functions. This allows for learning from both complete episode returns and TD errors \cite{Schulmanetal_ICLR2016, konda2003onactor}. This dual approach effectively balances the trade-off between the high variance typically associated with policy gradients and the bias inherent in value function approximation. Modern actor-critic implementations incorporate various advanced techniques, such as trust region optimization \cite{schulman2015trust}, target networks \cite{mnih2015human}, and generalized advantage estimation (GAE) \cite{Schulmanetal_ICLR2016}, showcasing the framework's adaptability and extensibility.

\subsubsection{Actor Critic}
\label{subsec:ac}

The actor critic (AC) algorithm \cite{konda1999actor} is composed of two essential elements: the policy function and the value function. This algorithm utilizes a neural network to approximate the state value function, which in turn improves policy updates by reducing gradient variance, a common issue in vanilla policy gradient methods.

Let $\theta$ denote the value function parameters and $\omega$ represent the policy function parameters. The policy parameter update follows $\omega \leftarrow \omega + \alpha_{\omega}(t) Q^{\theta}(s_{t}, a_{t}) \nabla_\omega \ln \pi^{\omega}(a_{t} \mid s_{t})$, where $\alpha_{\omega}(t)$ represents the actor learning rate. The parameters of action-state value are updated according to $\theta \leftarrow \theta + \!\alpha_{\theta}(t) \!\left( r_{t}(s_{t}, \!a_{t}) \!+\! \gamma Q^{\theta}(s_{t+1}, \!a_{t+1}) \!-\!\!Q^{\theta}\!(s_{t}, \!a_{t}\!)\right) \nabla_{\theta} Q^{\theta}\!(s_{t}, \!a_{t}\!)$, where $\alpha_{\theta}(t)$ denotes the critic learning rate.


Actor-critic (AC) methods offer robust stability and favorable convergence but face practical limitations, leading to enhanced variants like asynchronous advantage actor-critic (A3C) \cite{mnih2016asynchronous}, which enables parallel training with multiple actors synchronizing periodically with global parameters. The advantage actor-critic (A2C) \cite{10577995}, a synchronous refinement of A3C, ensures more consistent parameter updates by requiring all actors to complete tasks before global updates, enhancing stability and accelerating convergence.

\subsubsection{Deterministic policy gradient}
\label{subsec:dpg}

Policies can be classified as either stochastic or deterministic in RL. A stochastic policy generates a probability distribution over actions, denoted as $\pi(s_{t})$, from which actions are sampled: $a_{t} \sim \pi(s_{t})$. The actor parameter updates in this framework rely on the state-action value function $Q(s_{t}, a_{t})$. However, for stochastic policies, the computation of the policy gradient computation requires integration over both state and action spaces, which can be computationally demanding, especially in environments with high-dimensional action spaces.

The deterministic policy gradient (DPG) algorithm \cite{silver2014deterministic} addresses the computational challenge of RL by directly mapping states to specific actions: $a_{t} = \pi(s_{t})$. This deterministic approach simplifies the integration process significantly by eliminating the need to integrate over the action space. As a result, the computational complexity is reduced, leading to improved sample efficiency, particularly in problems with large or continuous action spaces. The objective function for the deterministic policy is formulated as $J(\omega)=\int_{\mathcal{S}} d^{\pi}(s) Q^{\theta} \left(s, \pi^{\omega}(s)\right) ds$, where $d^{\pi}(s)$ represents the stationary distribution of states under policy $\pi$. The DPG allows for more efficient estimation of the action-value function compared to its stochastic counterpart. By applying the chain rule to $J(\omega)$, we can derive eq. (\ref{dpg_loss_gradient}),

\begin{equation}\label{dpg_loss_gradient}
    \begin{split}
        \nabla_\omega J(\omega) & =\left.\int_{\mathcal{S}} d^\pi(s) \nabla_a Q^{\theta}(s, a) \nabla_\omega \pi^{\omega}(s)\right|_{a=\pi^{\omega}(s)} ds \\
        & =\mathbb{E}_{s \sim d^{\pi}}\left[\left.\nabla_a Q^{\theta}(s, a) \nabla_\omega \pi^{\omega}(s)\right|_{a=\pi^{\omega}(s)}\right].
    \end{split}
\end{equation}

Deterministic policies offer computational efficiency by evaluating $Q^{\theta}(s, \pi^{\omega}(s))$ directly, avoiding the need for computing expected values across actions and eliminating importance sampling, which enhances stability. However, deterministic policies limit exploration, especially in low-noise environments. This can be mitigated by adding noise to the policy during action selection or by using an off-policy learning strategy with a stochastic behavior policy for sample collection.

\subsubsection{Deep Deterministic Policy Gradient}
\label{subsec:ddpg}

Deep Deterministic Policy Gradient (DDPG) \cite{DBLPLillicrapHPHETS15} extends the DPG framework by using deep neural networks to approximate the policy $\pi^{w}$ and critic $Q^{\theta}$, addressing stability issues in actor-critic methods for complex control problems. To enhance stability, DDPG incorporates DQN techniques like experience replay and target networks and adapts the framework for continuous action spaces. However, the deterministic policy limits exploration, so DDPG adds exploration noise by modifying the policy as $\pi^{\prime}(s_{t})=\pi^{\omega}(s_{t})+\mathcal{N}$.

A key challenge in value function approximation is computing $Q^{\text{target}}$ with the same network, leading to instability. DDPG addresses this with "soft" target updates, instead of direct weight copying as in DQN. By maintaining target policy and state-action value networks, DDPG achieves more stable training. The critic update follows $\theta \leftarrow \theta + \alpha_{\theta}(t) \left( y_{t} - Q^{\theta}(s_{t}, a_{t})\right) \nabla_{\theta} Q^{\theta}(s_{t}, a_{t})$, where the target value is computed as $y_{t} = r_{t}(s_{t}, a_{t}) + \gamma Q^{\theta^{\text{target}}}(s_{t+1}, \pi^{\omega^{\text{target}}}(s_{t+1}))$. Experience replay, adopted from DQN, stabilizes training by breaking temporal correlations and providing more independent, identically distributed samples. The actor update follows the same form as in DPG, while target networks use a soft update mechanism, $\theta^{\text{target}} \leftarrow \zeta \theta + (1 - \zeta) \theta^{\text{target}}$ and $\omega^{\text{target}} \leftarrow \zeta \omega + (1 - \zeta) \omega^{\text{target}}$, with a slow rate $\zeta$ to ensure gradual adjustments, enhancing training stability.

\subsubsection{Twin Delayed Deep Deterministic}
\label{subsec:td3}

Twin Delayed Deep Deterministic (TD3) \cite{fujimoto2018addressing} mitigates overestimation bias in actor-critic architectures by adapting DDQN techniques within the DDPG framework. Although DDQN separates value estimation and action selection, actor-critic models with neural network approximations still face overestimation. To counter this, TD3 applies the current policy instead of the target policy $y_{t} = r(s_{t}, a_{t}) + \gamma Q^{\theta^{\text{target}}}\left(s_{t+1}, \pi^{\omega}(s_{t+1}) \right)$.

Building upon the principles of double Q-learning, TD3 utilizes dual deterministic actors ($\pi^{\omega_{1}}, \pi^{\omega_{2}}$) paired with corresponding critics ($Q^{\theta_{1}}, Q^{\theta_{2}}$). This architecture results in double Q-learning Bellman targets, as shown in eq. (\ref{double_q_bellman}),

\begin{equation}\label{double_q_bellman}
    \begin{split}
        y_{t}^{1} = r(s_{t}, a_{t}) + \gamma Q^{\theta^{\text{target}}_{2}} (s_{t+1}, \pi^{\omega_{1}}(s_{t+1})), \\
        y_{t}^{2} = r(s_{t}, a_{t}) + \gamma Q^{\theta^{\text{target}}_{1}} (s_{t+1}, \pi^{\omega_{2}}(s_{t+1})).
    \end{split}
\end{equation}

While optimizing policy $\pi^{\omega}$ using independent estimation in target updates helps avoid policy-induced bias, the unconstrained critics can lead to situations where estimation of one critic consistently exceeds the other: $Q^{\theta_{2}} (s_{t}, \pi^{\omega_{1}}(s_{t})) > Q^{\theta_{1}}(s_{t}, \pi^{\omega_{1}}(s_{t}))$. TD3 addresses it through the clipped double Q-Learning algorithm, which takes the minimum of both estimation, as shown in $y_{1} = r(s_{t}, a_{t}) + \gamma \min_{j=1,2} Q_{j}^{\theta^{\text{target}}} (s_{t+1}, \pi^{\omega_{1}}(s_{t+1}))$.


In actor-critic models, interdependent policy and value updates can lead to value overestimation and degraded policy quality. TD3 mitigates this by updating the policy network less frequently than the Q-function, similar to DQN’s target network approach, ensuring accurate value estimation before policy updates. TD3 also prevents overfitting in deterministic policies by adding clipped random noise to actions and averaging over mini-batches, as shown in eq. (\ref{td3_smooth}),

\begin{equation}\label{td3_smooth}
    \begin{split}
        & y = r(s_{t}, a_{t}) + \\
        & \!\gamma \!\min_{j=1,2}\!Q_{j}^{\theta^{\text{target}}}\!\left(s_{t+1}, \!\pi^{\omega_{1}} \!\left(s_{t+1}\right)+\operatorname{clip}(\mathcal{N}(0, \sigma),\!-c,+c)\right),
    \end{split}
\end{equation}
where $\sigma$ and $c$ control the noise characteristics. This approach adds clipped random noise to selected actions and averages over mini-batches. TD3 addresses challenges in continuous action spaces by applying double Q-Learning and stabilization techniques, effectively reducing overestimation bias and ensuring stable training by balancing policy and value network updates.

\subsubsection{Soft Actor-Critic }
\label{subsec:sac}

Soft Actor-Critic (SAC) \cite{haarnoja2018soft, 10354525} is an off-policy actor-critic algorithm in the maximum entropy RL framework. By incorporating policy entropy into the reward, SAC encourages exploration while maintaining task performance, seeking a policy that maximizes both expected return and entropy at each state, thus balancing exploration and exploitation. The augmented reward function is $r_{\text {soft }}\left(s_t, a_t\right)=r\left(s_t, a_t\right)+\beta \underset{s_{t+1} \sim p}{\mathbb{E}}\left[\mathcal{H}\left(\pi\left(\cdot \mid s_{t+1}\right)\right)\right]$, where $\beta$ is a temperature parameter balancing entropy, $p$ is the state transition distribution, and $\mathcal{H}(\pi(\cdot \mid s_{t}))$ is the policy entropy, increasing action randomness to promote exploration and prevent premature convergence. $d^{\pi}$ is the joint distribution of $(s, a)$ induced by the policy $\pi$, and $\beta$ is a coefficient that balances reward maximization and entropy maximization. Building upon the traditional Bellman equation in eq. (\ref{bell_eq_tra}),

\begin{equation} \label{bell_eq_tra}
    Q\left(s_t, a_t\right)=r\left(s_t, a_t\right)+\underset{\substack{s_{t+1} \sim p \\ a_{t+1} \sim \pi}}{\mathbb{E}}\left[Q\left(s_{t+1}, a_{t+1}\right)\right].
\end{equation}

The soft Q-function evaluation can be derived in eq. (\ref{sac_evaluation}),

\begin{equation}\label{sac_evaluation}
    \begin{split}
        & Q_{\text {soft }}\left(s_t, a_t\right) - r\left(s_t, a_t\right) \\
        & = \beta \!\underset{s_{t+1}}{\mathbb{E}}\!\left[H\left(\pi\!\left(\cdot \!\mid \!s_{t+1}\!\right)\right)\right]+ \underset{s_{t+1}, a_{t+1}}{\mathbb{E}}\left[Q_{\mathrm {soft }}\left(s_{t+1}, \!a_{t+1}\right)\right]  \\
        & = \beta \underset{s_{t+1}}{\mathbb{E}}\left[\underset{a_{t+1}}{\mathbb{E}}\left(-\log \pi\left(a_{t+1} \mid s_{t+1}\right)\right)\right] \\
        & \quad \quad + \underset{s_{t+1}, a_{t+1}}{\mathbb{E}}\left[Q_{\mathrm {soft }}\left(s_{t+1}, a_{t+1}\right)\right]  \\
        & = \underset{s_{t+1}, \!a_{t+1}}{\mathbb{E}}\!\left[Q_{\mathrm {soft }}\!\left(s_{t+1}, \!a_{t+1}\!\right)\!-\!\beta \!\log \left(\pi\left(a_{t+1} \!\mid \!s_{t+1}\right)\right)\!\right].
    \end{split}
\end{equation}

Next, we apply the soft policy improvement process. Based on the soft policy improvement theorem, a new policy $\tilde{\pi}$ is defined as $\tilde{\pi}(a_{t} \mid s_{t}) \propto \exp \left(\frac{1}{\beta} Q_{\mathrm {soft}}^\pi(s_{t}, a_{t})\right), \quad \forall s_{t}$. Assume $Q$ and $\int \exp (Q(s_{t}, a_{t}))$ are bounded for any state $s$. Then $Q_{s o f t}^{\tilde{\pi}}(s_{t}, a_{t}) \geq Q_{s o f t}^\pi(s_{t}, a_{t}) \forall s_{t}, a_{t} .$ We can prove this theorem according to the definition of KL divergence, leading to eq. (\ref{KL_soft_ac}),

\begin{equation}\label{KL_soft_ac}
    \begin{split}
        & -\mathrm{D}_{\mathrm{KL}}(\pi(\cdot \mid s) \| \tilde{\pi}(\cdot \mid s))=\int_a \pi(a \mid s) \log \frac{\tilde{\pi}(a \mid s)}{\pi(a \mid s)} d a \\
        & =\int_a \pi(a \!\mid\! s) \!\log \!\tilde{\pi}(a \!\mid\! s) d a-\int_a \pi(a \!\mid\! s) \log \pi(a \!\mid\! s) d a \\
        & \!=\int_a \!\pi(a \!\mid \!s) \!\log \!\frac{\exp \left(\frac{1}{\beta} \!Q_{\mathrm{soft }}^\pi(s, \!a)\right)}{\int_a \!\exp \!\left(\frac{1}{\beta} Q_{\mathrm{soft}}^\pi(s, \!a)\!\right) d a} d a \!+\! \mathcal{H}(\pi(a \!\mid\! s)\!) \\
        & =\int_a \pi(a \mid s) \frac{1}{\beta} Q_{\mathrm{soft}}^\pi(s, a) d a- \\
        &  \!\quad \!\int_a \!\pi\left(\log \int_a \exp \left(\!\frac{1}{\beta} \!Q_{\mathrm{soft}}^\pi(s, a)\!\right) d a\right) d a \!+\! \mathcal{H}(\pi(a \!\mid\! s) \\
        & =\mathbb{E}_{a \sim \pi}\left[\frac{1}{\beta} Q_{\mathrm{soft}}^\pi(s, a)\right]- \\ 
        & \quad \quad \quad  \log \int_a \exp \left(\frac{1}{\beta} Q_{\mathrm{soft}}^\pi(s, a)\right) d a+\mathcal{H}(\pi(a \mid s)).
    \end{split}
\end{equation}

Maintaining the current value estimation $Q_{\mathrm{soft}^{\pi}}(s, a)$, but acting according to the new policy $\tilde{\pi}$,

\begin{align}
& \mathcal{H}(\tilde{\pi}(\cdot \mid s))+\mathbb{E}_{a \sim \bar{\pi}}\left[\frac{1}{\beta} Q_{\mathrm {soft }}^\pi \left( s, a \right)\right] \nonumber \\
& =-\mathrm{D}_{\mathrm{KL}}(\tilde{\pi}(\cdot \mid s) \| \tilde{\pi}(\cdot \mid s))+\log \int \exp \left(\frac{1}{\beta} Q_{\mathrm{soft}}^\pi(s, a)\right) d a \nonumber \\
& =0+\log \int \exp \left(\frac{1}{\beta} Q_{\mathrm{soft}}^\pi(s, a)\right) d a.
\end{align}

Since $\mathrm{D}_{\mathrm{KL}}(\tilde{\pi}(\cdot \mid s) \| \tilde{\pi}(\cdot \mid s)) \geq 0$ with equality holding only if $\pi = \tilde{\pi}$, we have $\mathcal{H}(\pi(\cdot \mid s)) +\mathbb{E}_{a \sim \pi}\left[\frac{1}{\beta} Q_{\mathrm {soft }}^\pi(s, a)\right] \leq \mathcal{H}(\tilde{\pi}(\cdot \mid s))+\mathbb{E}_{a \sim \tilde{\pi}}\left[\frac{1}{\beta} Q_{\mathrm {soft }}^\pi(s, a)\right]$. Multiplying both sides by $\beta$, the relationship between the soft $V$ values under the two policies is obtained as $\beta \mathcal{H}(\pi(\cdot \mid s)) + \mathbb{E}_{a \sim \pi}\left[Q_{\mathrm {soft }}^\pi(s, a)\right] \leq \beta \mathcal{H}(\tilde{\pi}(\cdot \mid s))+\mathbb{E}_{a \sim \tilde{\pi}}\left[Q_{\mathrm {soft }}^\pi(s, a)\right]$. Then, we can derive eq. (\ref{soft_improve_theorm}),

\begin{align}& Q_{\mathrm {soft }}^\pi(s, a) \nonumber \\
& =\mathbb{E}_{s_2}\left[r_1+\gamma\left(\mathcal{H}\left(\pi\left(\cdot \mid s_2\right)\right)+\mathbb{E}_{a_2 \sim \pi}\left[Q_{\mathrm {soft }}^\pi\left(s_2, a_2\right)\right]\right)\right] \nonumber \\
& \leq \mathbb{E}_{s_2}\left[r_1+\gamma\left(\mathcal{H}\left(\tilde{\pi}\left(\cdot \mid s_2\right)\right)+\mathbb{E}_{a_2 \sim \tilde{\pi}}\left[Q_{\mathrm {soft }}^\pi \left(s_2, a_2\right)\right]\right)\right] \nonumber \\
& = \mathbb{E}_{s_2}\left[r_1+\gamma\left(\mathcal{H}\left(\tilde{\pi}\left(\cdot \mid s_2\right)\right)+r_2\right)\right] \nonumber \\
& \quad \quad \quad \quad + \gamma^2 \mathbb{E}_{s_3}\left[\mathcal{H} \left(\pi\left(\cdot \mid s_3\right)\right)+ \mathbb{E}_{a_3 \sim \pi}\left[Q_{\mathrm {soft }}^\pi\left(s_3, a_3\right)\right]\right] \nonumber \\
& \vdots \nonumber \\
& \leq \mathbb{E}_{\tau \sim \tilde{\pi}}\left[r_0+\sum_{t=1}^{\infty} \gamma^t\left(\mathcal{H}\left(\tilde{\pi}\left(\cdot \mid s_t\right)\right)+r_t\right)\right] \nonumber \\
& =Q_{\text {soft }}^{\tilde{\pi}}(s, a) .
\label{soft_improve_theorm}
\end{align}

Therefore, our task is to find a new policy that satisfies the condition described in eq. (\ref{sac_update_policy}),

\begin{align}
    \pi_{\text {new}}\!=\!\underset{\pi^{\prime}}{\arg \operatorname{min}} \mathrm{D}_{\mathrm{KL}}\!\left(\pi^{\prime}\left(\!\cdot \!\mid\! s_t\right) \!\| \!\frac{\exp \left(\beta^{-1} Q_{\text {soft }}^{\pi_{\text {old}}}\!\left(s_t, \!\cdot \!\right)\!\right)}{Z^{\pi_{\text {old }}}\left(s_t\right)}\!\right).
    \label{sac_update_policy}
\end{align}

Eq. (\ref{sac_update_policy}) is equivalent to eq. (\ref{sac_update_policy_equivalent}),

\begin{align}
    & \underset{\omega}{\operatorname{min}} D_{K L}\left(\pi^\omega\left(\cdot \mid \!s_t\right) \| \exp \left[\beta^{-1} \!Q^\theta\left(s_t, \!a_t\right)-\log \left(Z^\theta\left(s_t\right)\right)\!\right]\!\right) \nonumber \\
    & = \underset{\omega}{\operatorname{min}} \!\underset{a_t \!\sim\! \pi^\omega}{\mathbb{E}}\left[\log \pi^\omega\left(a_t \!\mid\! s_t\right)\!-\!\beta^{-1} \!Q^\theta\!\left(s_t, \!a_t\!\right)\!+\!\log \!\left(Z^\theta\left(s_t\right)\right)\right] \nonumber \\
    & =\underset{\omega}{\operatorname{min}} \underset{a_t \sim \pi^\omega}{\mathbb{E}}\left[\beta \log \pi^\omega\left(a_t \mid s_t\right)-Q^\theta\left(s_t, a_t\right)\right].
    \label{sac_update_policy_equivalent}
\end{align}

Combining clip double Q-learning and experience replay, the loss function of the actor can be written in eq. (\ref{sac_actor_loss}),

\begin{align}
    & L_\pi(\omega) \!= \! \underset{s_t \!\sim d}{\mathbb{E}}\!\left[\!\underset{a_t \!\sim \pi^\omega}{\mathbb{E}}\!\left[\!\beta \!\log \!\pi^\omega\!\left(a_t \!\mid\! s_t\!\right)\!-\!\min _{\theta_i} \!Q^{\theta_i}\!\left(s_t, \!a_t\!\right)\!\right]\!\right].
    \label{sac_actor_loss}
\end{align}

The temperature parameter $\beta$ plays a key role in balancing entropy maximization with reward optimization. An improper $\beta$ can significantly impact performance; for instance, a high $\beta$ turns the algorithm into a standard AC method, overemphasizing entropy at the cost of reward signals. To address this, SAC introduces an adaptive temperature mechanism with a target entropy $\hat{H}$, acknowledging that optimal temperature varies by task and learning stage. The suggested value for the target entropy, $\hat{H}$, is determined to be $(A)$, and the temperature is adjusted automatically by optimizing an entropy-based loss function, as defined in eq. (\ref{entropy_alpha_loss}), 

\begin{equation}\label{entropy_alpha_loss}
    \begin{split}
        L(\beta) & =\underset{\substack{s_t \sim d \\
        a_t \sim \pi}}{\mathbb{E}}\left[-\beta \log \pi\left(a_t \mid s_t\right)-\beta \hat{H}\right] \\
        & =\beta \underset{s_t \sim D}{\mathbb{E}}\left[H\left(\pi\left(\cdot \mid s_t\right)\right)-\hat{H}\right].
    \end{split}
\end{equation}
The update formula of $\beta$ is given in eq. (\ref{entropy_alpha_update}),

\begin{equation}\label{entropy_alpha_update}
    \begin{split}
        \beta & =\beta- \alpha_{e}(t) \cdot \nabla_\beta L(\beta) \\
        & =\beta- \alpha_{e}(t) \underset{s_t \sim D}{\mathbb{E}}\left[H\left(\pi\left(\cdot \mid s_t\right)\right)-\hat{H}\right],
    \end{split}
\end{equation}
where $\alpha_{e}(t)$ represents a learning rate for the temperature.

\subsubsection{Others}
\label{subsec:ac_others}

Recent advancements in actor-critic methods have focused on computational efficiency and scalability through distributed architectures and optimized training. The importance-weighted actor-learner architecture (IMPALA) \cite{espeholt2018impala} demonstrates high scalability with its V-trace off-policy correction, enabling stable learning across thousands of machines. Other architectures, like GPU A3C (GA3C) \cite{babaeizadeh2016reinforcement} and Parallel Advantage Actor-Critic (PAAC) \cite{alfredo2017efficient}, enhance efficiency through hybrid CPU/GPU approaches and single-machine parallelization. Unsupervised Reinforcement and Auxiliary Learning (UNREAL) \cite{jaderberg2016reinforcement} improves data efficiency with auxiliary tasks.

Hybrid methods address limitations in policy gradients, such as  policy gradient and Q-learning (PGQ) \cite{o2016combining}, which combines policy gradients with Q-learning, and the reactor architecture \cite{gruslys2017reactor}, which improves sample efficiency using multi-step returns. Off-policy corrections, like retrace($\lambda$) \cite{munos2016safe}, combine importance sampling and tree-backup methods for low variance and efficiency, further enhanced in the reactor framework with a $\beta$-leave-one-out policy gradient for practical sample efficiency.

\subsection{Multi-agent DRL}
\label{sec:MA}

\subsubsection{Overview}

While single-agent DRL has achieved success in various domains, it often falls short in complex tasks, leading to increased interest in multi-agent systems. These systems mirror natural collective behaviors like birds migrating or wolves hunting in packs, offering advantages through cooperation, flexibility, adaptability, and improved safety through redundancy.

Traditional single-agent RL algorithms cannot be directly applied to multi-agent contexts due to unique challenges in agent interactions and coordination. This has motivated the development of multi-agent reinforcement learning (MARL), which includes approaches like multi-actor critic, multi-critic multi-DQN, multi-agent DDPG (MADDPG), and multi-agent PPO (MAPPO) to address different aspects of multi-agent learning challenges.

\subsubsection{Multi-Actor Critic}
\label{subsec:mac}

\begin{figure*}[!t]
\centering
\subfloat[Markov Games]{\includegraphics[width=1.5in]{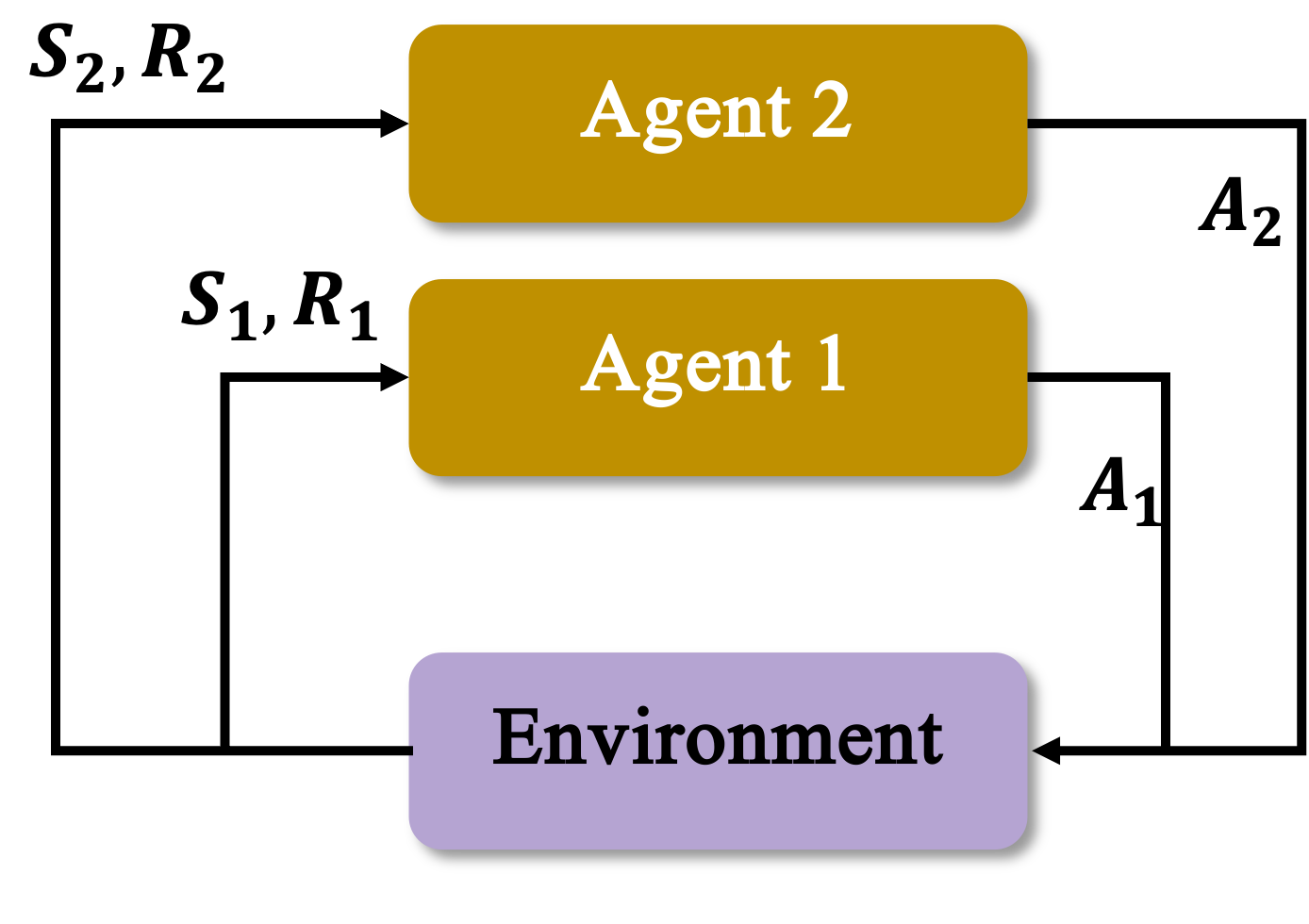}%
\label{ma_1}}
\hfil
\subfloat[POMG]{\includegraphics[width=1.5in]{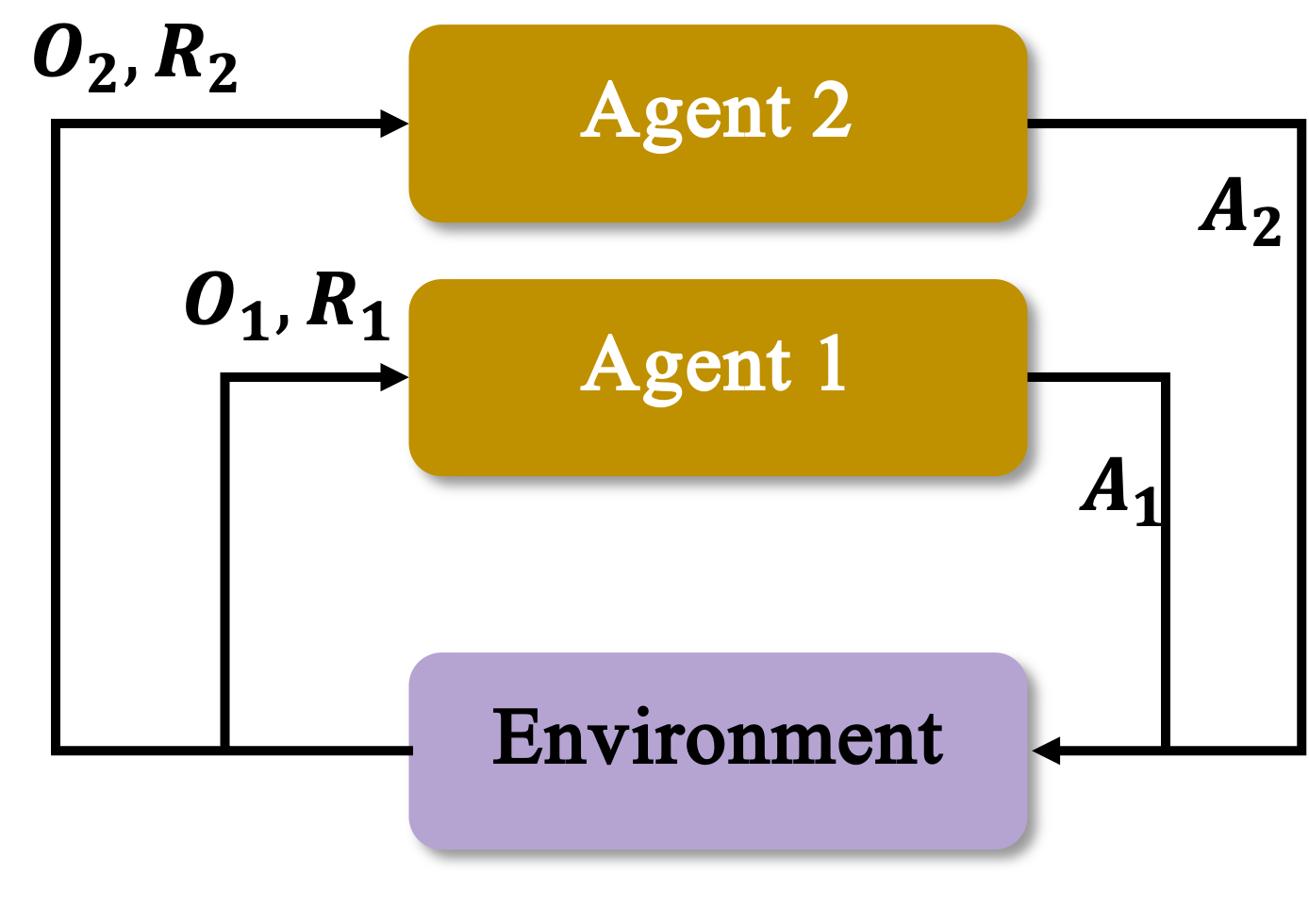}%
\label{ma_2}}
\hfil
\subfloat[Dec-POMDP]{\includegraphics[width=1.5in]{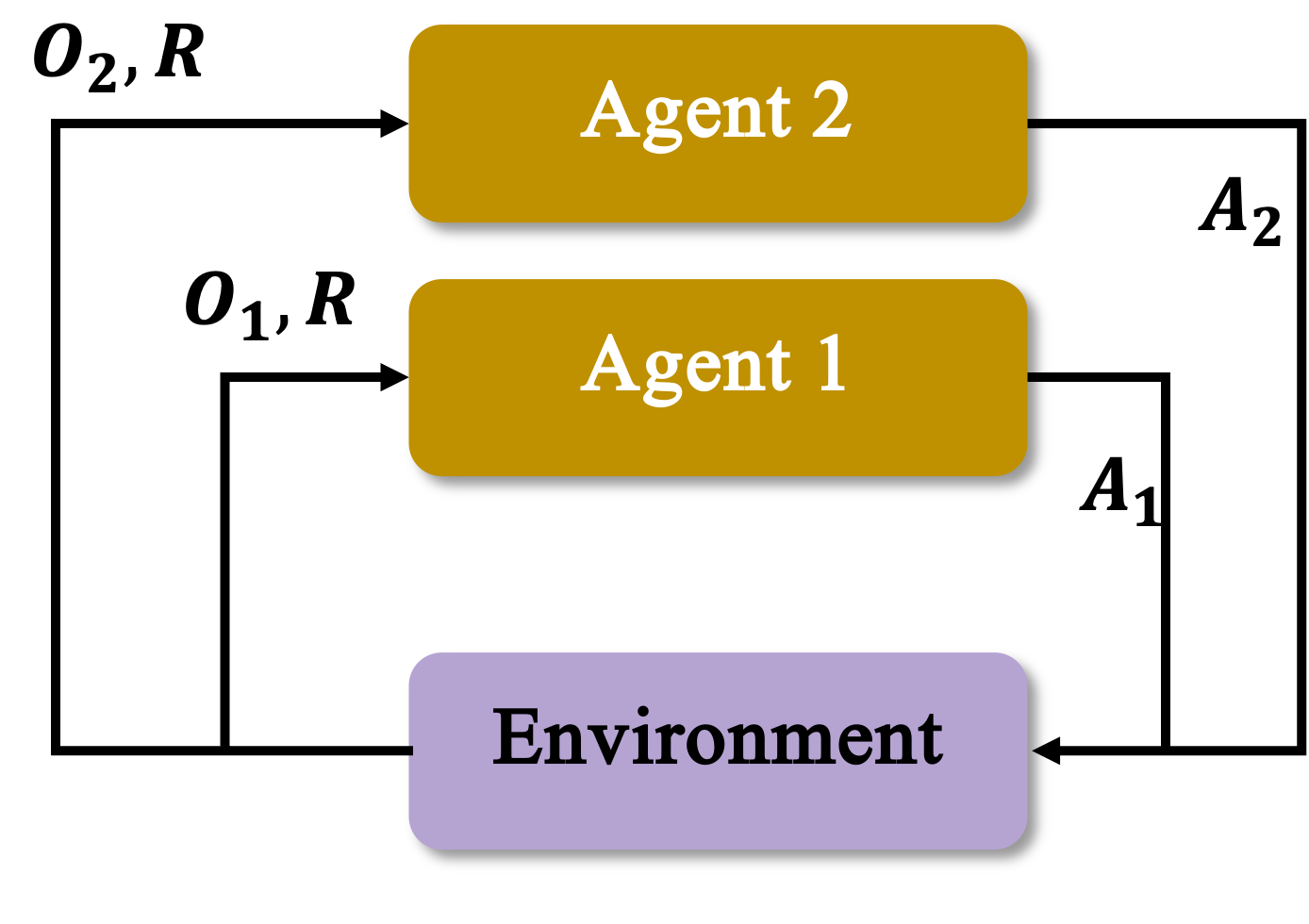}%
\label{ma_3}}
\caption{Different frameworks for MARL problems.}
\label{ma}
\end{figure*}

Markov games, also known as stochastic games, provide a theoretical framework for MARL by studying multiple interacting agents in a fully observable environment, applicable in cooperative, collaborative, and mixed settings. In decentralized partially observable MDPs (Dec-POMDPs), all agents aim to maximize a joint reward function, even though they have different individual goals. Partially observable Markov games (POMG), also referred to as partially observable stochastic games, correspond to Dec-POMDPs. However, instead of a joint reward function, agents optimize their respective reward functions in a partially observable environment. The comparison is illustrated in the Fig \ref{ma}.

POMG implicitly models the distribution of the belief states of other agents and is formally defined as a tuple $\langle \mathcal{I}, \mathcal{S}, {\mathcal{A}_{i}}, {\mathcal{O}_{i}}, d^0, \mathcal{P}, \mathcal{R} \rangle$, where $\mathcal{I}$ represents the set of $N$ agents, $\mathcal{S}$ the global state set, $\mathcal{A}_{i}$ the action set for agent $i$, and $\mathcal{A} = \mathcal{A}_{1} \times \mathcal{A}_{2} \times \ldots \times \mathcal{A}_{N}$ the joint action set. $\mathcal{O}_{i}$ defines the local observation set for agent $i$ at global state $s$, while $\mathcal{O} = \mathcal{O}_{1} \times \mathcal{O}_{2} \times \ldots \times \mathcal{O}_{N}$ forms the joint observation set. $d^0$ represents the initial state distribution, and $\mathcal{P}$ defines the state transition and observation probability set, where $\mathcal{P}(s_{t+1}, o_{t+1} \mid s_{t}, a_{t})$ represents the probability of transitioning to state $s_{t+1}$ and receiving observation $o_{t+1}$ after taking joint action $a_{t}$ in state $s_{t}$. The reward function $r_{i}: \mathcal{S} \times \mathcal{A} \to \mathbb{R}$ determines the reward for agent $i$.
Within this framework, agents utilize policies $\pi_{\omega_{i}}(a_{i} \mid o_{i})$ that are parameterized by $\omega_{i}$ to generate actions $a_{i}$ based on their individual observations $o_{i}$. Bold font is used to denote joint quantities across all agents, with the subscript $-i$ indicating quantities for all agents except agent $i$ (e.g., $\mathbf{a}=\left[a^{i}, \mathbf{a}^{-i}\right]$).

MARL algorithms typically fall into three frameworks: centralized learning, decentralized learning, and centralized training with decentralized execution (CTDE). Centralized methods learn a single policy mapping joint observations to actions \cite{claus1998dynamics}, but face challenges with exponential growth in observation and action spaces. Decentralized learning allows agents to optimize rewards independently \cite{littman1994markov}, reducing the action space from $|\mathcal{A}|^{n}$ to $n|\mathcal{A}|$, but may suffer from instability \cite{foerster2017learning}.

CTDE bridges these approaches, with methods including actor-critic structures using centralized critics \cite{lowe2017multi, foerster2018counterfactual} and value-decomposition methods \cite{sunehag2017value,rashid2020monotonic,son2019qtran}. These methods have achieved state-of-the-art results in MARL benchmarks \cite{wang2021rode, wang2021qplex}. MARL research spans four main areas \cite{claus1998dynamics, hernandez2018multiagent}: \textbf{behavior analysis}, where single-agent RL algorithms are applied to multi-agent systems; \textbf{agent communication}, which enhances learning through explicit information exchange; \textbf{cooperative learning}, which typically uses centralized training with distributed execution to enable collaboration; and \textbf{agent modeling}, which focuses on improving cooperation by modeling other agents' strategies. Agent communication enhances performance, agent modeling addresses imperfect information, behavior analysis is straightforward to implement, and cooperative learning achieves high training efficiency. This work primarily emphasizes behavior analysis and cooperative learning due to their widespread use in multi-agent systems.

\subsubsection{Multi-Critic Multi-DQN}
\label{subsec:mdqn}

Independent learning is explored in multi-agent environments through Q-learning \cite{tan1993multi}, particularly in cooperative games \cite{lauer2000algorithm}. Independent Q-learning (IQL) enables agents to learn policies individually while treating others as part of the environment \cite{tan1993multi}, reducing computational requirements compared to centralized learning.

The multi-DQN framework expands DQN by incorporating IQL, where each agent $i$ maintains its own DQN to calculate a local action-value function $Q_{i}(o_{i}, a_{i})$ based on observation $o_{i}$ and action $a_{i}$. This function estimates the expected cumulative future rewards for taking action $a_{i}$ from state $o_{i}$ under the joint policy $\boldsymbol{\pi} = [\pi_{i}, \boldsymbol{\pi}_{-i}]$. In fully observable settings, the Q-function takes the form of eq. (\ref{ma_q_function}),

\begin{equation}
    Q_i^{\boldsymbol{\pi}}\left(s_{t}, a_{t}^{i}\right)=E_{\boldsymbol{\pi}}\left\{\sum_{t=1}^{\infty} \gamma^{t-1} r_{i, t} \mid s_t=s, a_{t} = a^{i}_{t} \right\}.
    \label{ma_q_function}
\end{equation}

Despite non-stationarity and uncertain convergence, IQL has shown robust practical performance \cite{matignon2012independent}. The corresponding Bellman optimality for a single agent, considering other agents' policies, is expressed in eq. (\ref{ma_bellman_optimality}),

\begin{equation}\label{ma_bellman_optimality}
    \begin{split}
        & Q_{i}^*\left(s_{t}, \!a^{i}_{t} \!\mid\! \mathbf{a}^{-i}_{t} \right)=\sum_{\mathbf{a}^{-i}_{t}} \mathbf{a}^{-i}_{t} \!\left(\mathbf{a}^{-i}_{t} \!\mid\! s_{t}\right)[ r \!\left(s_{t}, \!a^{i}_{t}, \mathbf{a}^{-i}_{t} \!\right) \!+\! \\
        & \quad \gamma \sum_{s_{t+1}} \mathcal{P} \left(s_{t+1} \mid s_{t}, a^{i}_{t}, \mathbf{a}^{-i}_{t} \right) \max _{a^{i}_{t+1}} Q_{i}^*\left(s_{t+1}, a_{t+1}^{i} \right)].
    \end{split}
\end{equation}

One of the main challenges in this framework is the non-stationarity component $\boldsymbol{\pi}_{-i}\left(\mathbf{a}_{t}^{-i} \mid s_{t}\right) = \Pi_{-i \in N} \boldsymbol{\pi}_{-i} \left(a_{t}^{-i} \mid s_{t} \right)$, which changes as other agents' policies evolve during the learning process. This change creates a non-stationary environment for each agent, which goes against the Markov assumptions necessary for Q-learning convergence. The non-stationarity has a significant impact on experience replay in multi-DQN, as previous data becomes outdated and potentially misleading. This challenge can be formally expressed as $\mathcal{P} \left(s_{t+1} \mid s_{t}, a_{t}, \boldsymbol{\pi}_1, \ldots, \boldsymbol{\pi}_N\right) \neq \mathcal{P} \left(s_{t+1} \mid s_{t}, a_{t}, \tilde{\boldsymbol{\pi}}_1, \ldots, \tilde{\boldsymbol{\pi}}_N \right)$.

Without replay memory, IQL can still learn by tracking other agents' policies, but replay memory hinders this process. Previous work has attempted to address the issue by inputting other agent’s policy parameters to Q function \cite{tesauro2003extending}, explicitly adding the iteration index to the replay buffer. Therefore, there are many works that extend the normal Q-function of state-action pairs $Q(s, a)$ to a function of states and joint actions of all agents $Q(s, \Vec{a})$ \cite{littman1994markov, bowling2000convergence,hu1998multiagent,littman2001friend} or all other mixed strategy \cite{tesauro2003extending}. Off-environment importance sampling \cite{ciosek2017offer} can be used to stabilise experience replay \cite{foerster2017stabilising}. Therefore to enable importance sampling, at the time of collection $t_{c}$, $\pi_{-i}^{t_{c}}(\mathbf{a}^{-i}_{t} \mid s_{t})$ is stored in the replay memory, forming an augmented transition tuple $\left\langle s_{t}, a_{t}^{i}, r_{t}, \pi\left(\mathbf{a}_{t}^{-i} \mid s_{t}\right), s_{t+1}\right\rangle^{\left(t_c\right)}$. At the time of replay $t_{c}$, we train off-environment by minimising an importance weighted loss function in eq. (\ref{w_L_f_2}),

\begin{align}\label{w_L_f_2}
    \mathcal{L}(\theta_{i})\!=\!\sum_{i=1}^b \!\frac{\pi_{-i}^{t_c}\left(\mathbf{a}_{t}^{-i} \mid s_{t} \right)}{\pi_{-i}^{t_i}\left(\mathbf{a}_{-i} \!\mid\! s_{t}\right)}\left[\left(y^{i}_{target} - Q^{\theta_{i}}(s_{t}, \!a_{t}^{i}) \right)^2\right],
\end{align}
where $t_i$ represents the collection time of the $i$-th sample, and $b$ the batch size.

\subsubsection{Multi-Actor-Critic MADDPG}
\label{subsec:maddpg}


In multi-agent RL, centralized learning leads to exponential growth in state space with increasing agents. Decentralized learning faces two challenges: multi-critic approaches must address environmental non-stationarity, while policy gradient methods encounter higher variance with more agents, as each agent's reward is influenced by others' actions. However, a key insight reveals that if we know the actions taken by all agents, the environment remains stationary even as policies change. This can be formally expressed through the eq. (\ref{non_stationarity}), for any $\pi_i \neq \tilde{\pi}_i$,

\begin{equation}\label{non_stationarity}
    \begin{split}
        & \mathcal{P} \left(s_{t+1} \mid s_{t}, a_{t}^{1}, \cdots, a_{t}^{N}, \pi_{1}, \cdots, \pi_{N} \right) \\
        & = \mathcal{P} \left(s_{t+1} \mid s_{t}, a_{t}^{1}, \cdots, a_{t}^{N} \right) \\
        & = \mathcal{P} \left(s_{t+1} \mid s_{t}, a_{t}^{1}, \cdots, a_{t}^{N}, \tilde{\pi}_1, \cdots, \tilde{\pi}_N \right).
    \end{split}
\end{equation}

To address non-stationarity in IQL, hyper Q-learning \cite{tesauro2003extending} allows agents to learn policies based on estimation of other agents' policies, though this increases Q-function dimensionality. MADDPG \cite{lowe2017multi} tackles this dimensionality issue through neural network approximation and employs a centralized critic for each agent, enabling learning of continuous policies in both cooperative and competitive scenarios. In formal terms, let us consider a game involving $N$ agents, each with their own policies represented by the parameters $\boldsymbol{\omega}={\omega_1, ..., \omega_N}$. The set of policies is denoted as $\boldsymbol{\pi}={\pi_1, ..., \pi_N}$. For agent $i$, the gradient of the expected return $J(\omega_i)=\mathbb{E}[r_i]$ can be expressed in eq. (\ref{maddpg_pg_loss}),

\begin{align} \label{maddpg_pg_loss}
    & \nabla_{\omega^{i}} J\left(\omega^{i} \right)= \\
    & \mathbb{E}_{s \sim d^{\pi}, a_{t}^{i} \sim \pi_{i}^{\omega^{i}}}\left[\nabla_{\omega^{i}} \log \pi_{i}^{\omega^{i}}\left(a_{t}^{i} \mid o_{t}^{i} \right) Q^{\theta_{i}}\left(\mathbf{x}_{t}, a_{t}^{1}, \ldots, a_{t}^{N} \right)\right], \nonumber
\end{align}
where $Q^{\theta_i}(\mathbf{x}_t, a_t^1, ..., a_t^N)$ represents a centralized action-value function that takes as input the actions of all agents $a_t^1, ..., a_t^N$ at time step $t$, along with state information $\mathbf{x}_t$, and outputs the Q-value for agent $i$. Since each $Q^{\theta_i}$ is learned separately, agents can have arbitrary reward structures, including conflicting rewards in competitive settings. Now we can extend the stochastic policy $a_{t}^{i} \sim \pi^{\omega^{i}}$ to work with deterministic policies $a_{t}^{i} = \pi^{\omega^{i}}$ in eq. (\ref{d_s_p_2}),

\begin{align}\label{d_s_p_2}
    & \nabla_{\omega^{i}} J\left(\omega^{i}\right)=  \\
    & \mathbb{E} \left[\left.\nabla_{\omega^{i}} \pi^{\omega^{i}}\left(a_{t}^{i} \mid o_{t}^{i} \right) \nabla_{a_{t}^{i}} Q^{{\theta_{i}}}\left(\mathbf{x}_{t}, a_{t}^{1}, \ldots, a_{t}^{N} \right)\right|_{a_{t}^{i} = \pi^{\omega^{i}} \left(o_{t}^{i} \right)}\right]. \nonumber
\end{align}

The experience replay buffer $\mathcal{D}$ stores tuples in the form of $\left(\mathbf{x}_{t}, \mathbf{x}_{t+1}, a_{t}^{1}, \ldots, a_{t}^{N}, r_{t}^{1}, \ldots, r_{t}^{N} \right)$ , capturing the interactions of all agents. Each tuple includes the information state at time step $t$, $\mathbf{x}_{t}$, the subsequent state $\mathbf{x}_{t+1}$, the actions taken by all agents $a_{t}^{1}, \ldots, a_{t}^{N}$, and the rewards received $r_{t}^{1}, \ldots, r_{t}^{N}$. The centralized state-action value function $Q^{\theta_{i}}$ is then updated according to eq.  (\ref{maddpg_value_loss}),

\begin{align} \label{maddpg_value_loss}
& \mathcal{L}\left(\theta_i\right)=\mathbb{E}_{\mathbf{x}_{t}, a_{t}, r_{t}, \mathbf{x}_{t+1}}\left[\left(Q^{\theta_{i}} \left(\mathbf{x}_{t}, a_{t}^{1}, \ldots, a_{t}^{N}\right)-y_{t}\right)^2 \right]  \\
& y_{t}=r_{t}^{i} + \left.\gamma Q^{\boldsymbol{\theta_{i}}^{\text{target}}}\left(\mathbf{x}_{t+1}, a_{t+1}^{1}, \ldots, a_{t+1}^{N} \right)\right|_{a_{t+1}^{j} = \pi^{\omega_{j}^{\text{target}}} \left(o_{t+1}^{j} \right)}. \nonumber
\end{align}

The set of target policies with delayed parameters, denoted as $\omega^{\text{target}}$, comprises of individual policies for each agent: $\pi^{\omega_{j}^{\text{target}}} = \left\{\pi^{\omega_{1}^{\text{target}}}, \ldots, \pi^{\omega_{N}^{\text{target}}}\right\}$. In order to eliminate the need for prior knowledge for policies of the other agents, each agent $i$ can also maintain an estimated policy, denoted as $\hat{\pi}^{\omega_{i}^{j}}$, for the actual policy $\pi^{\omega^{j}}$ of agent $j$. This estimated policy is refined by optimizing the log-likelihood of agent $j$'s action, along with an entropy regularization term as described in eq. (\ref{maddpg_other_agent_policy}),

\begin{equation}
    \mathcal{L}\left(\omega_i^j\right)=-\mathbb{E}_{o_j, a_j}\left[\log \hat{\pi}^{\omega_i^j} \left(a_j \mid o_j\right)+\lambda \mathcal{H} \left( \hat{\pi}^{\omega_i^j} \right)\right],
    \label{maddpg_other_agent_policy}
\end{equation}
where $\mathcal{H}$ represents the entropy of the policy distribution.  $y_{t}$ in eq. (\ref{maddpg_value_loss}) can be replaced with estimated value $\hat{y}_{t}$ in eq. (\ref{maddpg_approximate_y}),

\begin{align} \label{maddpg_approximate_y}
    & \hat{y}_{t} = r_{t}^{i} + \\
    & \gamma Q^{\theta_{i}^{\text{target}}} \left(\mathbf{x}_{t+1}, \hat{\pi}^{\omega_{1}^{\text{target}}}\left(o_1\right), \ldots, \pi^{\omega_{i}^{\text{target}}}\left(o_i\right), \ldots, \hat{\pi}^{\omega_{N}^{\text{target}}}\left(o_N\right)\right) \nonumber,
\end{align}
where $\hat{\pi}^{\omega_{j}^{\text{target}}}$ represents the target network corresponding to the approximate policy $\pi^{\omega_{j}^{\text{target}}}$. To tackle high variance in competitive or cooperative environments, MADDPG introduces policy ensembles, maintaining and training $K$ distinct policies per agent. A policy is randomly chosen for rollouts, while all $K$ policies contribute to gradient updates. Extending DDPG with a centralized critic and decentralized actors, MADDPG enables agents to learn from estimated policies of others and reduces variance, making it effective in multi-agent settings.

\subsubsection{Multi-Actor-Critic MAPPO}
\label{subsec:mappo}

PPO has been adapted to multi-agent environments through two main approaches: Independent PPO (IPPO) \cite{de2020independent} and MAPPO \cite{yu2022surprising}. IPPO trains separate policies for each agent and has shown success in SMAC, but faces scalability issues and generally underperforms compared to specialized methods like QMix. MAPPO, building on parameter sharing principles \cite{gupta2017cooperative}, enables agents to learn from collective experiences while maintaining behavioral diversity through unique observations. MAPPO's development incorporates Orthogonal initialization \cite{hu2019simplified} and follows implementation insights from single-agent RL studies \cite{andrychowicz2020matters, engstrom2019implementation, Ilyas2020A, tucker2018mirage}. Research suggests using larger batches for gradient estimation, often dividing training data into no more than two mini-batches \cite{Ilyas2020A}. The actor network $\pi^{\omega}$ processes agent observations $o_t$ to generate appropriate action distributions.

The actor network $\pi^{\omega}$ processes agent observations $o_t$ to generate appropriate action distributions. The actions are then sampled from these distributions. The training objective of the network is to maximize the function defined in eq. (\ref{mappo_actor_loss}),

\begin{align} \label{mappo_actor_loss}
    & L(\omega)=\mathbb{E} \left[ \sum_{i=1}^n \min \left(r^{(i)} A^{(i)}, \operatorname{clip}\left(r^{(i)}, 1-c, 1 +c \right) A^{(i)}\right)\right] \nonumber \\
    & + \alpha_{e} \sum_{i=1}^n \mathcal{H} \left(\pi^{\omega}\left(o^{(i)}\right)\right),
\end{align}
where $r^{(i)}$ is the probability ratio between the current policy $\pi^{\omega}$ and the old policy $\pi^{\omega_{\text{old}}}$, $A^{(i)}$ and $o^{(i)}$ the actions and observations, respectively. The advantage estimation $A^{(i)}$ is calculated by GAE method. The term $\mathcal{H}$ refers to the entropy, $\alpha_{e}$ the entropy coefficient hyper-parameter. The critic network is optimized by minimizing loss function in eq.  (\ref{mappo_critic_loss}),

\begin{align} \label{mappo_critic_loss}
    & L(\theta) = \sum_{i=1}^n\left(\max \left\{\left(V_\theta\left(s^{(i)}\right)-\hat{R}_i\right)^2, \right. \right. \\
    & \left. \left. \!\left(\!\operatorname{clip}\!\left(V_\theta\left(s_i^{(k)}\!\right), V_{\theta_{\text{old}}}\left(s^{(i)}\!\right)\!-\! c, \!V_{\theta_{\text{old}}}\!\left(s^{(i)}\!\right)\!+\! c \!\right)\!-\!\hat{R}_i\right)^2\!\right\}\!\right) , \nonumber
\end{align}
where $\hat{R}_i$ denotes the discounted reward-to-go, and $n$ is the number of agents. MAPPO proposes that clipping the policy and value functions helps reduce non-stationarity from policy updates, while value normalization stabilizes value function learning.



\section{Value-based DRL for Power Allocation Optimization}

\begin{figure*}
    \centering
    \includegraphics[width=0.7\linewidth]{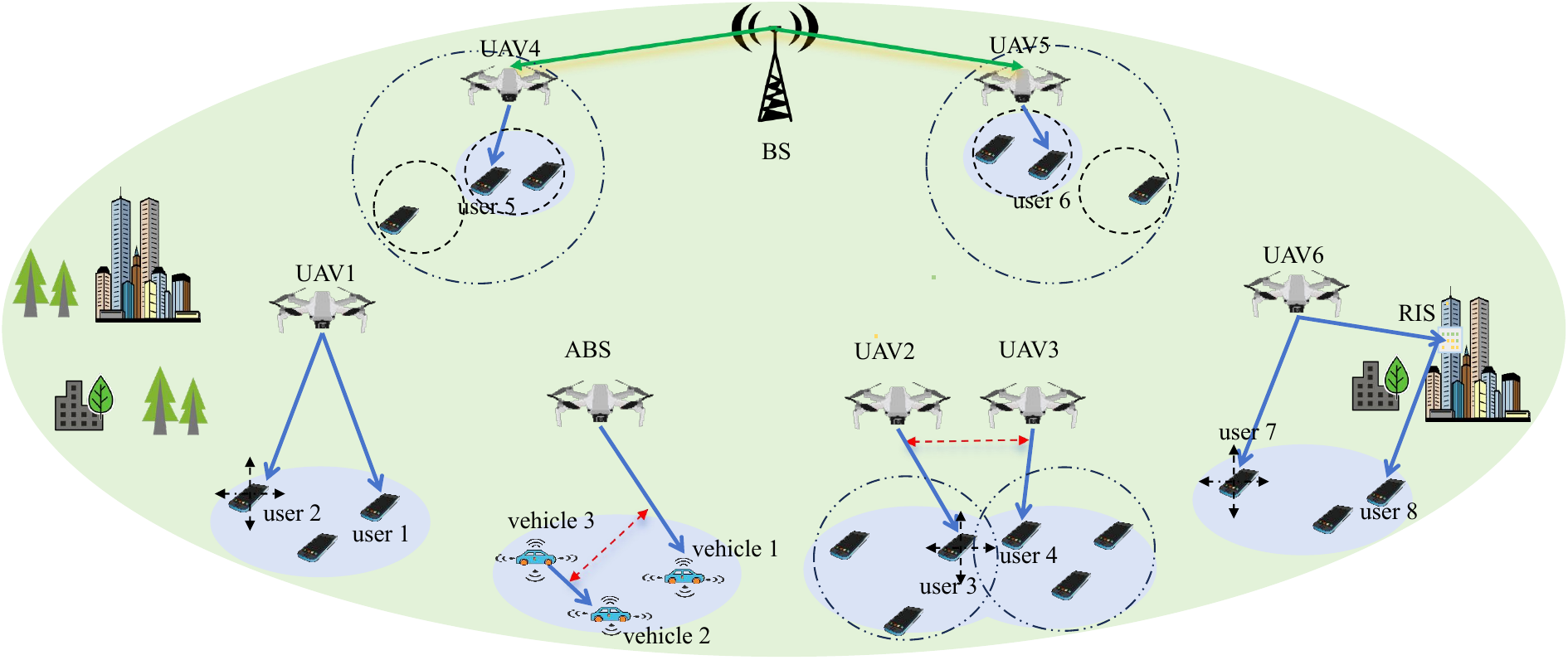}
    \caption{Power Allocation in different scenarios.}
    \label{Power Allocation scenario}
\end{figure*}

We investigate DRL applications in the field of power allocation into two categories, one-hop \cite{li, LiuYitong, Zhong, Qi} and multi-hop \cite{wang, tang, ZhangQian, LiXuanheng} scenarios to highlight the distinct challenges and approaches associated with each type of network topology.
In one-hop scenarios, communication occurs directly between a UAV and its base station or other communication nodes. The optimization problem typically focuses on maximizing power efficiency, reducing interference, and ensuring reliable communication while minimizing energy consumption. In this scenario, the problem is more straightforward, as the optimization is localized to the interaction between two nodes.
On the other hand, multi-hop scenarios involve communication across multiple UAVs or between a UAV and several relay stations, creating a more complex network topology. Here, power allocation must consider the cooperative interaction between several UAVs, accounting for the energy consumption of intermediate nodes, interference from neighboring nodes, and the need for maintaining signal strength across extended distances. The optimization becomes more complex due to the multi-agent nature of the problem, requiring coordination and distributed decision-making. The summary of DRL implementation for problem formulations of power allocation is provided in Table \ref{Power Allocation}.

\begin{table*}[]
\begin{center}
\caption{Power Allocation}
\label{Power Allocation}
\begin{tabular}{|c|c|c|c|ccc|}
\hline
\multirow{2}{*}{\textbf{Scenario}}               & \multirow{2}{*}{\textbf{Work}} & \multirow{2}{*}{\textbf{Alg.}} & \multirow{2}{*}{\textbf{Rewards}} & \multicolumn{3}{c|}{\textbf{Action Constraints}}                                                                                     \\ \cline{5-7} 
                                                 &                                &                                &                                   & \multicolumn{1}{c|}{\textbf{Direct Range}} & \multicolumn{1}{c|}{\textbf{0-1 Combination}} & \textbf{Others} \\ \hline
\multirow{4}{*}{One-hop}                         & \cite{li}                             & DQN                            & Inconsistent                      & \multicolumn{1}{c|}{NM}                     & \multicolumn{1}{c|}{NM}            & NM   \\ \cline{2-7} 
                                                 & \cite{LiuYitong}                      & DQN                            & Consistent                        & \multicolumn{1}{c|}{NM}                     & \multicolumn{1}{c|}{NM}            & NM             \\ \cline{2-7} 
                                                 & \cite{Zhong}                        & DQN                            & Consistent                        & \multicolumn{1}{c|}{PR}                                & \multicolumn{1}{c|}{/}                        & NM   \\ \cline{2-7} 
                                                 & \cite{Qi}                             & DDQN                           & Consistent                        & \multicolumn{1}{c|}{PN, NS}                            & \multicolumn{1}{c|}{/}                        & NM   \\ \hline
\multicolumn{1}{|l|}{\multirow{4}{*}{Multi-hop}} & \cite{wang}       & DQN                            & Consistent                        & \multicolumn{1}{c|}{/}                                 & \multicolumn{1}{c|}{NM}            & PR              \\ \cline{2-7} 
\multicolumn{1}{|l|}{}                           & \cite{tang}      & DDQN                           & Inconsistent                      & \multicolumn{1}{c|}{PR}                                & \multicolumn{1}{c|}{NM}            & NM   \\ \cline{2-7} 
\multicolumn{1}{|l|}{}                           & \cite{ZhangQian}                      & DQN                            & Inconsistent                      & \multicolumn{1}{c|}{NM}                     & \multicolumn{1}{c|}{/}                        & NM   \\ \cline{2-7} 
\multicolumn{1}{|l|}{}                           & \cite{LiXuanheng}                     & DDQN                           & Consistent                        & \multicolumn{1}{c|}{NM}                     & \multicolumn{1}{c|}{NM}            & NM   \\ \hline
\end{tabular}
\parbox{\textwidth}{\centering
\vspace{0.5em}
Inconsistent/Consistent: The reward is inconsistent/consistent with the objective in the problem formulation. Direct Range: There is a constraint on ranges for actions. 0-1 Combination: The actions are binary variables, and their linear sum remains a binary result. PD: Direct Penalty; PR: Penalty in Rewards; NS: Network Structure;  NM: Not mentioned how to implement constraints in DRL proposals.}
\end{center}
\end{table*}

\subsection{Value-based DRL in One-hop Scenarios}

In the one-hop scenario, \cite{li, LiuYitong, Zhong} adopt the DQN algorithm, while \cite{Qi} applies the DDQN algorithm. Traditional resource allocation methods face challenges in ensuring stable communication coverage due to the high mobility of UAVs and the instability of air-to-ground channels \cite{li}. 
Fig.~\ref{Power Allocation scenario} depicts the scenario analyzed in \cite{li}, which involves UAV 1 and users 1 and 2.
To ensure the continuity of communication coverage between UAV 1 and users 1 and 2, the channel allocation and power allocation of UAV 1 are optimized.
The paper proposes a joint allocation of subchannels and power to meet the coverage requirements of ground users, employing a DQN algorithm to solve this problem. 
The problem formulation aims to maximize spectral efficiency and minimize rate variance at each time slot, while the DRL framework focuses on maximizing cumulative rewards over all time slots. As a result, the solution derived from the DRL algorithm is not entirely aligned with the original problem's objective.

The study does not detail methods for incorporating constraints within the DQN framework, but several strategies can be proposed. For the transmit power range constraint, a ReLU activation function can ensure non-negative outputs. To enforce the sum of transmit power constraint, a penalty term can be added to the reward function, where exceeding the power threshold incurs a negative reward. Sub-channel allocation can be treated as a binary classification problem, with a sigmoid activation function used in the output layer to restrict values to [0, 1], which is then rounded to binary outputs. Similarly, the connection indication symbol constraint can be implemented by applying a penalty-based approach to the reward function, ensuring compliance while guiding the agent's learning process. These methods enhance the DQN framework's ability to address multiple constraints effectively.

A similar study to \cite{li} explores the optimization of bandwidth allocation, transmission power, and UAV scheduling in \cite{LiuYitong}. The mathematical model in this work aims to maximize the long-term average total data transmission, while its DQN-based approach seeks to maximize cumulative rewards. The solution obtained through the DQN algorithm aligns with the objective of the problem formulation, with the reward function defined as the amount of data transmitted. However, the paper does not detail methods for implementing the constraints outlined in the formulation model.To address the first constraint on the cluster service indicator for each UAV, the third constraint on the total bandwidth, and the fourth constraint on power range, penalty terms can be added to the reward function. If a constraint is violated, the corresponding penalty term is set to a negative value to reduce the reward; otherwise, it is set to zero. For the second constraint, where the cluster service indicator is binary (0 or 1), this can be enforced through the network structure by using a softmax function to generate a probability distribution over possible values, followed by one-hot encoding to set the highest-probability item to 1 and others to 0.
The fifth constraint, related to battery capacity, and the sixth constraint, concerning the energy of sensor nodes, can be implemented by introducing coefficients for the respective variables. The actual value of each variable is calculated as the product of its upper bound and the coefficient, which is constrained to the range [0, 1] using a sigmoid activation function. These methods provide a systematic way to incorporate the constraints into the DQN framework.

Additionally, trajectory design and power allocation in \cite{Zhong} are jointly coupled to enhance spectral efficiency and optimize system throughput while employing non-orthogonal multiple access (NOMA) in UAV-assisted cellular networks.
As depicted in Fig.~\ref{Power Allocation scenario}, this study involves UAVs 2 and 3, along with users 3 and 4. By jointly optimizing the trajectories and power allocation of UAVs 2 and 3, it effectively minimizes channel interference between the UAVs, thereby enhancing overall throughput.
The objective in the formulation model is to maximize system throughput across multiple time slots, aligning with the DQN goal of maximizing cumulative rewards. The solution derived from the DRL algorithm matches the optimal solution of the model's objective.
To implement the sixth constraint, a penalty term is incorporated into the reward function as part of the denominator. When the constraint on the data transmission rate is violated, the penalty term takes a large value, significantly reducing the reward and guiding the agent to adhere to the constraint.

However, the study does not detail the implementation methods for the remaining constraints. These constraints can be addressed by adding penalty terms to the reward function. If a constraint is violated, the penalty term takes a negative value; otherwise, it remains zero, leaving the reward unaffected. Alternatively, a direct penalty approach can be adopted, where any constraint violation replaces the reward with a negative penalty value.
For the first constraint regarding UAV trajectories, adjustments to the DQN network structure can be made. For example, a height coefficient can be introduced for UAV altitude, with the actual altitude calculated as the product of the maximum allowable height and the coefficient. The height coefficient is constrained to the range [0,1] using a sigmoid activation function, ensuring non-negative altitudes. A similar approach can be applied to the UAV's horizontal and vertical trajectory coordinates, using a tanh function to map the coefficients to the range [-1,1].

Finally, energy consumption is another consideration in UAV networks.
\cite{Qi} addresses this challenge by jointly optimizing content placement, spectrum allocation, co-channel link pairing, and power control to enhance UAV energy efficiency while ensuring user quality of service (QoS).
As shown in Fig.~\ref{Power Allocation scenario}, the study involves an aerial base station (ABS) and vehicles 1, 2, and 3. 
By optimizing power allocation between the ABS and vehicles 1 and 2, as well as between vehicle 3 and vehicle 2, the energy efficiency of the ABS is significantly enhanced.
Its objective in the problem formulation is to maximize the average energy efficiency over multiple time slots, while the goal of DRL is to maximize cumulative rewards. The solution obtained by its DDQN algorithm is consistent with the solution for the formulation objective. 
Additionally, to implement the first constraint on the range of the transmission rate in its DDQN algorithm, a penalty term is added to the reward function. When the transmission rate fails to meet the first constraint, the penalty term takes a negative value; otherwise, the penalty term is 0.
To implement the fourth and fifth constraints on the range of the transmission power in its DDQN algorithms, the actions are discretized and their range is confined within the bounds specified by the constraints.

However, the proposal does not outline specific methods for implementing the remaining constraints within the DRL framework.
To implement the second constraint on the reliability requirement and the third constraint on the average delay, we can follow the approach used for the first constraint in its DDQN algorithm. Specifically, a penalty term is added to the reward function for both the second and third constraints.

\subsection{Value-based DRL in Multi-hop Scenarios}

In multi-hop scenarios, typical problems, such as power allocation, spectral assignment, and UAV trajectory optimization \cite{wang, ZhangQian}, can benefit from the application of DRL algorithms to enhance network performance. Multi-hop communication inherently involves larger complexity due to the need to manage interactions across multiple nodes, making DRL's ability to handle high-dimensional state and action spaces particularly advantageous.
Moreover, multi-hop networks support a variety of applications, such as video transmission \cite{tang} and traffic forwarding \cite{LiXuanheng}, which require efficient resource management and low-latency communication. DRL-based approaches can dynamically allocate resources and adapt to changing network conditions, ensuring reliable data delivery and optimal performance in these scenarios.

\textbf{Typical Issues in Multi-hop Scenarios}

To handle communication disruptions caused by infrastructure damage in disaster scenarios \cite{wang}, UAVs can be deployed as information relays between the macro base station (BS) and users, helping to restore communication links.
As shown in Fig.~\ref{Power Allocation scenario}, the key elements in this example include the BS, UAVs 4 and 5, and users 5 and 6.
The BS achieves communication with users 5 and 6 by treating UAVs 4 and 5 as relay nodes. The overall spectral efficiency is improved by optimizing the transmission power allocation of the BS and the service area allocation of UAVs 4 and 5.
The objective of the problem formulation is to maximize the average spectral efficiency, while the DQN algorithm focuses on maximizing cumulative rewards. By using spectral efficiency as the reward function in the DQN algorithm, the solution produced by the algorithm aligns with the original problem objective.
To implement the first constraint, which limits the total allocated power, a direct penalty method is applied. Specifically, if the constraint is violated, the reward is set to a negative value. Otherwise, the reward remains non-negative.

However, the methods for implementing the second and third constraints on the zone selection are not mentioned in the paper.
The second and third constraints are about the 0-1 variable, which can be implemented by configuring the network structure accordingly.
Specifically, a softmax activation function is used to transform all outputs into a probability distribution, ensuring the probability sum to 1. The action with the highest probability is then selected as the output action.

Reconfigurable Intelligent Surface (RIS) is a key component in 6G and future networks. When integrated with UAV networks, the complexity of the optimization increases, particularly when jointly optimizing UAV trajectories, power allocation, and RIS phase shifts \cite{tang}.
For instance, in Fig.~\ref{Power Allocation scenario}, the elements considered include the RIS, UAVs 2, 3, and 6, along with users 3, 4, 7, and 8.
When there are obstacles preventing direct communication between UAV 6 and users 7 and 8, the RIS can act as a relay node to enable communication between UAV 6 and users 7 and 8. By jointly optimizing the movement trajectories and power allocation of UAVs 2, 3, and 6, as well as the RIS phase shifts, signal interference between UAVs is minimized and the overall throughput is improved.

Its objective of the problem formulation is to maximize the system throughput in each time slot, while the DRL algorithm is designed to maximize cumulative rewards over all time slots. This creates an inconsistency between the problem objective and the solution obtained by its DRL algorithm.
Additionally, to enforce the first to third constraints on the UAV's flying area and the eighth constraint on the equivalent channel coefficient, penalty terms are added to the reward function. If any of these constraints are violated, a significant penalty is subtracted from the reward to discourage such actions and guide the learning process toward feasible solutions.

However, the implementation methods for the remaining constraints are not explicitly discussed in the paper.
The seventh constraint on the feasible RIS phase shift range and the ninth constraint on the total UAV energy consumption can be easily addressed within the DRL algorithm. These constraints are straightforward to incorporate as they involve simple boundary checks or adjustments to ensure compliance.
For the fifth constraint, which involves the service indication mark between users and UAVs, it is recommended to follow the approach outlined in the paper. This involves introducing penalty values into the reward function. By assigning different penalty values based on the specific violations of various constraints, the DRL algorithm can effectively discourage actions that violate the constraint and encourage feasible solutions.
Furthermore, for constraints directly related to DRL actions, such as the fourth constraint to prevent UAV collisions and the sixth constraint on the total allocated power, the action mask approach is an effective solution. In this approach, action options that violate the relevant constraints are filtered out before an action is selected. This ensures that the DRL agent operates within a valid action space, avoiding constraint violations and improving both the reliability and efficiency of the learning process.

\textbf{Multi-hop Applications}

\cite{ZhangQian} investigates the challenge of efficient video transmission in UAV-assisted wireless networks by jointly optimizing video quality level selection and power allocation. The objective is to maximize the ratio of video quality to power consumption, thereby improving energy efficiency while ensuring secure video transmission. While the problem formulation aims to maximize energy efficiency at each time slot, the DRL algorithm is designed to optimize cumulative rewards across all time slots, leading to a potential misalignment with the formulation objective. 

However, the article does not detail the implementation of constraints. For the first constraint regarding the secrecy timeout probability, a penalty-based approach can be adopted by incorporating a penalty term into the reward function. Violations of this constraint result in a significant reduction in the reward, guiding the DRL agent toward actions that satisfy the constraint. The second constraint, which relates to selecting suitable layers from a group of pictures encoded from a video, can be addressed by structuring the output layer of the DRL action network such that each neuron corresponds to a possible action. Applying a softmax activation function to this layer transforms the outputs into a probability distribution, and the action with the highest probability is selected.
To handle the third and fourth constraints concerning the power allocation range, the action variables can be modified using an action coefficient mechanism. This involves scaling each action variable by its corresponding action coefficient, where the true action value is the product of the coefficient and the maximum possible value. A sigmoid activation function ensures that these coefficients remain within the given range, thereby enforcing compliance with the power allocation constraints. By adopting these methods, the DRL algorithm can effectively integrate the constraints into its optimization process, ensuring improved video transmission efficiency and security.


When UAVs are integrated with cognitive radio technology, energy efficiency of traffic forwarding becomes a critical concern. \cite{LiXuanheng} addresses this issue by exploring the use of cognitive radio technology-enhanced UAVs (CUAVs) to assist in data traffic forwarding within IoT networks. In this configuration, UAVs act as mobile relay stations, collecting data from ground-based IoT devices and forwarding it to BSs. By exploiting idle spectrum resources, the UAV establishes wireless backhaul links, enabling efficient data transmission.
The objective of this paper is to maximize energy efficiency over multiple time slots, while the goal of the DRL algorithm is to maximize cumulative rewards. The solution obtained through the DRL algorithm aligns with the target solution of the problem, as both aim to optimize energy efficiency in a long time. To apply the DRL algorithm, the reward is set as energy efficiency, allowing the agent to learn optimal policies that balance energy use with network performance.

However, the solution does not provide explicit methods for implementing the constraints. To address this, we propose the following solutions for each constraint.
For the first and sixth constraints that pertain to the service area selection for CUAVs, we can implement them by defining the network structure. Specifically, a softmax function can be used to generate a probability distribution over all possible values of the decision variable.  One-hot encoding can be applied to select the value with the highest probability to 1, and the remaining values can be set to 0. The action corresponding to the selected value of 1 is then chosen as the output action, ensuring the selection of valid service areas for the CUAVs.
For the second and seventh constraints, which relate to spectral selection, and the third constraint on the remaining energy, penalty terms can be incorporated into the reward function. If any of these constraints are violated, the corresponding penalty is added to the reward. The penalty reduces the reward value, discouraging the agent from selecting actions that lead to violations of these constraints.
To implement the fourth constraint, which restricts the power allocation range, a power coefficient can be introduced. The actual power value is determined by multiplying the maximum power limit by the power coefficient. This transforms the power allocation problem into setting the power coefficient, and a sigmoid activation function can be used to constrain the power coefficient to the range [0, 1], ensuring that the power allocation is within the specified bounds.
For the fifth constraint, which involves limiting certain decision variables to the range [0, 1], we can directly use a sigmoid activation function to restrict the variable to the desired range, ensuring the constraints are respected during the DRL process.

\section{Policy-based DRL for Channel Assignment Optimization}

\begin{figure*}
    \centering
    \includegraphics[width=0.7\linewidth]{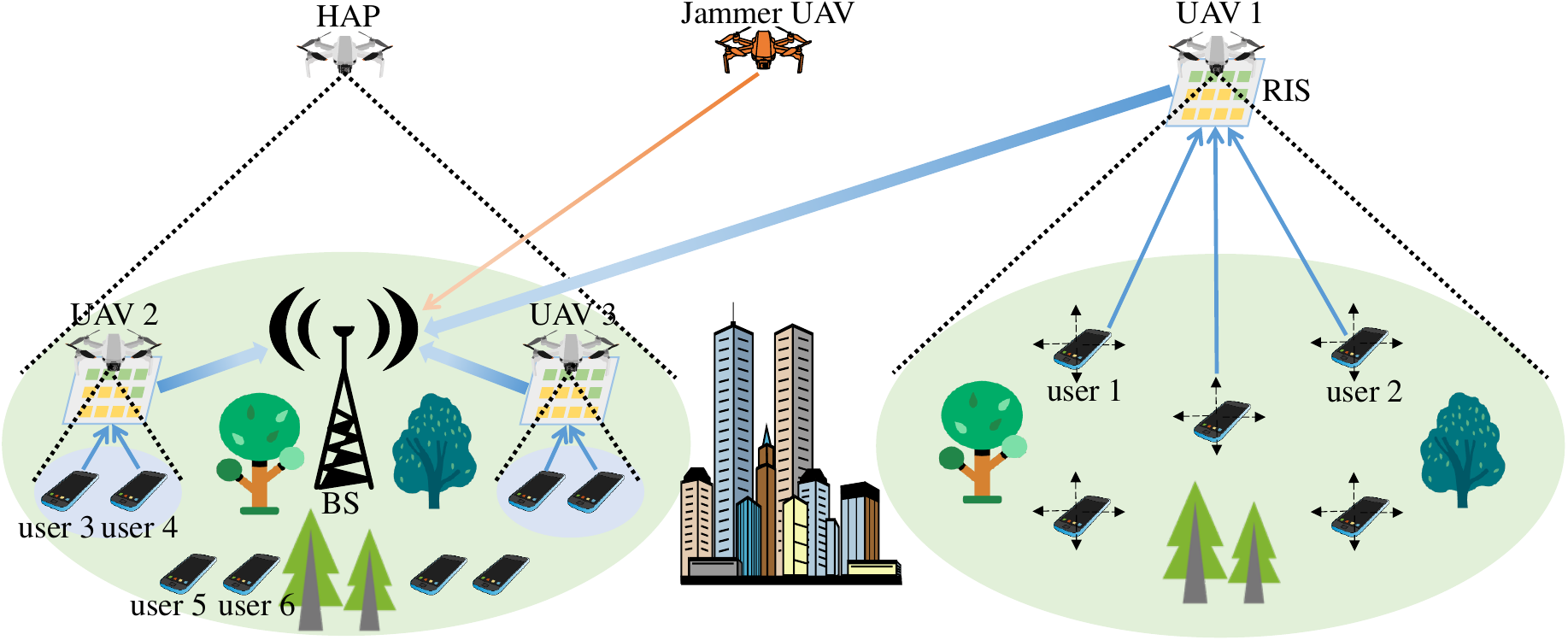}
    \caption{Channel assignment in different scenarios.}
    \label{channel assignment scenario}
\end{figure*}


\begin{table*}[]
\begin{center}
\caption{Channel Assignment}
\label{Channel Assignment}
\begin{tabular}{|c|c|c|c|ccc|}
\hline
\multirow{2}{*}{\textbf{Scenario}}               & \multirow{2}{*}{\textbf{Work}}        & \multirow{2}{*}{\textbf{Alg.}} & \multirow{2}{*}{\textbf{Rewards}} & \multicolumn{3}{c|}{\textbf{Action Constraints}}                                                                                     \\ \cline{5-7} 
                                                 &                                       &                               &                                   & \multicolumn{1}{c|}{\textbf{Direct Range}} & \multicolumn{1}{c|}{\textbf{0-1 Combination}} & \textbf{Others} \\ \hline
\multirow{5}{*}{One-hop}                         & {\cite{channelFairEE}} & DDPG                          & Consistent                        & \multicolumn{1}{c|}{PR}                                & \multicolumn{1}{c|}{/}                        & NS              \\ \cline{2-7} 
                                                 & {\cite{channel5}}      & DDPG                          & Consistent                        & \multicolumn{1}{c|}{NM}                     & \multicolumn{1}{c|}{NS}                       &   /              \\ \cline{2-7} 
                                                 & {\cite{channel6}}      & DDPG                          & Consistent                        & \multicolumn{1}{c|}{/}                                 & \multicolumn{1}{c|}{NM}            & NM   \\ \cline{2-7} 
                                                 & {\cite{channelQos}}   & DDPG                          & Consistent                        & \multicolumn{1}{c|}{NS}                                & \multicolumn{1}{c|}{NM}            & NM   \\ \cline{2-7} 
                                                 & {\cite{channel8}}      & DDPG                          & Consistent                        & \multicolumn{1}{c|}{NM}                     & \multicolumn{1}{c|}{NM}            & NM   \\ \hline
\multicolumn{1}{|l|}{\multirow{3}{*}{Multi-hop}} & {\cite{channel7}}      & TRPO                          & Consistent                        & \multicolumn{1}{c|}{PD,NS}                             & \multicolumn{1}{c|}{NM}            & NM  \\ \cline{2-7} 
\multicolumn{1}{|l|}{}                           & {\cite{channelEE}}     & PPO                           & Consistent                        & \multicolumn{1}{c|}{PR,NS}                             & \multicolumn{1}{c|}{/}                        & /               \\ \cline{2-7} 
\multicolumn{1}{|l|}{}                           & {\cite{channelAAoI}}   & PPO                           & Inconsistent                      & \multicolumn{1}{c|}{NM}                     & \multicolumn{1}{c|}{NM}            & PD              \\ \hline
\end{tabular}
\end{center}
\end{table*}

We explore the application of policy-based DRL techniques to optimize channel assignment in UAV communication networks. This section is divided into two subsections: DRL in one-hop scenarios and DRL in multi-hop scenarios, each addressing the unique challenges and solutions for channel assignment in different communication settings.
We focus on problems in the former where UAVs communicate directly with ground users or base stations using a single-hop communication model. The objective is to efficiently allocate communication channels between the UAV and users, ensuring that network performance is optimized in terms of throughput, fairness, and interference mitigation. We explore how policy-based DRL can be applied to dynamically assign channels, adapting to changes in the environment, and making decisions that maximize system overall efficiency.
The communication is more complex in the later, involving multiple UAVs and users, where data is forwarded across several hops before reaching its destination. This scenario presents additional challenges, such as managing inter-UAV interference, coordinating the channel assignment among multiple UAVs, and handling the dynamic topology of the network. We investigate how policy-based DRL can be used to optimize channel allocation in multi-hop networks, focusing on improving network throughput, minimizing delays, and ensuring efficient resource utilization in more complex communication environments.  The summary of DRL implementation for formulations of channel assignment problem is listed in Table \ref{Channel Assignment}. 

\subsection{Policy-based DRL in One-hop Scenarios}

We explore two key categories of research focusing on the DRL application in optimizing channel assignment in UAV communication networks under one-hop scenarios.
The first category addresses typical issues encountered in one-hop communication settings, including energy efficiency, fairness, and channel interference. These issues are crucial for ensuring efficient and reliable communication between a UAV and ground users or base stations. DRL techniques are applied to optimize channel allocation, reduce power consumption, balance fairness among users, and mitigate interference, thereby enhancing system throughput and overall network performance \cite{channelFairEE, channel5, channel6}.
The second category focuses on one-hop applications within UAV networks, such as data forwarding and task offloading \cite{channelQos, channel8}. Although these applications involve one-hop communication, they present unique challenges like QoS requirements for applications. DRL can be leveraged to dynamically manage UAV resources, optimize the forwarding of data to ground stations or users, and determine efficient offloading strategies for computational tasks.  

\textbf{Typical Issues in One-hop Scenarios}

The issue of balancing energy efficiency and fairness is addressed in UAV-assisted communication systems \cite{channelFairEE}. The main problem arises from the fact that a UAV, while optimizing its communication performance, also maintains fairness in resource allocation to all users over time. In Fig.~\ref{channel assignment scenario}, UAV 1 provides communication services for users 1 and 2.
To address these challenges, the study proposes a DDPG-based approach that aims to optimize the 3-D trajectory design and channel assignment for the UAV. 
The DDPG reward in the paper captures the trade-offs involved in optimizing both fairness and energy efficiency, with the UAV adjusting its trajectory and resource allocation accordingly. 
Furthermore, the study introduces the concept of penalty rewards to implement various constraints within the DRL framework. These penalty rewards are designed to ensure that all operational constraints, such as UAV acceleration limits, are respected throughout the optimization process, ensuring that the UAV acceleration stays within specified bounds. Similar penalty terms are used for other constraints, such as those related to power allocation and communication quality.

In the design of the reward function, the study does not clearly explain how some parameters, like distance metrics and UAV coordinates, are updated during the optimization process. Specifically, there is no detailed description of how the UAV speed and directional components in the action space affect its position in the state space.
Importantly, the paper provides comprehensive guidance on implementing constraints effectively within the DDPG framework, ensuring robust compliance with all system requirements.

The fairness and network throughput influenced by UAV mobility and channel assignment, similar to the work \cite{channelFairEE}, are considered in \cite{channel5}, but without information of user locations and channel parameters. Taking UAVs 2 and 3, and users at left in Fig.~\ref{channel assignment scenario} as an example, optimizing channel assignment for the two UAVs can mitigate channel interference while multiple orthogonal channels are shared. 
The authors propose a DDPG framework that incorporates a fairness indicator to optimize throughput while ensuring fairness in resource distribution.
The objective in the problem formulation is to maximize both the overall throughput and the fairness across consecutive time slots, which is consistent with the reward in the DDPG proposal.
Regarding the constraints, the paper focuses on converting discrete vector actions into a decimal integer, addressing specific operational limitations of the UAVs in terms of action selection. 

However, some constraints in the problem formulation are not fully addressed.
One significant physical limitation is the restriction on UAV movements to ensure they do not exceed the maximum allowable distance during their trajectories. This can be done by normalizing the movement vector based on the maximum allowable distance.
Another important constraint concerns power allocation, specifically ensuring that the total power across all channels does not exceed the maximum power capacity available to the system. To implement this constraint in the DDPG proposal, a power allocation sub-network can be integrated into the policy framework. The sub-network outputs power values that are then processed by a normalization layer.

Policy-based DRL for channel assignment is also studied in \cite{channel6} with a similar scenario to \cite{channel5}. The main difference lies in two assumptions. The first is that users are grouped into clusters. The second is that the cluster movement and UAV trajectories are described by a given model. Both of them are determined prior to channel assignment. 
The approach jointly optimizes channel assignment, power control, and user association to enhance cumulative throughput, demonstrating superior performance and effective coverage through simulations. The objective in the problem formulation is to maximize the cumulative throughput while ensuring uninterrupted services and meeting QoS requirements across multiple time slots. Meanwhile, the goal of the proposed DDPG scheme is to maximize cumulative network throughput. 

However, the work \cite{channel6} does not explicitly detail how constraints are implemented within the DRL framework. Therefore, we propose methods to incorporate these constraints effectively. The first constraint is related to the minimum height separation between UAVs to prevent collisions. To implement the constraint in the DRL framework, a penalty mechanism can be introduced: when this constraint is violated, a significant negative reward or penalty is applied. It discourages the agent from selecting actions that lead to violations.
Another constraint defines the binary variable for channel assignment and user association. DDPG can include a mechanism where selecting invalid combinations results in a penalty or reduced reward. Furthermore, the constraint ensuring that each subcarrier is assigned to only one mobile user at a given time can be implemented in DRL by validating actions before execution. Penalties are applied when overlapping assignments are detected.
For the constraint on the maximum transmit power of the UAVs, this can be integrated into DRL by monitoring power allocation during actions and imposing a negative reward when the power exceeds the given threshold. This discourages the agent from over-allocating power.
Finally, a constraint requires that the SINR between the UAV and mobile users meets a minimum threshold. The reward function can be designed to include SINR checks: if an action results in an SINR below the threshold, a penalty is assigned. In summary, these methods ensure that all constraints are effectively implemented in the DRL framework.

\textbf{One-hop Applications}

UAVs and high-altitude platforms (HAPs) can function as computing servers, allowing ground users with limited resources to offload tasks for processing \cite{channelQos}. For instance, as shown on the left side of Fig.~\ref{channel assignment scenario}, users sharing the channel resources can offload their tasks to the HAP or UAVs 2 or 3 for processing.
The primary goal in the problem formulation is to maximize the number of IoT devices whose tasks are completed within deadline constraints while minimizing overall energy consumption. The goal in the problem formulation focuses on long-term optimization, which aligns with the reward design in its DDPG proposal.

Although the study outlines the implementation of a specific constraint within the DRL framework, it overlooks some additional constraints that are essential for optimal performance. To implement the constraint related to task association, one solution is to use action discretization techniques. This approach converts the continuous outputs into discrete actions, ensuring that only feasible associations are considered when assigning IoT devices to UAVs or HAP. For the constraints on the total bandwidth allocation within a given threshold, one approach is to use a projection operator. The operator adjusts the bandwidth allocation at each decision step to ensure that the selected bandwidth stays within the allowable threshold. Additionally, the constraint related to energy consumption can be managed by integrating energy efficiency directly into the reward function. This integration encourages the model to prioritize energy savings while still meeting the other performance goals, such as completing tasks on time.

One assumption in \cite{channelQos} is that UAVs follow their predefined trajectories based on the pre-programmed flight plans. System performance can be improved by designing UAV trajectories. 
The study \cite{channel8} proposes a joint design strategy focusing on channel availability and UAV trajectories while taking into account traffic demands and energy refreshment over time slots in UAV-assisted cognitive radio IoT networks.
The DRL-based solution aims to maximize cumulative rewards, which implements the formulation objective. 

The proposal does not provide a clear explanation to implement the constraints within the DRL framework.
For example, the proposal imposes a constraint to ensure that a UAV only selects one serving area at a time. This can be implemented in the DRL framework by generating a probability vector and selecting the action with the highest probability. Another constraint involves ensuring that a UAV accesses a specific number of sub-channels from the available pool. This can be achieved in the DRL framework by applying a mask during action selection. Additionally, the paper includes a constraint requiring that the remaining energy of a UAV remains above a threshold. This can be implemented by incorporating a penalty into the reward function.
Further, there is a constraint on the transmission power of UAVs. This can be enforced in the DRL framework by using clipping techniques to ensure that the chosen transmission power stays within bounds. Another constraint ensures that some variables remain within a defined range, which can be achieved through normalization or bounded activation functions. Lastly, constraints related to binary decision variables for area selection and channel access can be implemented in the DRL framework by incorporating a softmax layer and using binarization logic during the post-processing of the action vector to ensure binary outputs.

\subsection{Policy-based DRL in Multi-hop Scenarios}
In this subsection, we also categorize studies into two groups: typical issues and multi-hop applications.
The first category addresses typical issues that arise in multi-hop scenarios, such as channel interference and trajectory design. 
The second category focuses on multi-hop applications while considering channel assignment, which are more complex and involve advanced network functionalities. Examples include channel jamming and age of information (AoI) in industrial IoT. Channel jamming presents a significant challenge in multi-hop networks, where malicious interference from a mobile UAV can degrade communication performance. On the other hand, UAVs often assist in data collection and transmission, and managing the age of information is crucial to ensure timely data delivery while delays are influenced by channel assignment. This subsection demonstrates how policy-based DRL can be effectively applied to solve these issues.

\textbf{Typical Issues in Multi-hop Scenarios}

Both energy efficiency and collision avoidance in UAV networks are closely linked to channel assignment and trajectory design \cite{channel7}, especially when UAVs make decisions in a distributed manner.
In Fig.~\ref{channel assignment scenario}, the network consists of UAVs 2 and 3, users 3 and 4.
The objective in its problem formulation is to maximize energy efficiency by optimizing UAV trajectories and channel allocation across multiple time slots in a THz-enabled UAV network, which is aligned with the reward in its DRL framework. 
The work specifically implements collision-related constraints through direct penalties and adjustments to the network structure. However, the paper does not fully address how to implement the constraint related to the minimum QoS for IoT devices in the DRL framework. 
To address the minimum QoS levels, we introduce a dual-component reward function that encourages the agent to meet requirements from IoT devices, penalizing deviations from thresholds.

\textbf{Multi-hop Applications}

Jamming threats are mitigated by considering the impact of channel assignment within a local area in hybrid RIS-mounted UAV networks \cite{channelEE}. For example, UAV 1, by utilizing RIS, enhances the anti-jamming performance for communications between the BS and users on the left side of Fig.~\ref{channel assignment scenario}.
The objective of the problem is to maximize long-term energy efficiency by jointly optimizing channel assignment for the RIS system, which is achieved through the reward design in its PPO-based approach. There are constraints on transmission power, channel assignment, and SINR requirements, which are incorporated into the PPO-based approach through a single penalty in the reward function. While this approach theoretically works, it has some practical drawbacks, as the impact of these constraints on energy efficiency varies. Moreover, UAV mobility is not considered in the model. For example, adjusting UAV trajectories closer to the receiver can help mitigate jamming threats.


Hybrid RIS-mounted UAV networks are also considered in order to manage AoI in industrial IoT  \cite{channelAAoI}. The network consists of the BS, HAP, UAVs 2 and 3, and users on the left side of Fig.~\ref{channel assignment scenario}. IoT devices can transmit data directly to the BS or via reflection through RIS-mounted UAVs. The channel assignment for links between IoT devices and the BS, along with UAV trajectories and RIS configuration, significantly impacts the objective of minimizing data AoI.
The main objective in its problem formulation is to minimize the maximum AoI across all devices over time, which is closely aligned with the long-term cumulative reward in its PPO algorithm.


In terms of implementing its constraints within the DRL framework, the paper primarily addresses one constraint, where violations lead to the termination of learning. However, for the implementation of the other constraints, several methods can be applied.
For UAV trajectories, constraints are introduced to ensure that UAV movement stays within predefined areas. This can be achieved by rescaling or clipping the output of the policy network to enforce boundary conditions on the UAV's paths.
The proposal can implement power constraints by normalizing the power allocation variables to ensure that they stay within the prescribed bounds. Similarly, to ensure proper power scaling for each user, the power output is scaled down proportionally to meet the required constraints.
For binary decisions, such as subcarrier assignment, the system can employ a binary encoding technique. The policy network outputs an integer within a defined range, which is then mapped to a binary vector. Each bit of this vector represents the state (0 or 1) of the corresponding element in the subcarrier assignment.
To ensure that RIS phase shift stays within the allowable range, a nonlinear activation function like tanh can be applied, and it is scaled and shifted accordingly. Likewise, to keep the reflection amplitude within the desired range, a sigmoid function can be used to enforce the amplitude constraint.


Besides, it is worth noting that data can be relayed among multiple UAVs in this scenario. However, the study primarily focuses on how to channel assignment and configure RIS settings, without addressing the design of routing paths for data transmission.

\section{Actor-Critic based DRL for Caching Optimization}

\begin{figure*}
    \centering
    \includegraphics[width=0.6\linewidth]{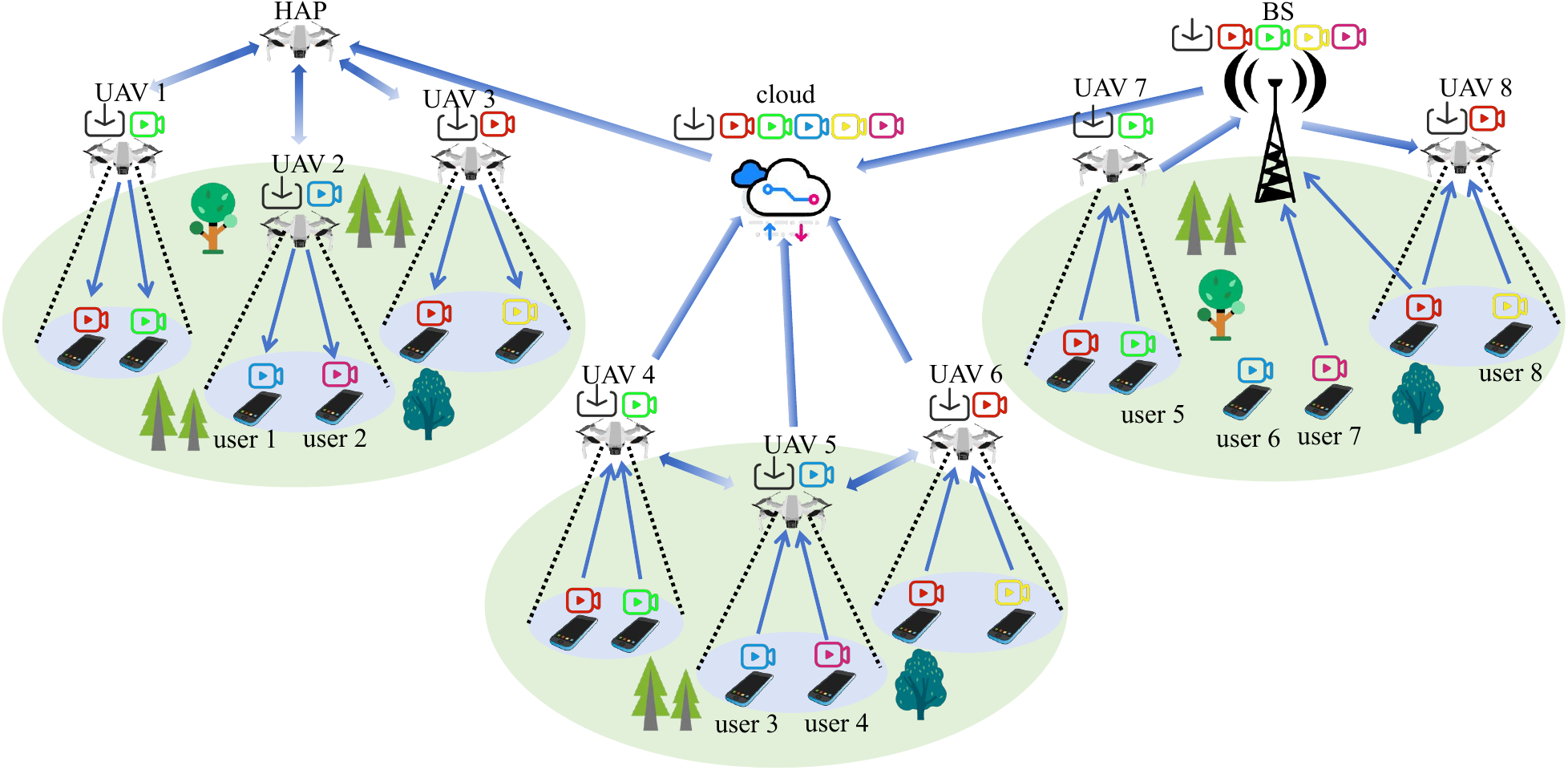}
    \caption{Caching optimization scenario}
    \label{Caching Optimization scenario}
\end{figure*}

We investigate DRL applications in the field of caching into two categories, caching without UAV interaction and caching with UAV interaction scenarios. This classification is based on whether UAVs operate independently or collaborate to optimize caching strategies. In caching without UAV interaction, individual UAVs focus on optimizing their own caching decisions to enhance content delivery, without explicit coordination with other UAVs, as explored in \cite{huang2024joint}, \cite{lin2024deep}, \cite{ji2022trajectory}, \cite{chen2024caching}, \cite{ren2024joint}. On the other hand, caching with UAV interaction involves multiple UAVs collaborating to share cached content, adjust their trajectories, and manage resources collectively, as discussed in  \cite{qin2024drl}, \cite{qin2024collaborative}, \cite{muslih2024cache}. This separation highlights the differing complexities and performance benefits associated with independent and cooperative UAV caching strategies. The summary of DRL implementation for problem formulations of caching optimization is provided in Table \ref{Caching}.

\begin{table*}[]
\begin{center}
\caption{Caching}
\label{Caching}
\begin{tabular}{|c|c|c|c|ccc|}
\hline
\multicolumn{1}{|c|}{\multirow{2}{*}{\textbf{Scenario}}} & \multirow{2}{*}{\textbf{Work}} & \multirow{2}{*}{\textbf{Alg.}} & \multirow{2}{*}{\textbf{Rewards}} & \multicolumn{3}{c|}{\textbf{Action Constraints}}                                                                                     \\ \cline{5-7} 
\multicolumn{1}{|c|}{}                                   &                                &                               &                                   & \multicolumn{1}{c|}{\textbf{Direct Range}} & \multicolumn{1}{c|}{\textbf{0-1 Combination}} & \textbf{Others} \\ \hline
\multirow{5}{*}{Caching without UAV Interaction}         & {\cite{huang2024joint}}                               & TD3                           & Consistent                        & \multicolumn{1}{c|}{PR}                                & \multicolumn{1}{c|}{NM}            & PR              \\ \cline{2-7} 
                                                         & {\cite{lin2024deep}}                               & TD3                         & Inconsistent                      & \multicolumn{1}{c|}{NM}                     & \multicolumn{1}{c|}{NM}            & NM   \\ \cline{2-7} 
                                                         & {\cite{ji2022trajectory}}                               & PPO                           & Consistent                        & \multicolumn{1}{c|}{NM}                     & \multicolumn{1}{c|}{NM}            & NM   \\ \cline{2-7} 
                                                         & {\cite{chen2024caching}}                               & PPO                           & Consistent                        & \multicolumn{1}{c|}{NS}                                & \multicolumn{1}{c|}{NS}                       & NM   \\ \cline{2-7} 
                                                         & {\cite{ren2024joint}}                               & PPO                           & Consistent                        & \multicolumn{1}{c|}{NM}                     & \multicolumn{1}{c|}{NM}            & NM   \\ \hline
\multirow{3}{*}{Caching with UAV Interaction}            & {\cite{qin2024drl}}                               & PPO                           & Consistent                        & \multicolumn{1}{c|}{/}                                 & \multicolumn{1}{c|}{NM}            & NM   \\ \cline{2-7} 
                                                         & \cite{qin2024collaborative}                              & DDPG                          & Consistent                        & \multicolumn{1}{c|}{PR}                                & \multicolumn{1}{c|}{PR}                       & PR              \\ \cline{2-7} 
                                                         & \cite{muslih2024cache}                               & DDPG                          & Inconsistent                      & \multicolumn{1}{c|}{NM}                     & \multicolumn{1}{c|}{NM}            & NM   \\ \hline
\end{tabular}
\end{center}
\end{table*}

\subsection{Caching without UAV Interaction}
In the caching without UAV interaction scenarios, \cite{huang2024joint} and \cite{lin2024deep} apply the TD3 algorithm, while \cite{ji2022trajectory}, \cite{chen2024caching}, and \cite{ren2024joint} utilize the PPO algorithm. These studies explore different approaches to optimize caching decisions in wireless networks, with each algorithm offering unique benefits in terms of stability, convergence, and learning efficiency.

\textbf{TD3 based Caching}

In the Internet of Vehicles (IoV) environment, mobile edge computing (MEC) systems face challenges in ensuring QoS and efficiently managing data caching. \cite{huang2024joint} proposes a joint UAV-assisted IoV architecture that leverages the mobile flexibility and line-of-sight communication capabilities of UAVs, combined with DRL methods, to optimize data caching and computation offloading, and reduce task processing delays. As illustrated in Fig. \ref{Caching Optimization scenario}, UAV 8, users (vehicles) 7 and 8, and the BS are involved in the scenario. User 8, located outside the service area of the BS, can benefit from QoS by accessing cached information provided by UAV 8. 
The objective in its problem formulation is to maximize UAV cache hit ratio and minimize average task processing delay, which is consistent with the reward in its TD3 algorithm. 
In addition, when the environmental constraints on cache capacity and processing delay are not met, a large negative penalty is added to the reward function to guide the agent towards feasible solutions.

However, the paper does not specify methods for implementing resource allocation and binary decision constraints. Below are several approaches to implement these constraints in the DRL framework.
For the resource allocation constraint, continuous action spaces can be used, where the action values represent decisions for resource allocation. The output can be constrained within the range of [0, 1] using clipping or activation functions such as sigmoid or tanh. Additionally, a penalty term can be added to the reward function, which applies a negative reward if the total resource allocation exceeds the allowed threshold. 
For the binary decision constraint, a common method is to define the action space as binary, where each action (such as task offloading or data caching) can only take two values: 0 (no action) or 1 (action taken). This ensures that decisions regarding task offloading, data caching, or other discrete actions are binary, as required by the problem formulation. The binary decisions can be implemented using techniques such as the Gumbel-softmax trick, which allows continuous relaxation of binary decisions, enabling the DRL algorithm to optimize these choices efficiently.


The UAV-assisted dynamic data caching and computation offloading strategy proposed in \cite{huang2024joint} aims to minimize task processing delays and energy consumption, thus enhancing system performance. However, in multi-UAV-assisted edge computing systems, the optimization of service caching and task offloading  becomes even more challenging due to network dynamics, storage and energy restrictions. To overcome this, \cite{lin2024deep} proposes a joint UAV-based architecture that integrates task popularity with a greedy algorithm. 
As illustrated on the right side of Fig. \ref{Caching Optimization scenario}, UAVs 7 and 8, users, and the BS are involved in the scenario. The request delay of the users benefits from the cached data of UAVs 7 and 8, while the BS handles the small amount of data not cached by the UAVs.
The problem objective minimizes the overall system delay in each time slot, while its DRL algorithm aims to maximize cumulative rewards.  The solution obtained by PPO differs from that in each time slot. 

However, the paper does not specify exact methods in its PPO framework for implementing the constraints.
To implement energy consumption, resource capacity, and geographic coverage constraints, a general approach is to use the penalty method. In this method, a penalty term is added to the reward function in DRL to discourage constraint violations. For example, to ensure that energy consumption limits are respected, if the energy consumed by a device or UAV exceeds its specified maximum energy budget, a penalty is applied to the reward. This penalty reduces the agent reward, guiding it to avoid actions that violate the energy constraints.
For binary constraints, such as offloading decisions and data caching, the sigmoid activation function can be applied to constrain the output within the range of [0, 1]. A threshold (e.g., 0.5) can then be set, where outputs greater than 0.5 are set to 1 (action taken), and outputs less than or equal to 0.5 are set to 0 (no action). This binary decision-making approach ensures that the DRL algorithm only selects discrete actions while adhering to the problem constraints.
To handle continuous constraints, such as resource allocation and task offloading, the sigmoid function can also be applied. The sigmoid function maps continuous variables to the range of [0, 1], making it suitable for normalizing and restricting values within the acceptable limits for continuous decision variables.

\textbf{PPO based Caching}

Although the service caching optimization strategy proposed in \cite{lin2024deep} provides significant improvements, the challenge of dealing with uncertainties in user requests and network dynamics still remains, especially in multi-access environments. To address these issues, \cite{ji2022trajectory} integrates edge caching and UAV communication by incorporating UAV trajectory optimization. It reduces content acquisition delays and enhances the performance of UAV systems in highly dynamic environments, by means of PPO to adapt to the unpredictable network states. The formulation objective is to minimize the total content retrieval delay for all users in the cell over the entire time period. The solution obtained by the PPO algorithm aligns with the objective. To apply DRL and encourage agent exploration, the reward consists of both extrinsic and intrinsic components, where the objective function influences the extrinsic reward, and the agent's exploration behavior affects the intrinsic reward. 

However, the paper does not specify the exact methods for implementing the constraints in its PPO framework. Below are several approaches for implementing these constraints. For the binary decision variable constraints related to cache capacity, cache diversity, and user association, penalty terms can be added to the reward function for violations. For the constraints concerning transmission power and UAV movement, the PPO output values are mapped using sigmoid or tanh functions to ensure they remain within the corresponding physical constraint ranges.

While UAV trajectory optimization has been shown to effectively reduce content acquisition delays \cite{ji2022trajectory}, the need for real-time updates to the caching strategy remains critical for addressing the dynamic and unpredictable nature of user requests. To solve this problem, the study in \cite{chen2024caching} proposes a joint design of caching strategy and resource allocation in UAV networks. Specifically, the problem is formulated as an optimization problem aimed at minimizing the average delay of content downloading, which is proven to be NP-hard. To address this, the problem is first modeled as a MDP, and then dual-clip PPO is proposed. The algorithm adaptively determines the caching strategy, UAV trajectory, and transmission power, balancing exploration and exploitation by optimizing the strategies. Its problem objective is to minimize the average content download delay, while the goal of DRL is to maximize cumulative rewards. The solution obtained by the DRL algorithm is consistent with the objective. 


Dynamic caching optimization for general content delivery is addressed in \cite{chen2024caching}. With the growing demand for video content, \cite{ren2024joint} extends the work by introducing a video-adaptive caching solution. By combining LSTM networks with PPO, this study optimizes UAV trajectories and caching decisions specifically for video content, responding to the highly dynamic nature of video requests. The proposed method significantly improves system performance, ensuring faster content delivery and reduced latency in video streaming. The objective in the problem formulation is to minimize the total system service delay over time slots, while the goal of DRL is to maximize cumulative reward. The solution obtained by the DRL algorithm is consistent with the objective solution of this paper. 
\subsection{Caching with UAV Interaction}

For caching optimization with UAV interaction, \cite{qin2024drl} utilizes PPO algorithms, while \cite{qin2024collaborative} and \cite{muslih2024cache} apply DDPG algorithms. 

\textbf{PPO based Caching}

In \cite{qin2024drl}, in the complex environment of a multi-UAV collaborative caching network, traditional content retrieval methods face challenges such as high latency and bandwidth bottlenecks. To address this challenge, this paper proposes a multi-UAV collaborative caching architecture in a NOMA manner, utilizing the mobility and cooperation capabilities of UAVs. The architecture, combined with DRL methods, optimizes caching decisions and flight trajectory planning to reduce content retrieval delay and improve content hit ratio. As illustrated on the left side of Fig. \ref{Caching Optimization scenario}, HAP, UAVs 1-3, users, and the cloud are involved in the scenario. Based on the data cached by the UAVs, users 1 and 2 have three request paths: local UAV 2, neighboring UAVs 1 and 3, and the cloud, with the HAP acting as an intermediary for data transmission.

Its objective in the problem formulation is to maximize the long-term weighted content hit ratio. The solution obtained by the PPO based proposal is consistent with the in the formulation. The reward is set as the total content retrieval delay for all MUs in each time slot. However, the paper does not specifically describe methods for enforcing the constraints. Here are several approaches to implementing these constraints. To implement the binary caching decision constraint, a discrete action space is defined where caching decisions take binary values (0 or 1). This is achieved using activation functions like Sigmoid or softmax. To ensure the caching capacity constraint, penalty terms are applied for exceeding the UAV's capacity, or caching capacity is modeled as a state variable adjusted via optimization. For the content retrieval mode constraint, softmax functions in the action space ensure that only one retrieval mode (edge, cooperative, or cloud) is selected per user per time slot.

\textbf{DDPG based Caching}

The study in \cite{qin2024drl} addresses the optimization of caching and UAV trajectory planning in multi-UAV collaborative networks, but it mainly focuses on fixed spectrum resource allocation. \cite{qin2024collaborative} introduces an air-to-air-to-ground edge computing network supporting clustered NOMA, which faces challenges such as spectrum resource scarcity and dynamic network environments. To tackle these issues, \cite{qin2024collaborative} proposes an architecture based on clustered non-orthogonal multiple access, utilizing satellites and multiple UAVs as collaborative edge servers. This approach, combined with DRL methods, optimizes data caching and computation offloading, thereby reducing system latency. As illustrated in the middle of Fig. \ref{Caching Optimization scenario}, UAVs 4-6, users, and the cloud (satellite) are involved in the scenario. Users offload tasks to UAV 5, or to UAVs 4 and 6, or the cloud, based on cached strategies. 

The objective in the problem formulation is to minimize the average system delay over the entire time period. The solution obtained by the DRL algorithm aligns with the objective. Considering the constraints on task offloading, program caching, and UAV distance, penalty terms are added to the reward function to minimize system delay. The task offloading and program caching constraints both involve binary decisions.  The task offloading constraint ensures that each terminal can offload to only one UAV, while the program caching constraint ensures that the total cached program size at each UAV does not exceed its storage capacity. These constraints can be implemented using a discrete action space, where the network output is mapped to a range of zero to one using a softmax or sigmoid function.
 
The framework presented in \cite{qin2024collaborative} effectively addresses the challenges posed by dynamic network conditions and task offloading. However, it assumes static UAV positions, limiting its flexibility. \cite{muslih2024cache} explores a highly dynamic environment in which UAVs adjust their positions in real-time while also sharing their cached content, offering a more adaptable solution for reducing transmission delays in UAV-based networks. 

The objective \cite{muslih2024cache} is to minimize the total transmission delay for ground users in each time slot. The solution obtained by the DRL algorithm differs from the maximum reward per time slot. In addition, the paper does not specify the exact methods in the DRL framework for implementing these constraints. For binary constraints, such as cache hit and user association decisions, use a discrete action space or binary selection, allowing the agent to choose the appropriate variables at each step.  For linear constraints like cache capacity, transmission delay, maximum speed, and user limits, penalty terms are added to the reward function for violations.  Position and coverage constraints are handled by a boundary filter, keeping UAVs within the service area.  For collision avoidance, the DRL policy monitors UAV distances and adjusts trajectories to maintain a safe separation.

The objective in \cite{muslih2024cache} is to minimize the total transmission delay for ground users in each time slot. However, the solution obtained by the DRL algorithm maximizes the cumulative reward across all time slots, which may differ from the objective of minimizing delay in a single time slot. Additionally, the paper does not specify the exact methods for implementing constraints within the DRL framework.
For binary constraints, such as cache hit and user association decisions, a discrete action space or binary selection approach can be used. This allows the agent to choose the appropriate actions at each step, ensuring that decisions are made in a way that adheres to the binary nature of the problem.
For linear constraints, such as cache capacity, transmission delay, maximum speed, and user limits, penalty terms are incorporated into the reward function to penalize constraint violations. These penalties guide the agent towards feasible solutions by discouraging actions that lead to violations.
Position and coverage constraints are typically handled by using a boundary filter, which ensures that UAVs remain within the designated service area.
For collision avoidance, it can be implemented by using a distance-based reward or penalty system.

\section{Multi-agent DRL for UAV-assisted Networks}

In UAV-assisted communication networks, MADRL has emerged as a powerful framework for addressing various optimization challenges, including UAV trajectory planning, energy minimization, task offloading, etc. The dynamic and complex nature of these networks has led to three distinct MADRL implementation approaches. In the basic approach, UAVs are modeled as homogeneous agents, focusing primarily on UAV-specific optimization objectives. However, given the complexity of communication systems involving multiple interacting entities (UAVs, users, and sometimes remote infrastructure like satellites), many studies adopt a heterogeneous agent modeling approach. This approach treats different system components as independent intelligent agents with distinct characteristics, objectives, and observational capabilities. For instance, UAVs may focus on trajectory and energy optimization, while users prioritize data transmission quality. Such heterogeneous modeling enables more realistic and efficient system-level optimization, though it requires advanced techniques like shared policies with individual adaptability and modular training strategies to balance different entities' constraints while ensuring coordination.

To further enhance training efficiency and convergence, researchers have developed a hybrid approach that decomposes complex problems into two sub-problems. The first subproblem is typically addressed using traditional optimization methods such as ant colony optimization (ACO) or Dynamic programming (DP), which are particularly effective for specific constraints or static scenarios. The second subproblem then employs MADRL-based approaches to handle dynamic and multi-agent interactions, building upon the solutions obtained from traditional algorithms. This hybrid approach combines the strengths of classical optimization and RL, achieving superior performance in highly dynamic and complex environments. Table \ref{MADRL} provides a comprehensive overview of recent MADRL approaches, categorizing them according to agent homogeneity or heterogeneity, reward consistency, and action constraints, highlighting the flexibility of MADRL to address various challenges in UAV-assisted communication networks. As action constraints have been thoroughly examined in the preceding sections, this section focuses primarily on analyzing reward mechanisms and their implications for system optimization.

\begin{table*}[ht]
\centering
\caption{MADRL Comparison}
\label{MADRL}
\resizebox{\textwidth}{!}{%
\begin{tabular}{|c|c|c|c|c|c|c|c|}
\hline
\multirow{2}{*}{\textbf{Work}} & \multicolumn{2}{|c|}{\textbf{Algorithm}} & \multirow{2}{*}{\textbf{Agent Type}} & \multirow{2}{*}{\textbf{Rewards}} & \multicolumn{3}{|c|}{\textbf{Action Constraints}} \\ \cline{2-3} \cline{6-8} 
                               & \textbf{DRL} & \textbf{Traditional}     &                                                           &                                   & \textbf{Direct Range} & \textbf{0-1 Combination} & \textbf{Others} \\ \hline
\cite{kim2024cooperative}      & CommNet-based MADRL & / & Homogeneous                          & Inconsistent                        & NS                                & NM       & NM               \\ \hline
\cite{doi:10.1142/S0218126625501129} & MAPPO  & /        & Homogeneous                           & Inconsistent                      & NS                                & /                     & NM   \\ \hline
\cite{fi16070245}              & MAPPO        & /         & Homogeneous                          &  Inconsistent                      & NS                                & /                     & NM   \\ \hline
\cite{10559211}                & TPCD-ICAGC (Actor-attention-critic)   & / & Heterogeneous                  & Inconsistent                        & PR                                & /                     & NM   \\ \hline
\cite{10198525}                & b-MAPPO      & /           & Heterogeneous                          & Inconsistent                      & NS                     & PR           & PR, NS   \\ \hline
\cite{10680055}                & MADQN       &  DM-ACO                 & Heterogeneous                       & Inconsistent                        & NS                            & PR, NS                     & /   \\ \hline
\cite{10312789}                & MA-DuelingDQN & DIM         & Homogeneous                        & Consistent                      & NS                                & /                     & NM   \\ \hline
\cite{10367769}                & UA-MADDPG & GBCRA         & Heterogeneous                     & Inconsistent                      & NS, PR                                & NS                     & NM   \\ \hline
\end{tabular}%
}
\vspace{0.5em}
\end{table*}

\subsection{MADRL-based Solutions} \label{sub-madrl}

\cite{kim2024cooperative} investigates dynamic and uncertain UAVs-assisted communication environments to optimize both energy efficiency and service reliability within three-dimensional network spaces. It proposes a MADRL framework based on CommNet architecture for optimizing UAV positioning and resource allocation. The state space design encompasses critical parameters including three-dimensional UAV positions, remaining energy levels, and service status indicators like UE connectivity and QoS metrics. 

The reward function is structured with two primary components: individual rewards and collective rewards. Individual rewards incentivize energy-efficient UAV operations and high-quality service delivery through the consideration of remaining energy levels and user QoS metrics, while collective rewards enhance system-wide reliability by maximizing user coverage rates and minimizing inter-UAV coverage overlap. The framework implements penalties for inefficient energy utilization and redundant coverage overlap to ensure system stability and optimal performance. The action space comprises seven discrete movements, including variations in x, y, and z coordinates and hovering capabilities, with action constraints directly embedded within the network architecture. Although the study does not explicitly address UAV collision avoidance mechanisms, it approaches this objective indirectly through a comprehensive state space design incorporating positional information of all UAVs and penalty components in the reward function for coverage overlap. However, this approach presents two significant limitations: the proposed design can not completely prevent UAV collisions, and the discrete action space implementation is inconsistent with the continuous action requirements of UAVs in real-world environments.

In contrast to \cite{kim2024cooperative}, both \cite{doi:10.1142/S0218126625501129} and \cite{fi16070245} implement continuous action spaces in their approaches. \cite{doi:10.1142/S0218126625501129} introduces an emergency scenario force model (ESFM) for post-disaster emergency communication scenarios. In this model, GUs navigate towards their destinations along optimal paths while considering obstacle avoidance, injury treatment events, and so on. Under this dynamic ESFM scenario, the authors employ the multi-MAPPO algorithm to determine UAV movements, aiming to maximize GU coverage. The study further evaluates the network performance trained with MAPPO using two distinct reward functions: one considering system fairness and another without fairness considerations. \cite{fi16070245} additionally incorporates optimization of UAV energy consumption. Both studies implement flight zone restrictions for UAVs within their network structures, with \cite{doi:10.1142/S0218126625501129} specifically incorporating penalties in the reward function for UAVs that are too close to each other.

Unlike \cite{kim2024cooperative}, \cite{doi:10.1142/S0218126625501129}, and \cite{fi16070245}, both \cite{10559211} and \cite{10198525} extend their consideration beyond UAV systems to encompass user systems, modeling UAVs and users as two heterogeneous types of agents. In \cite{10559211}, GUs primarily focus on network QoS, with their reward function designed as a function of throughput, while UAVs consider both throughput and system fairness. Furthermore, UAVs and GUs operate with distinct observation and action spaces. GUs only need to determine which UAV to connect to, whereas UAVs consider both their movement velocity and bandwidth allocation for GU service provision. Regarding constraints on UAV flight boundaries, inter-UAV distances, and UAV-GU distances, this study directly incorporates them as penalty terms within the reward function. A particularly noteworthy aspect is that due to the divergent objectives between UAVs and GUs, combined with the state complexity, the study introduces an attention architecture, forming an actor-attention-critic network architecture. This enhancement helps agents (either UAVs or GUs) focus on information from other agents relevant to their decision-making, thereby optimizing collaborative capabilities.

In \cite{10198525}, the authors address a task offloading problem in UAV-assisted MEC networks, where GUs with limited computational capabilities frequently need to offload tasks to edge clouds for auxiliary computing. The study aims to optimize energy consumption across the entire system comprising both UAVs and GUs. For GU agents, energy consumption occurs primarily in the computation and communication aspects, which constitute the main components of the GU reward function. Additionally, the study incorporates QoS considerations by introducing a penalty term in the reward function for tasks that fail to meet latency requirements. For UAVs, energy consumption predominantly involves computation and mobility. The UAV reward function design includes not only energy consumption but also incorporates three penalty terms: failure to meet GU latency requirements, insufficient inter-UAV spacing, and UAV flight beyond service area boundaries. Regarding action spaces, GUs must decide both which UAV to allocate tasks to and the quantity of tasks to allocate, while UAVs must determine their movement trajectories and CPU frequency allocation for processing different tasks. To address this complex problem involving both binary and continuous variables, the authors employ the b-MAPPO algorithm. They enhance the traditional MAPPO's Gaussian distribution output by implementing Beta distribution to improve boundary condition adaptability and policy flexibility.

\subsection{Hybrid MADRL Solutions: Combining Traditional Optimization}

In contrast to \ref{sub-madrl}, \cite{10680055} not only employs heterogeneous agent modeling but also decomposes the problem into two sub-problems during the modeling process. They first solve one sub-problem using traditional optimization methods, then incorporate its results along with the second sub-problem into a MADRL-based algorithm. \cite{10680055} investigates content caching optimization in autonomous vehicular networks supported by a non-terrestrial network comprising low earth orbit (LEO) satellites and UAVs. In dynamic and complex communication environments, the study addresses the optimization of caching strategies under heterogeneous cache hardware and frequent network dynamics to reduce system propagation delay and improve cache hit ratios. 

The authors propose a MADRL-based hierarchical caching and asynchronous update scheme. This approach combines a discrete multi-objective ant colony optimization (DM-ACO) algorithm for selecting appropriate caching satellites with MADRL for hierarchical caching decisions and asynchronous updates. In implementing the MADQN algorithm, UAVs and LEO satellites are modeled as two distinct types of heterogeneous agents, with their action spaces encompassing decisions between content caching and replacement. To simplify the process, the study implements an asynchronous update mechanism, splitting content switching into two steps: discarding existing content and caching new content. Regarding reward function design, the UAV-layer reward function encourages UAVs to prioritize caching content with high latency requirements and optimizes caching decisions based on the squared ratio of content transmission delay to threshold. The LEO layer, conversely, weights content access efficiency through cache hit ratio. It is worth noting that these reward functions are inconsistent with its objective function in the mathematical model.

Unlike \cite{10680055}, \cite{10312789} addresses the joint optimization of task offloading and UAV trajectory in UAVs-assisted MEC networks, with particular emphasis on dynamic demand and service fairness optimization. The study proposes a multi-agent energy-efficient joint trajectory and computation offloading scheme, which achieves joint optimization of trajectory control and task offloading decisions through an optimization-embedding MADRL (OMADRL) algorithm. It employs MA-Dueling DQN to learn UAV trajectory policies, while offloading decisions are determined through a mixed integer non-linear programming approach (called Dinkelbach iteration method). These offloading decisions subsequently guide trajectory control, thereby reducing action space dimensionality and enhancing convergence efficiency. The study models all UAVs as homogeneous agents, with a discrete action space comprising 8 directional options and 4 velocity levels to describe flight trajectories in three-dimensional space. Additionally, the authors develop a ``collision avoidance action adjustment mechanism`` to prevent UAVs from approaching too closely to each other. Regarding task offloading, user tasks are categorized into delay-sensitive and delay-tolerant tasks, with sensitive tasks executed by the nearest UAV while tolerant tasks can be relayed to BS. The reward function, aligned with the objective function, incorporates fairness weights, delay tolerance, and energy efficiency, incentivizing UAVs to cover hotspot areas and optimize task allocation. 

\cite{10367769} focuses on a MEC network with air-ground collaboration, specifically addressing task offloading for user vehicles (UVs) and resource allocation for MEC servers in ultra-reliable low latency communications (URLLC). The paper introduces an air-ground vehicular collaborative computing network model and employs Lyapunov optimization to decompose the original problem into two subproblems: transmission cost optimization and computational resource allocation optimization. Consequently, the authors propose a URLLC-aware MADDPG (UA-MADDPG) algorithm and a greedy resource allocation method. By utilizing UA-MADDPG to jointly optimize UAV trajectory, task offloading, and spectrum allocation, the approach effectively addresses uncertainties in global state information. The study considers all UAVs, UVs, and OPs as homogeneous agents, with agent actions encompassing task offloading decisions for UVs, UAV flight distance and direction decisions, and spectrum resource allocation decisions for OPs. The action space incorporates both continuous and discrete variables. The reward function simultaneously considers system transmission costs and URLLC queue stability constraints, with additional penalties imposed for scenarios.


\section{Challenges and Open Issues}
\label{Challenges and Open Issues}

While DRL has demonstrated significant potential in optimizing UAV communication networks, there remain several challenges that need to be addressed for its widespread implementation and further advancement. These challenges arise from the complexity of real-world environments, the computational demands of DRL algorithms, and the need for adaptable and scalable solutions.

\vspace{-10pt}
\subsection{High Dimensionality and Computational Complexity}
UAV networks typically involve large-scale \cite{A1}, dynamic environments with numerous interconnected nodes \cite{A2&B} and fluctuating network conditions. This complexity results in high-dimensional state and action spaces, which can challenge the efficiency of DRL algorithms. Training DRL models in such environments often requires significant computational resources and extended training times, making real-time applications difficult. Addressing this requires developing more computationally efficient algorithms and employing techniques such as function approximation, transfer learning, or model compression to reduce the overhead.

\vspace{-10pt}
\subsection{Scalability and Multi-agent Coordination}
As UAV networks grow in size, the need for scalable DRL algorithms becomes more pressing. MADRL approaches are required for coordination among multiple UAVs to optimize tasks like power allocation, resource allocation \cite{A2&B}, channel assignment, and task offloading. However, as the number of agents increases, so does the complexity of learning effective policies due to the growing action spaces and interactions between agents. Solutions such as decentralized learning or hierarchical DRL architectures are necessary to ensure that these systems remain scalable without overwhelming computational resources \cite{B&D}.

\vspace{-10pt}
\subsection{Modeling Realistic Environments and Uncertainty}
UAV communication networks operate in highly dynamic environments where factors such as mobility  \cite{C1&D}, signal interference, and changing weather conditions can significantly impact performance. Traditional DRL algorithms often assume stationary environments, which may not accurately reflect the real-world uncertainties faced by UAVs. Developing DRL methods that can robustly adapt to dynamic, non-stationary conditions remains a key challenge \cite{E}. Approaches such as meta-RL and robust DRL have the potential to improve adaptability but require further research and validation.

\vspace{-10pt}
\subsection{Reward Function Design and Convergence}
The success of DRL relies heavily on the design of appropriate reward functions that reflect the objectives of the UAV communication tasks. Poorly designed rewards can lead to suboptimal policies or even divergence during training. In complex UAV networks, where multiple objectives (such as minimizing latency, maximizing throughput, and ensuring energy efficiency) are involved, balancing conflicting rewards poses an additional challenge \cite{B&D,C1&D}. Research is needed to explore better multi-objective optimization techniques and methods for reward shaping to guide DRL agents toward optimal solutions efficiently.

\vspace{-10pt}
\subsection{Data Efficiency and Public Datasets}
DRL algorithms often require a large volume of data to learn effective policies, making data efficiency a key challenge. Collecting real-world UAV network data is costly and time-consuming, and simulated environments may not always accurately reflect real-world conditions. Improving sample efficiency through techniques like model-based DRL or transfer learning can help address this issue by reducing the data needed for training \cite{E}.
Furthermore, the lack of publicly available datasets specific to UAV communication scenarios limits the ability to benchmark and compare DRL solutions. Creating shared datasets that capture UAV-specific tasks like power allocation and caching optimization is essential to accelerate research in this field.

\vspace{-10pt}
\section{Conclusion}
This survey has examined how DRL can be applied to solve critical mathematical optimization models in UAV communications and networking. Traditional approaches struggle with the dynamic and complex nature of UAV networks, while DRL offers a more adaptable solution by learning to optimize over time. By focusing on system modeling and progressing to DRL implementation, we have outlined how key optimization problems, such as power allocation, channel assignment, and task offloading, can be framed as mathematical models and efficiently solved using DRL techniques. This progression from modeling to implementation serves as a practical guide for researchers and practitioners. Additionally, we have identified future challenges and opportunities for further advancing the application of DRL in UAV networks, particularly in terms of scalability and real-time adaptability. The potential in DRL to continuously learn and adapt to evolving network conditions makes it a crucial tool for future UAV communication systems.

\noindent





\bibliographystyle{IEEEtran}
\bibliography{IEEEabrv,reference}

\vspace{-45pt}
\begin{IEEEbiography}
[{\includegraphics[width=1in,height=1.25in,clip,keepaspectratio]{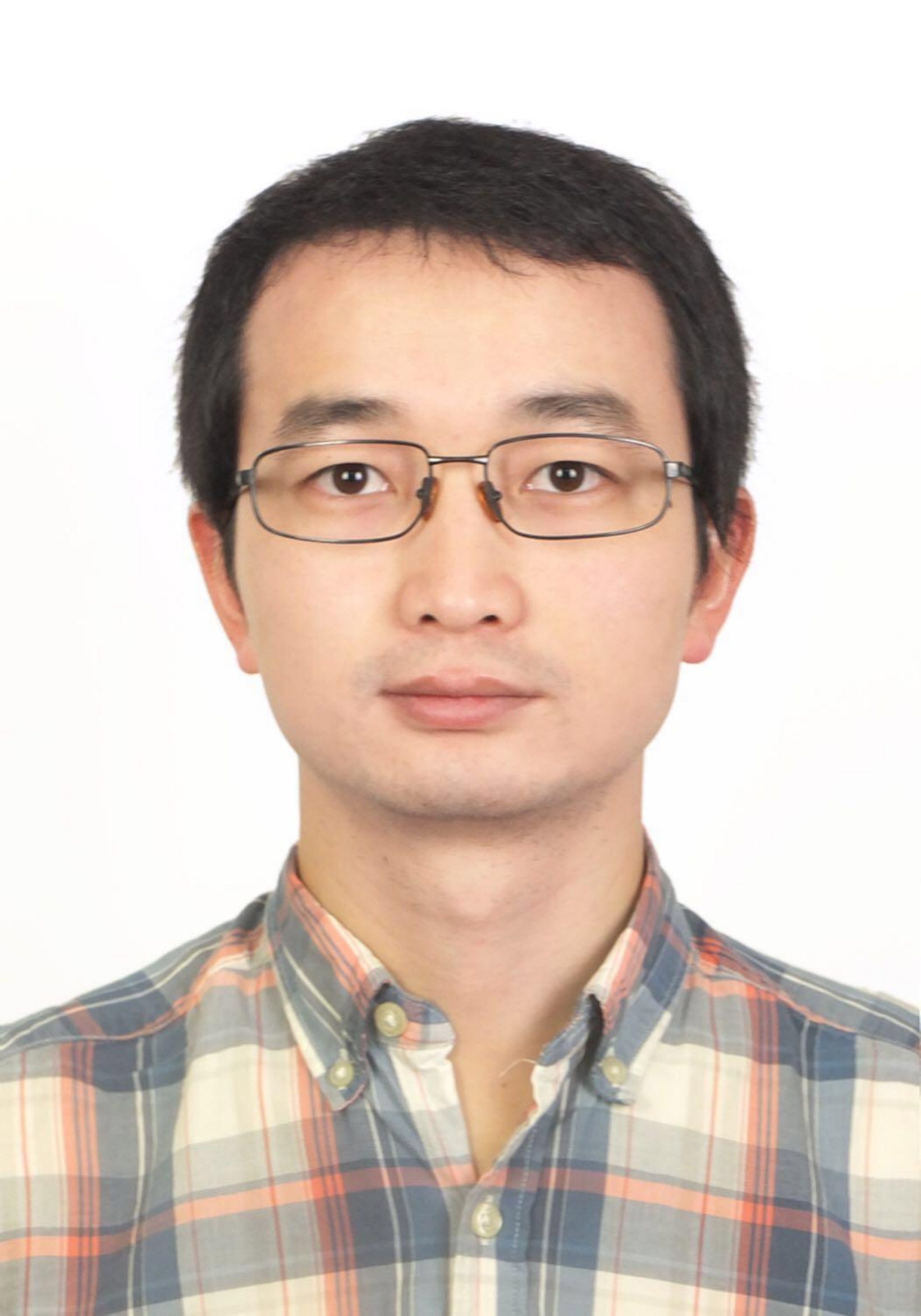}}]
{Wei Zhao} (S’12-M’16) received his Ph.D. degree in the Graduate School of Information Sciences, Tohoku University. He is currently an Associate Professor at the School of Computer Science and Technology, Anhui University of Technology. His research interests include deep reinforcement learning, edge computing, and resource allocation in wireless networks. He was the recipient of the IEEE WCSP-2014 Best Paper Award, and IEEE GLOBECOM-2014 Best Paper Award. He is a member of IEEE.
\end{IEEEbiography}

\vspace{-45pt}

\begin{IEEEbiography}
[{\includegraphics[width=1in,height=1.25in,clip,keepaspectratio]{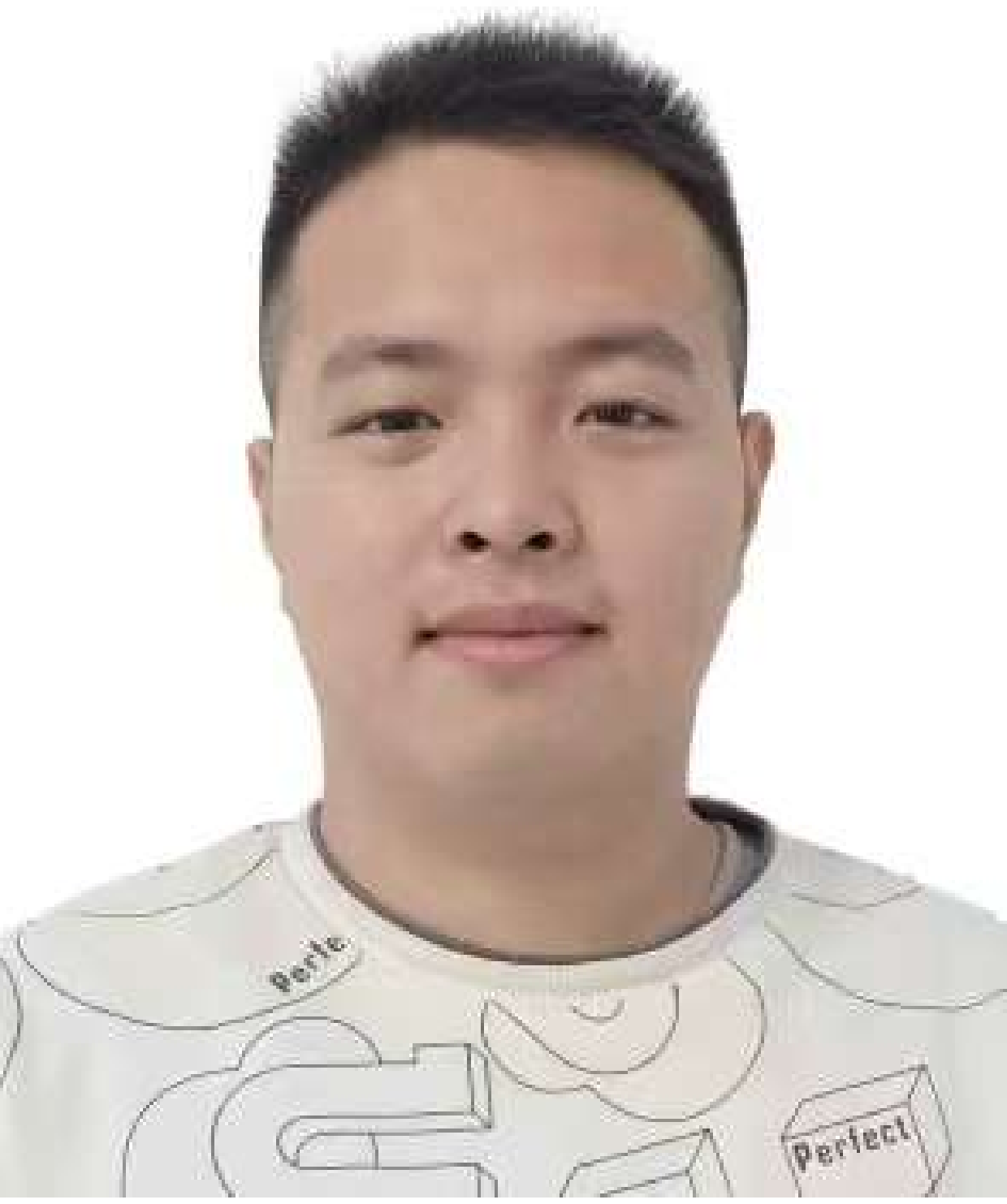}}]
{Shaoxin Cui} is currently pursuing a Master's degree in Engineering at Anhui University of Technology. His research interests center around deep reinforcement learning and edge computing, specifically in their application to unmanned aerial vehicles (UAVs). By leveraging advanced computing techniques, He aims to enhance UAV decision-making processes and overall performance for various applications.
\end{IEEEbiography}

\vspace{-45pt}

\begin{IEEEbiography}[{\includegraphics
[width=1in,height=1.25in,clip, keepaspectratio]{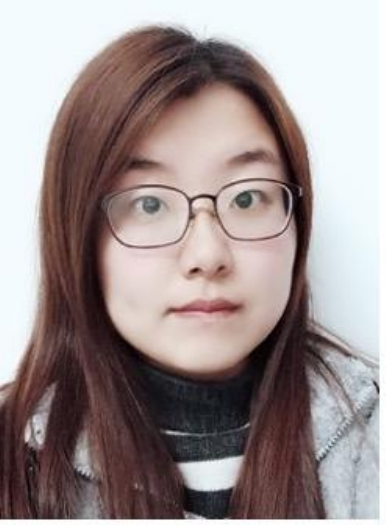}}]
{Wen Qiu} is currently a Research Assistant at Anhui University of Technology. She received her Ph.D. degree in Co-creative Engineering from Kitami Institute of Technology, Hokkaido, Japan. Her research interests encompass deep reinforcement learning and emergency wireless communication networks, focusing on developing intelligent networking solutions for disaster response scenarios. Her work specifically addresses challenges in network resilience, dynamic resource allocation, and system optimization using advanced machine learning techniques.
\end{IEEEbiography}

\vspace{-45pt}

\begin{IEEEbiography}[{\includegraphics
[width=1in,height=1.25in,clip,
keepaspectratio]{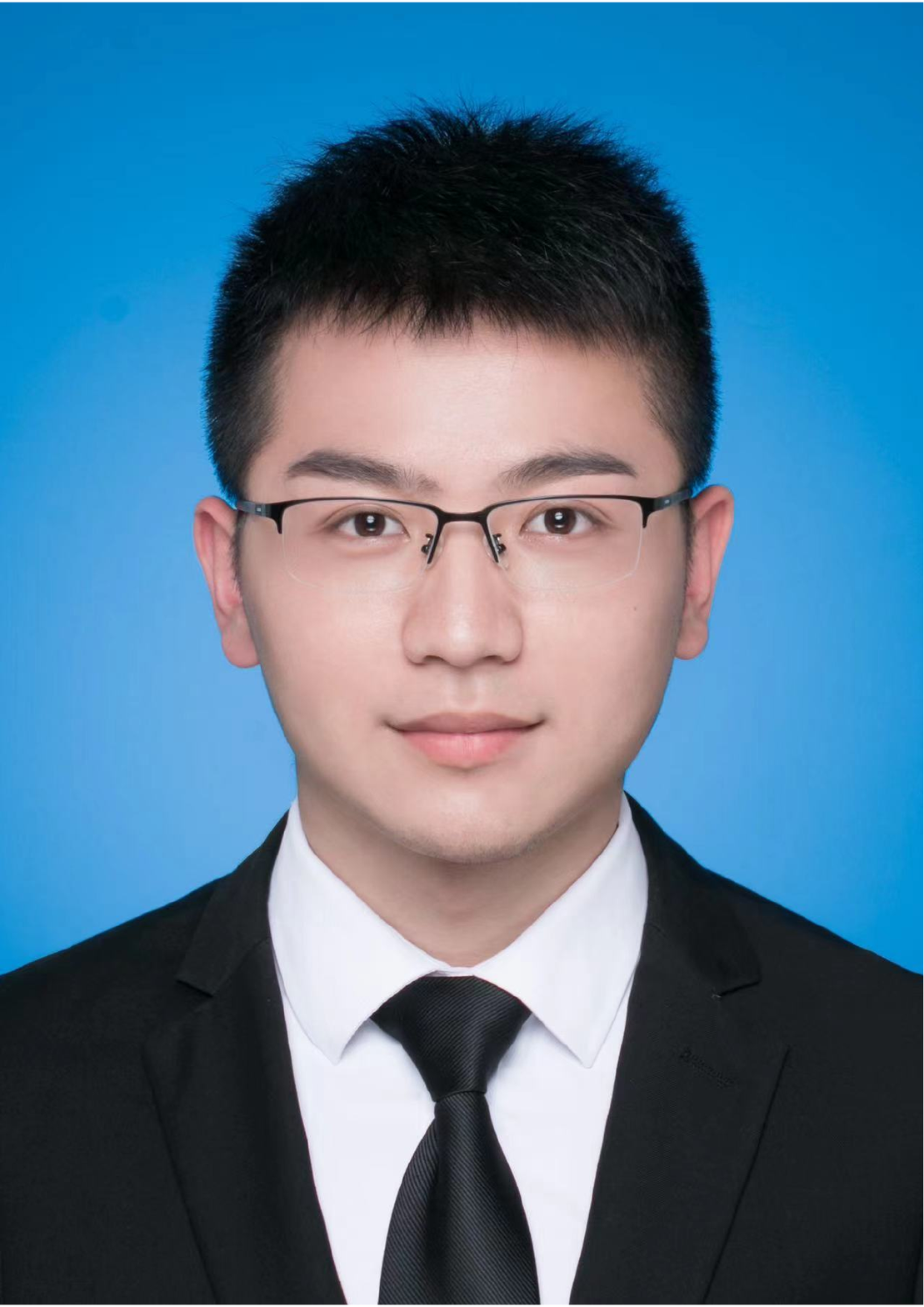}}]
{Zhiqiang He} is currently a Research Assistant at Anhui University of Technology. He received his MS degree in Control Science and Engineering from Northeastern University, Shenyang, China. His research interests focus on deep reinforcement learning and its control applications. He previously worked at Baidu and InspirAI, where he developed a master-level AI for the game Landlord that outperformed professional players.
\end{IEEEbiography}

\vspace{-45pt}

\begin{IEEEbiography}
[{\includegraphics[width=1in,height=1.25in,clip,keepaspectratio]{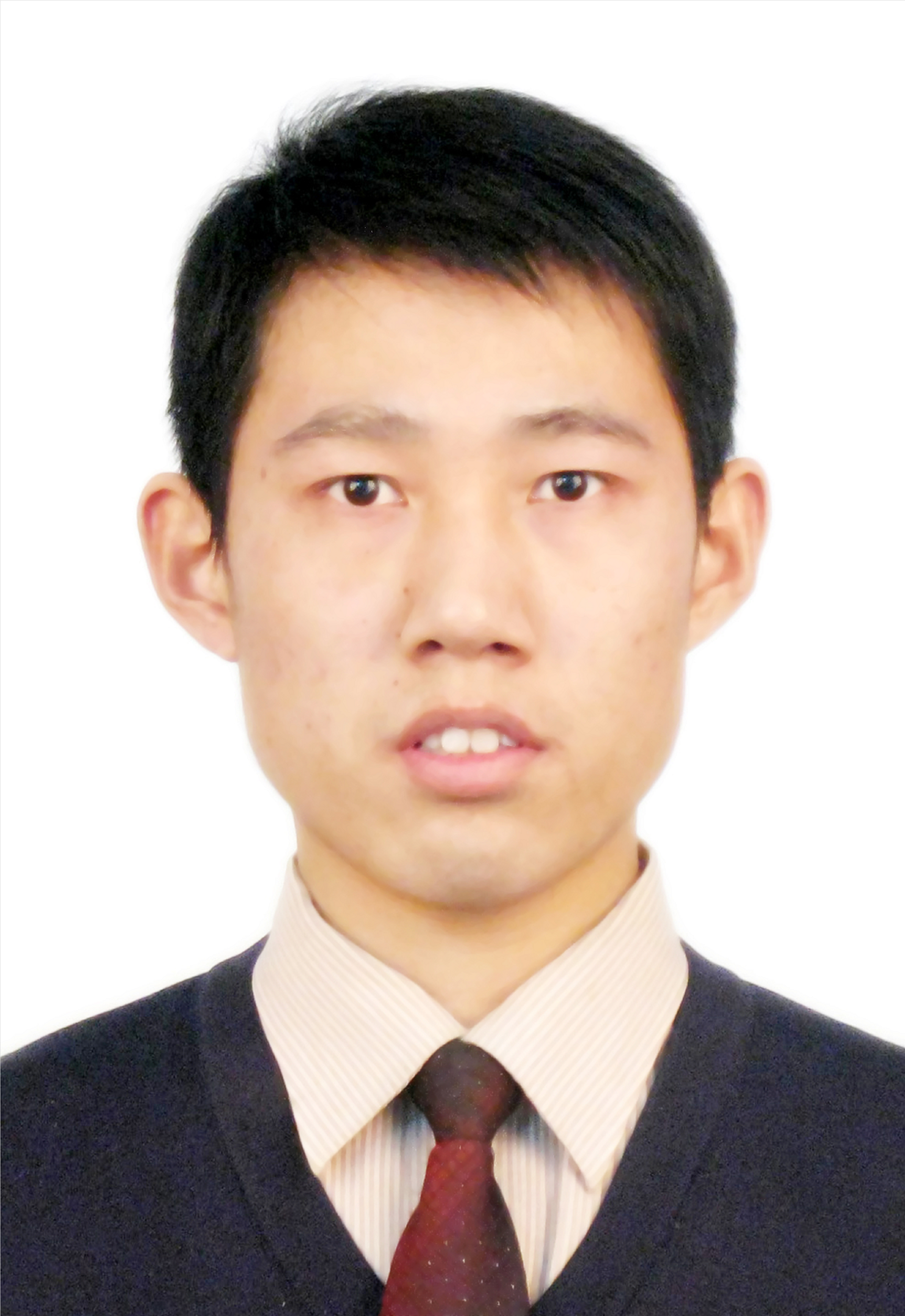}}]
{Zhi Liu} (S’11-M’14-SM’19) received the Ph.D. degree in informatics in National Institute of Informatics. He is currently an Associate Professor at the University of Electro-Communications. His research interest includes video network transmission and MEC. He is now an editorial board member of IEEE Transactions on Multimedia, IEEE Networks and Internet of Things Journal. He is a senior member of IEEE.
\end{IEEEbiography}

\vspace{-45pt}

\begin{IEEEbiography}
[{\includegraphics[width=1in,height=1.25in,clip,keepaspectratio]{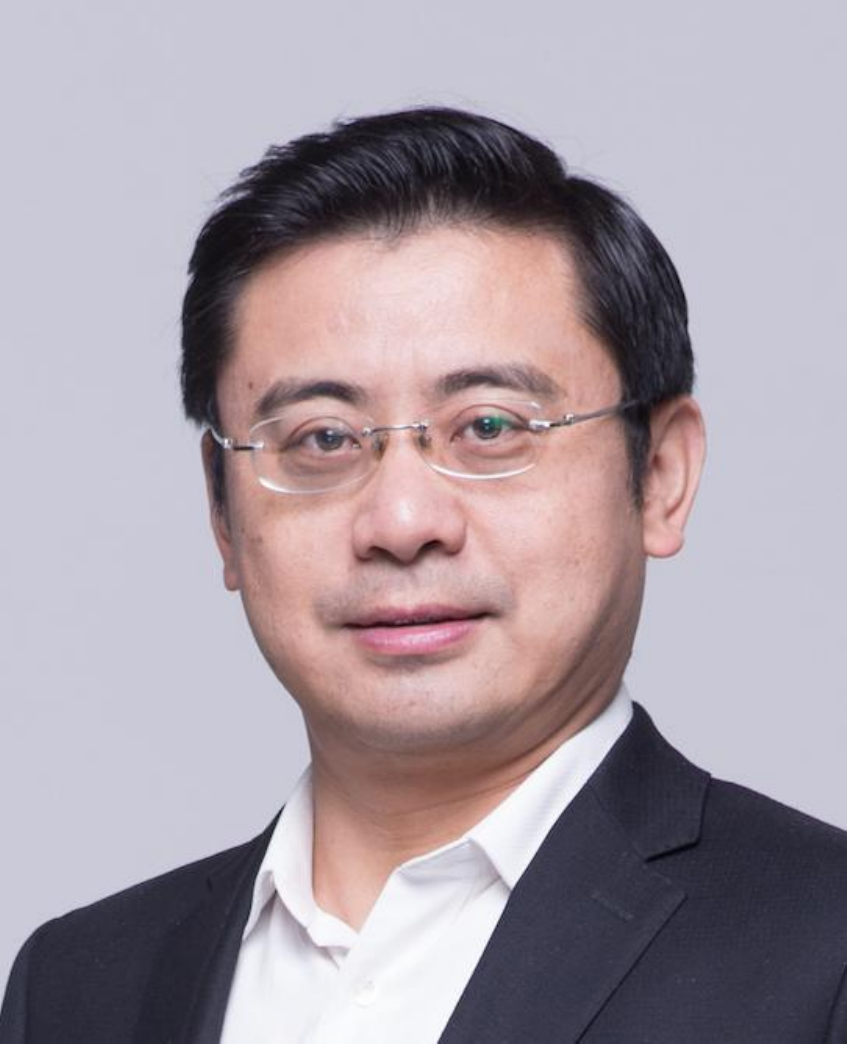}}]
{Xiao Zheng} received the Ph.D. degree in computer science and technology from Southeast University, China, in 2014. He is currently a professor with the School of Computer Science and Technology, Anhui University of Technology, Anhui, China. His research interests include service computing, mobile cloud computing and privacy protection. He has been a guest editor of IEICE Transactions on Communications. He is a senior member of CCF, and a member of the ACM.
\end{IEEEbiography}

\vspace{-45pt}

\begin{IEEEbiography}
[{\includegraphics[width=1in,clip,keepaspectratio]{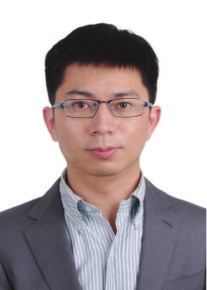}}]
{Bomin Mao} (Member, IEEE) is currently a Professor with the School of Cybersecurity, Northwestern Polytechnical University, China. His research interests are involving satellite networks, Internet of Things, edge computing, and vehicular networks. He received several Best Paper Awards from international conferences, namely IEEE GLOBECOM’17, GLOBECOM’18, IC-NIDC’18, ICC’23, and WOCC’23. He was a recipient of the prestigious IEEE COMSOC Asia Pacific Outstanding Paper Award (2020), Niwa Yasujiro Outstanding Paper Award (2019), and IEEE Computer Society Tokyo/Japan Joint Local Chapters Young Author Award (2020).
\end{IEEEbiography}

\vspace{-45pt}

\begin{IEEEbiography}
[{\includegraphics[width=1in,height=1.25in,clip,keepaspectratio]{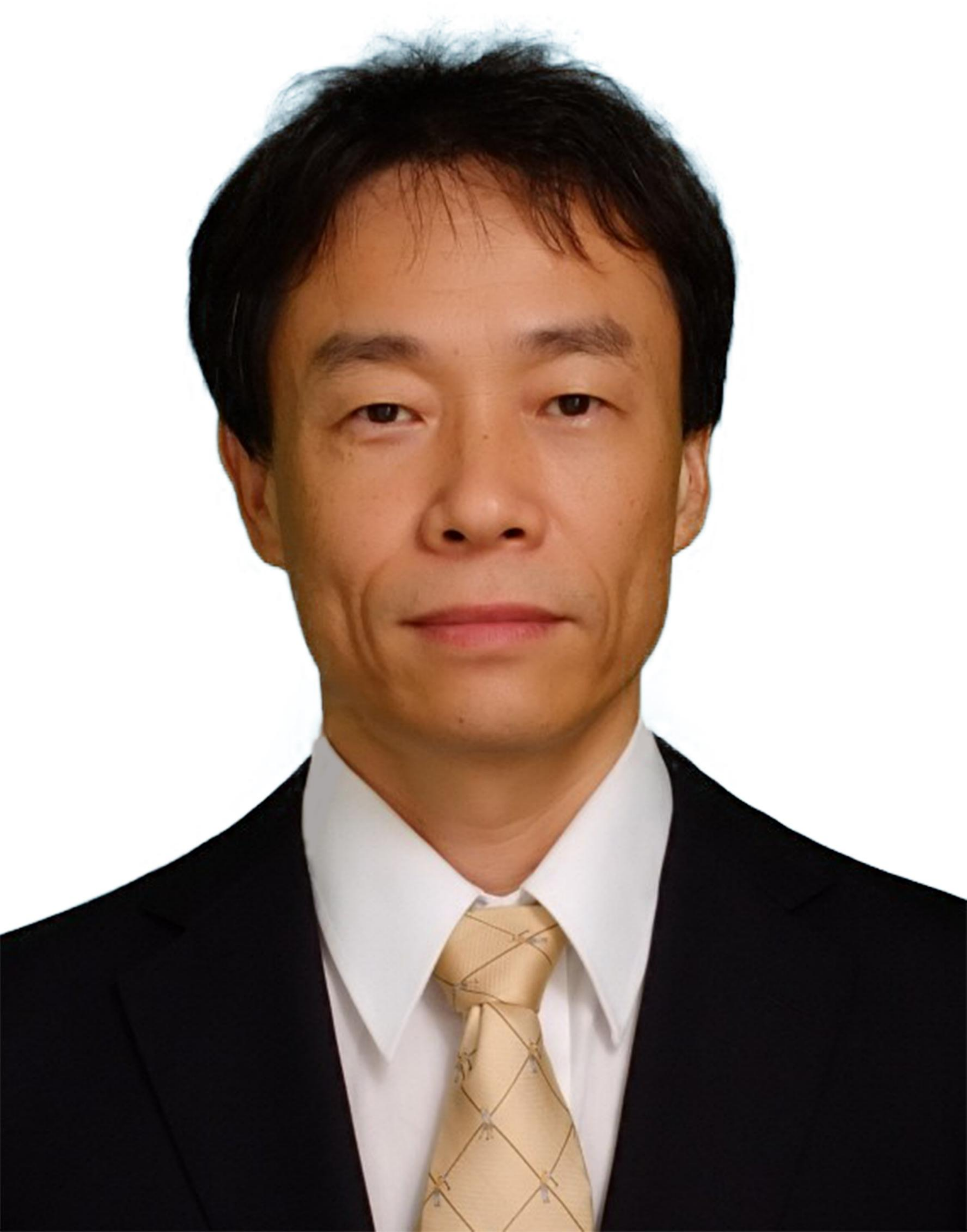}}]
{Nei Kato} (M’04, SM’05, F’13) is a full professor and the Dean of the Graduate School of Information Sciences, Tohoku University. He has been engaged in research on computer networking, wireless mobile communications, satellite communications, ad hoc and sensor and mesh networks, smart grid, AI, IoT, big data, and pattern recognition. He has published more than 500 papers in prestigious peer-reviewed journals and conferences. He was the Vice-President (Member \& Global Activities) of IEEE Communications Society (2018-2019), the Editor-in-Chief of IEEE Transactions on Vehicular Technology (2017-2020), and the Editor-in-Chief of IEEE Network (2015-2017). He is a Fellow of the Engineering Academy of Japan, IEEE and IEICE.
\end{IEEEbiography}

\end{document}